\documentclass[%
reprint,
amsmath,amssymb,
aps,
prb,
]{revtex4-2}
\usepackage{graphicx}
\usepackage{dcolumn}
\usepackage{bm}
\usepackage{hyperref}
\usepackage{xcolor}

\begin{document}
\title{Semiclassical theory for plasmons in two-dimensional inhomogeneous media}

\author{T. M. Koskamp}
\email{tjacco.koskamp@ru.nl}
\author{M. I. Katsnelson}
\email{m.katsnelson@science.ru.nl}
\author{K. J. A. Reijnders}
\email{k.reijnders@science.ru.nl}
\affiliation{
  Radboud University, Institute for Molecules and Materials, Heyendaalseweg 135, 6525 AJ Nijmegen, The Netherlands}
\date{\today}

\begin{abstract}
  The progress in two-dimensional materials has led to rapid experimental developments in quantum plasmonics, where light is manipulated using plasmons.
  Although numerical methods can be used to quantitatively describe plasmons in spatially inhomogeneous systems, they are limited to relatively small setups.
  Here, we present a novel semi-analytical method to describe plasmons in two-dimensional inhomogeneous media within the framework of the Random Phase Approximation~(RPA).
  Our approach is based on the semiclassical approximation, which is formally applicable when the length scale of the inhomogeneity is much larger than the plasmon wavelength.
  We obtain an effective classical Hamiltonian for quantum plasmons by first separating the in-plane and out-of-plane degrees of freedom and subsequently employing the semiclassical Ansatz for the electrostatic plasmon potential.
  We illustrate this general theory by considering scattering of plasmons by radially symmetric inhomogeneities. We derive a semiclassical expression for the differential scattering cross section and compute its numerical values for a specific model of the inhomogeneity.
\end{abstract}

\maketitle

\section{Introduction}\label{sec:introduction}

Plasmons~\cite{Vonsovsky89,Giuliani05,Platzman73,Nozieres99}, quantized collective oscillations of conduction electrons in metals or semiconductors, 
were theoretically predicted by Bohm and Pines~\cite{Pines52}.
One of their key properties is that they can strongly interact with light.
By transforming photons into plasmons and vice versa, one creates the possibility to manipulate light using heterostructures of nanometer size~\cite{Barnes03}. 
In the past, these plasmonic devices operated in the classical regime, where the plasmon wavelength is much larger than the Fermi wavelength of the electrons.
Moreover, these devices are traditionally based on the manipulation of surface plasmons, which are localized excitations on the surface of a three-dimensional material~\cite{Tame13,Fitzgerald16}.

Both of these aspects have essentially changed in recent years. First, the quantum regime in plasmonics has been reached~\cite{Scholl12,Tame13,Fitzgerald16}.
When characterizing this regime, it is often sufficient to take only the quantum mechanical nature of the electrons in the material into account.
The electric field can then still be treated classically, i.e., through Maxwell's equations.

Second, the discovery of two-dimensional materials had a tremendous impact on the field of plasmonics~\cite{Grigorenko12,Basov16,Low17}.
From an experimental point of view, engineering two-dimensional materials, e.g. via nanopatterning or nanostructuring, is much easier than three-dimensional materials. 
Moreover, one can combine different two-dimensional crystals into heterostructures, creating a tunable and versatile platform for experiments~\cite{Geim13}.
An added advantage is that in two dimensions the amplitude of the plasmonic electric field can be probed in real space, which is much more difficult in three dimensions.

From a theoretical point of view, plasmons in two-dimensional materials behave differently from their counterparts in three-dimensional materials~\cite{Giuliani05}.
The latter have a gapped energy spectrum, and their energy at zero momentum is given by the plasma frequency.
Plasmons in two-dimensional materials, on the other hand, have a gapless spectrum that goes to zero as the square root of the momentum at low energies.
A direct consequence is that these plasmons can be created and measured using infrared optics~\cite{Fei12,Chen12}.

Plasmonic devices that are interesting for applications, including waveguides~\cite{Tame13,Rosner16,Jiang21} and photodetectors~\cite{Grigorenko12}, are necessarily spatially inhomogeneous.
On a theoretical level, one can model this inhomogeneity using a spatially varying local charge density or background dielectric constant.
Within the Random Phase Approximation (RPA), one can then write down a set of equations for the plasmon modes in the system~\cite{Vonsovsky89}.
Unfortunately, the absence of translational invariance makes these equations hard to solve, even with numerical methods.
Nevertheless, this approach has been used to compute the plasmonic modes in fractals~\cite{Westerhout18}, waveguides~\cite{Jiang21} and supercells of twisted bilayer graphene~\cite{Westerhout21}.
However, these numerical methods are still limited to relatively small systems. 
In order to gain better understanding of the underlying physics, it is in any case useful to combine numerical methods with (semi-)analytical approaches.

Recently, a new analytical method~\cite{Reijnders22} was proposed to study the RPA equations for bulk plasmon modes in three-dimensional systems.
It employs the semiclassical approximation, which is valid when the wavelength of the plasmons is much smaller than the spatial scale of changes in the charge density and the background dielectric constant.
The origin of this approach can be traced back to Refs.~\cite{Ishmukhametov71,Ishmukhametov75}, where heuristic arguments based on the Wentzel-Kramers-Brillouin (WKB) approximation were used.
The semiclassical theory presented in Ref.~\cite{Reijnders22} is mathematically fully consistent, and makes extensive use of the correspondence between quantum mechanical operators and classical observables on phase space~\cite{Maslov81,Guillemin77,Martinez02,Zworski12}.
Within this approach, an effective classical Hamiltonian for bulk plasmons in three-dimensional inhomogeneous media was derived, and the plasmonic bound states in waveguides and spherical nanoparticles were studied with the semiclassical quantization procedure.

In the present paper, we extend the semiclassical approach of Ref.~\cite{Reijnders22} to plasmons in two-dimensional inhomogeneous media.
We consider the semiclassical approximation an ideal tool to study these systems. First of all, it gives quantitative results in its regime of validity, that is, when the plasmon wavelength is much smaller than the spatial scale of changes in the system parameters. These outcomes can be compared with experimental results. We note that a direct comparison with numerical calculations is very complicated, since these are limited to relatively small systems, whereas our theory can provide quantitative understanding of larger systems.
Second, the semiclassical approximation provides a deeper understanding of the underlying physics. These physical principles can often be applied well outside the formal regime of validity and can be used to qualitatively explain both numerical and experimental results. Even the quantitative predictions may still have some value outside the regime of applicability. For the Schr\"odinger equation one can e.g. think of the harmonic oscillator and the hydrogen atom, for which the semiclassical results coincide with the exact result~\cite{Griffiths05,Heading62}.

At this point, let us more carefully explain what we mean with the term semiclassical approximation in this article.
In conventional semiclassical approaches, e.g. for the Schr\"odinger equation~\cite{Griffiths05,Heading62}, one connects a quantum mechanical theory to a fully classical theory. In the case of plasmons, this would mean connecting the RPA equations for a quantum plasma with a theory based on the Vlasov-Poisson system. In the latter theory, the electron distribution function is given by the Maxwell distribution instead of the Fermi-Dirac distribution.
The connection between the RPA equations and the Vlasov-Poisson system has been well studied, and can be formalized through the so-called Wigner function~\cite{Benedikter22}. In this formalism, the relevant semiclassical parameter is the ratio between the Fermi wavelength of the electrons and the plasmon wavelength.

In this paper, we consider a different type of asymptotic expansion, 
in which we connect the RPA equations for a spatially inhomogeneous system to Lindhard theory with spatially varying parameters.
In other words, the role of the ``classical theory'' is played by Lindhard theory with a position-dependent Fermi momentum and background dielectric constant.
To understand this in a more intuitive way, let us make an analogy with the transition from wave optics to geometrical optics: 
we construct a solution to the RPA equations (wave optics) by starting from classical trajectories (light rays in geometrical optics) that are determined by the Lindhard function with spatially varying parameters. On top of these classical trajectories, we then add the wave-like behavior (which causes interference).
More formally, we establish these relations using the conventional semiclassical Ansatz. The relevant dimensionless parameter in this asymptotic expansion is the ratio between the plasmon wavelength and the spatial scale of changes in the system parameters, which we require to be much smaller than one. It is in this sense that we call our theory semiclassical: we construct an asymptotic solution by first computing the trajectories from an effective classical Hamiltonian that describes the motion of quantum plasmons in classical phase space and afterwards add the wave-like character on top of these trajectories.

From a technical point of view, there is a key difference between studying two-dimensional and three-dimensional systems with this approach.
In the latter case, both the charge density and the electrostatic potential are three-dimensional quantities. For two-dimensional systems, the electrostatic potential is still three dimensional, but the charge density is (effectively) two dimensional. 
Throughout this paper, we keep applications to van der Waals materials in mind~\cite{Basov16,Low17}, which are purely two-dimensional as they consist of a single layer.
We therefore consider electrons that are confined to a plane and use a purely two-dimensional charge density. In reality, the charge density always has a finite extension in the out-of-plane direction. Numerical calculations have shown that this extension is on the order of a few interatomic distances~\cite{Marzari12,Wehling11}. Since this is the smallest length scale in the system, a purely two-dimensional charge density is a good first approximation. One of the most characteristic features of two-dimensional plasmons, namely their square root dispersion at low energies, has also been observed in experiments~\cite{Fei11, Liu08}.

In our semiclassical derivation, we make a separation in slow and fast variables.
We assume that the in-plane variables are slow, meaning that the spatial scale of changes in the charge density and background dielectric constant is much larger than the plasmon wavelength. On the contrary, we just established that the out-of-plane variable is fast, in the sense that the charge density changes abruptly at the boundary of the charge layer.
This separation in slow and fast variables allows us to asymptotically separate the in-plane and out-of-plane degrees of freedom using a generalization of the adiabatic Born-Oppenheimer approximation. More formally, we employ a modification of the operator separation of variables method~\cite{Berlyand87,Belov06}, before using the semiclassical Ansatz.

Our first result is an effective classical Hamiltonian for quantum plasmons in two-dimensional inhomogeneous systems, which describes their classical dynamics in phase space. This classical Hamiltonian turns out to be proportional to the Lindhard expression for the two-dimensional dielectric function with a coordinate dependent Fermi momentum and background dielectric constant.
Furthermore, we obtain a semiclassical expression for the full electrostatic potential, which reveals the wave-like character of the plasmons. 
We also discuss the energy density of the electromagnetic field. Until this point, our considerations are completely general.

We then proceed to illustrate our theory by considering scattering of plasmons by a radially symmetric inhomogeneity in the local charge density, which can for instance be caused by defects or by local gating created for example by a tip. 
We first review how the scattering cross section, which can be measured in experiments, can be expressed in terms of the phase shift. We then derive a semiclassical expression for this phase shift from our result for the electrostatic potential.
Since we make extensive use of the concept of a Lagrangian manifold~\cite{Maslov81,Guillemin77,Arnold82,Reijnders18,Dobrokhotov03} in this derivation, it can also be viewed as a specific application of the abstract formalism of the Maslov canonical operator~\cite{Maslov81} to a practical physics problem.
After obtaining an expression for the scattering cross section, we show its numerical values for plasmon scattering by a Gaussian bump or well in the charge density distribution of a metallic system. 
We compare the outcome with the classical trajectories, and highlight the role of interference between different trajectories.

Our paper is organized as follows. Section~\ref{sec:derivation} contains the derivation of the semiclassical formalism.
To make our article more self-contained, we first review the most relevant results from Ref.~\cite{Reijnders22} in Sec.~\ref{subsec:derivation-EOM-first-steps}. In Sec.~\ref{subsec:derivation-separation}, we then discuss the separation of the in-plane and out-of-plane degrees of freedom. We subsequently apply the semiclassical Ansatz in Sec.~\ref{subsec:derivation-SC-Ansatz}, and obtain the effective classical Hamiltonian~(\ref{eq:effclassicalHam}). As this separation of degrees of freedom is somewhat mathematically involved, we also present a more straightforward, though less elegant, approach in appendix~\ref{ap:bruteforce}.
In Sec.~\ref{subsec:derivation-amplitude}, we obtain an expression for the potential, to which we give a physical interpretation by computing the energy density of the electromagnetic field in Sec.~\ref{subsec:EnergyDensity}. Finally, we discuss the formal applicability of the semiclassical approximation in terms of dimensionless parameters in Sec.~\ref{subsec:derivation-applicability}.
We develop our scattering theory for plasmons in Sec.~\ref{sec:scattering}. We first review the relation between the differential cross section and the phase shift in Sec.~\ref{subsec:crossec}, and subsequently obtain a semiclassical expression for this phase shift in Sec.~\ref{subsec:scatphase}. In Sec.~\ref{subsec:scattdiff}, we recast this expression into a form that is more suitable for numerical computations. 
We discuss our numerical implementation in Sec.~\ref{sec:numerical}, and show the total and differential scattering cross sections for parameters indicative of a metallic system.
Finally, we present our conclusions in Sec.~\ref{sec:conclusion}, together with possible directions for future research.

We finish this introduction with a remark on our notation. In the plasmonics literature, the letter $\mathbf{q}$ is conventionally used for the plasmon wavevector. However, within the semiclassical approximation, one considers the momentum rather than the wavevector. Throughout this article, the letter $\mathbf{q}$ will therefore be used to denote the plasmon momentum, whilst the letter $\mathbf{p}$ is used to denote the electron momentum in the expression for the polarization.

\section{Derivation of the formalism}
\label{sec:derivation}

In this section, we derive our semiclassical formalism for a two-dimensional inhomogeneous electron gas, assuming a parabolic energy spectrum for the conduction electrons. As in Ref.~\cite{Reijnders22}, we use the equations of motion approach~\cite{Vonsovsky89}, which consists of three steps. 
In the first step, we consider the electrons, which are confined to two spatial dimensions $\mathbf{x}=(x,y)$.
We use the Liouville-von Neumann equation to establish a relation between the one-particle density operator $\hat{\rho}$ and the induced potential $V_\mathrm{pl}(\mathbf{x},t)$ in the plane. In the second step, we compute the induced electron density $n(\mathbf{x},t)$ using this density operator. 
In the third step, we consider the electric field, which is a three-dimensional quantity that also pervades the out-of-plane dimension $z$. We relate the time-dependent induced electron density to the three-dimensional potential $V(\mathbf{x},z,t)$ through the Poisson equation. 
Finally, we apply a self-consistency condition: the potential $V(\mathbf{x},z,t)$ should equal the in-plane induced potential  $V_\mathrm{pl}(\mathbf{x},t)$ at $z=0$. We remark that the latter condition is not needed for a three-dimensional charge density~\cite{Reijnders22}, as one directly obtains a self-consistent solution for the induced potential in that case.

In Sec.~\ref{subsec:derivation-EOM-first-steps}, we set the scene for our derivation, and review the first two steps of the procedure outlined above based on the results from Ref.~\cite{Reijnders22}. Section~\ref{subsec:derivation-separation} subsequently discusses how we can separate the in-plane degrees of freedom from the out-of-plane degree of freedom in the Poisson equation. We use a modified version of the operator separation of variables technique~\cite{Berlyand87,Belov06}, which allows us to perform the decoupling order by order in $\hbar$. This separation facilitates the subsequent application of the semiclassical approximation, which we consider in Sec.~\ref{subsec:derivation-SC-Ansatz}. We discuss the aforementioned self-consistency condition, and obtain an effective classical Hamiltonian for the plasmons. In Sec.~\ref{subsec:derivation-amplitude}, we solve the transport equation for the amplitude in the semiclassical Ansatz, and obtain an expression for the plasmon potential. We discuss its interpretation in Sec.~\ref{subsec:EnergyDensity}, where we compute the energy density of the electromagnetic field. Finally, we consider the applicability of the semiclassical approximation in Sec.~\ref{subsec:derivation-applicability}

\subsection{Density operator and induced electron density}
\label{subsec:derivation-EOM-first-steps}
Since the electrons in our two-dimensional material are confined to the two in-plane dimensions $\mathbf{x}=(x,y)$, 
their motion is governed by a two-dimensional Hamiltonian.
In the equations of motion approach~\cite{Vonsovsky89,Reijnders22}, we write this Hamiltonian as
\begin{align}  \label{eq:Hamiltonian-full}
  \hat{H} = \hat{H}_0 + V_\mathrm{pl}(\mathbf{x},t),
\end{align}
where the operator $\hat{H}_0$ describes the motion of the individual electrons and $V_\mathrm{pl}(\mathbf{x},t)$ is a scalar potential that expresses the electron-electron interaction within the system.
We remark that it is quite peculiar that we can capture the electron-electron interaction with a scalar potential $V_\mathrm{pl}(\mathbf{x},t)$, which we call the induced potential.
This is due to the nature of the RPA, in which you have a closed system of local equations for the single-particle density operator~\cite{Vonsovsky89,Giuliani05}.
Considering this system of equations is fully equivalent to the diagrammatic approach to the RPA, that is, the empty loop approximation for the polarization operator~\cite{Vonsovsky89,Giuliani05,Platzman73}.

We consider electrons with a quadratic dispersion that move in a spatially varying scalar potential $U(\mathbf{x})$, i.e.,
\begin{align}  \label{eq:Hamiltonian-bare}
  \hat{H}_0 = \frac{\hat{\mathbf{p}}^2}{2 m} + U(\mathbf{x}),  \hspace{1 cm}
  \hat{p}_x = - i \hbar \frac{\partial }{\partial x} ,
\end{align}
where $m$ is the effective electron mass of the system.
The potential $U(\mathbf{x})$ encodes the spatially varying electron density~$n^{(0)}(\mathbf{x})$.
Within the Thomas-Fermi approximation~\cite{Vonsovsky89,Giuliani05,Lieb81}, these quantities are related by
\begin{equation}  \label{eq:TF-approx}
  p_\mathrm{F}(\mathbf{x}) =  \hbar \left(\frac{4 \pi}{g_\mathrm{s}} n^{(0)}(\mathbf{x})\right)^{1/2}, \qquad U(\mathbf{x}) = \mu - \frac{p_\mathrm{F}^2(\mathbf{x})}{2 m},
\end{equation}
where $p_\mathrm{F}(\mathbf{x})$ is the position-dependent Fermi momentum, $\mu$ is the chemical potential of the system and $g_\mathrm{s}$ is the spin degeneracy.

We assume that the Hamiltonian $\hat{H}_0$ does not depend on time $t$, which allows us to write the induced potential as
\begin{align}  \label{eq:Vpl-time-dep}
  V_\mathrm{pl}(\mathbf{x},t) = V_\mathrm{pl}(\mathbf{x}) e^{-i E t / \hbar}.
\end{align}
Since we compute the retarded response function of the electrons, it would be more correct to write $E+i\eta$ and consider the limit $\eta \to 0^+$. However, we will implicitly assume this throughout the article.

In the remainder of this subsection, we summarize the main points of the derivation performed in Ref.~\cite{Reijnders22}, to which we refer for more detailed arguments.
The motion of the system of interacting electrons is described by the von Neumann equation for the density operator, i.e.
\begin{equation} \label{eq:von-Neumann}
  i \hbar \frac{\partial \hat{\rho}}{\partial t} = [ \hat{H} , \hat{\rho} ] .
\end{equation}
We can decompose the density operator as $\hat{\rho} = \hat{\rho}_0 + \hat{\rho}_1$, where $\hat{\rho}_0$ corresponds to the equilibrium situation originating from the Hamiltonian $\hat{H}_0$. The perturbation $\hat{\rho}_1$ to this equilibrium is caused by the electron-electron interaction, which implies that $\hat{\rho}_1$ should be of the same order of magnitude as $V_\mathrm{pl}$. Moreover, $\hat{\rho}_1$ should have the same time dependence as in Eq.~(\ref{eq:Vpl-time-dep}).

Since $V_\mathrm{pl}$ is assumed to be small, we can linearize the von Neumann equation~(\ref{eq:von-Neumann}). The operator $\hat{\rho}_0$ is time independent, because it corresponds to the equilibrium. The zeroth-order terms therefore give $[ \hat{H}_0 , \hat{\rho}_0 ] = 0$. The terms that are first order in $V_\mathrm{pl}$ lead to
\begin{equation}  \label{eq:von-Neumann-linearized}
  E \hat{\rho}_1 = [ \hat{H}_0, \hat{\rho}_1 ] + [ V_\mathrm{pl}, \hat{\rho}_0 ] ,
\end{equation}
where we have taken the time dependence of $\hat{\rho}_1$ into account.

At this point, we apply the semiclassical approximation. We construct an asymptotic solution for $V_\mathrm{pl}(\mathbf{x})$ in the form of the semiclassical Ansatz
\begin{align}  \label{eq:scansatz2d}
  V_{\mathrm{pl}}(\mathbf{x}) = \varphi(\mathbf{x},\hbar)  e^{i S(\mathbf{x})/ \hbar} ,
\end{align}
where $S(\mathbf{x})$ is the classical action and $\varphi(\mathbf{x},\hbar)$ is the amplitude. The latter
is a series in powers of $\hbar$, that is,
\begin{align}  \label{eq:sc-amplitude-expansion}
  \varphi(\mathbf{x},\hbar) = \varphi_0(\mathbf{x}) + \hbar \varphi_1(\mathbf{x}) + \mathcal{O} (\hbar^2).
\end{align}
In the semiclassical treatment of the Schr\"odinger equation~\cite{Griffiths05,Heading62}, the same Ansatz~(\ref{eq:scansatz2d}) is used for the wavefunction. In the latter context, the method is usually called the Wentzel-Kramers-Brillouin (WKB) approximation and consists of several steps.
First, one uses the semiclassical Ansatz to derive a Hamilton-Jacobi equation from the Schr\"odinger equation. 
One then extracts the classical Hamiltonian, which gives rise to classical trajectories through Hamilton's equations.
One finally uses the semiclassical Ansatz to add the wave-like character of the particles to these classical trajectories and to capture e.g. quantum interference.

Unfortunately, we cannot use the same semiclassical approach for the linearized Liouville-von Neumann equation~(\ref{eq:von-Neumann-linearized}), since it is an operator equation rather than a standard eigenvalue problem.
We therefore need a more advanced toolbox to derive an effective classical Hamiltonian, or, more fundamentally, to relate quantum operators on Hilbert space to classical observables on phase space. 
Very naively, one could think of quantum mechanical operators as functions of $\mathbf{x}$ and $\hat{\mathbf{p}}$ and replace all momentum operators $\hat{\mathbf{p}}$ by variables $\mathbf{p}$ to obtain classical observables on phase space. Whilst this approach gives the correct results to the lowest order in $\hbar$, it hardly seems like a well-defined mathematical procedure and gives wrong results for the higher-order corrections.

The most elegant way to express the relation between quantum mechanical operators on Hilbert space and functions on classical phase space $(\mathbf{x},\mathbf{p})$ is through the formalism of pseudodifferential operators~\cite{Maslov81,Martinez02,Zworski12}.
Within so-called standard quantization, one obtains a function $\sigma(\hat{a})=a(\mathbf{x},\mathbf{p},\hbar)$ on classical phase space from a quantum operator $\hat{a}$ with the formula
\begin{align}  \label{eq:standard-quantization-symbol}
  a(\mathbf{x},\mathbf{p},\hbar) = \sigma\left(\hat{a}\right) = e^{-i \left\langle\mathbf{p},\mathbf{x}\right\rangle/\hbar}\left(\hat{a} e^{i\left\langle\mathbf{p},\mathbf{x}\right\rangle/\hbar}\right) ,
\end{align}
where $\langle\mathbf{p},\mathbf{x}\rangle = \sum_j p_j x_j$ is the standard inner product on $\mathbb{R}^2$. The function $a(\mathbf{x},\mathbf{p},\hbar)$ is commonly called a symbol.
As an example, we may apply this formula to $\hat{H}_0$ in Eq.~(\ref{eq:Hamiltonian-bare}). We then find that $\sigma(\hat{H}_0) = H_0(\mathbf{x},\mathbf{p}) = \mathbf{p}^2/2 m + U(\mathbf{x})$, which is exactly what one obtains when replacing $\hat{\mathbf{p}}$ by $\mathbf{p}$ in $\hat{H}_0$.

Generalizing this previous example, we note that most of the symbols that we consider in this text are so-called classical symbols. These are symbols $a(\mathbf{x},\mathbf{p},\hbar)$ that have an asymptotic expansion in terms of $\hbar$~\cite{Martinez02}, i.e.,
\begin{equation}  \label{eq:symbol-classical-expansion}
  a(\mathbf{x},\mathbf{p},\hbar) = a_0(\mathbf{x},\mathbf{p}) + \hbar a_1(\mathbf{x},\mathbf{p}) + \mathcal{O}(\hbar^2) .
\end{equation}
One can think of the leading-order term $a_0(\mathbf{x},\mathbf{p})$ as the classical observable (on phase space) corresponding to the quantum operator $\hat{a}$. If one replaces the momentum operators $\hat{\mathbf{p}}$ by coordinates $\mathbf{p}$ in the operator $\hat{a}$, one precisely obtains $a_0$, which formalizes the very naive procedure that we sketched before.

Within standard quantization, Hermitian operators may have complex symbols. For instance, the operator $\tfrac{1}{2}(\mathbf{x}\cdot\hat{\mathbf{p}} + \hat{\mathbf{p}}\cdot\mathbf{x} )$ has symbol $\mathbf{x}\cdot\mathbf{p} - i\hbar$. 
One can show that the symbol of a Hermitian operator satisfies the relation
\begin{equation}  \label{eq:symbol-Hermitian-operator-condition}
  a_1(\mathbf{x},\mathbf{p}) = -\frac{i}{2}\sum_j \frac{\partial a_0}{\partial p_j \partial x_j} (\mathbf{x},\mathbf{p}) ,
\end{equation}
see e.g. Ref.~\cite{Reijnders22}. These complex symbols should therefore be seen as an artifact of the procedure, and by no means imply that we are dealing with non-Hermitian operators.
In order to avoid these complex symbols, one may use Weyl quantization~\cite{Martinez02,Zworski12}, in which the relation between operators and their symbols differs from Eq.~(\ref{eq:standard-quantization-symbol}) and Hermitian operators correspond to real symbols. 
Although both quantization schemes are formally equivalent, Weyl quantization typically makes the calculations more complicated. We therefore use standard quantization throughout this text.

Starting from a symbol $a(\mathbf{x},\mathbf{p},\hbar)$, one obtains the corresponding pseudodifferential operator $\hat{a}$, within standard quantization, with the Fourier transform $\mathcal{F}$, namely~\cite{Maslov81,Martinez02,Zworski12}
\begin{equation}  \label{eq:standard-quantization-fourier}
  \left( \hat{a} \, f \right) (\mathbf{x}) = \mathcal{F}^{-1}_{\mathbf{p} \to \mathbf{x}} a(\mathbf{x},\mathbf{p},\hbar) \mathcal{F}_{\mathbf{y} \to \mathbf{p}} f(\mathbf{y}) .
\end{equation}
The operator constructed in this way corresponds to the situation where the momentum operator always acts first, and the position operator acts second. For instance, quantization of $\mathbf{x}\cdot\mathbf{p}$ with this procedure gives $\mathbf{x}\cdot\hat{\mathbf{p}}$, which is not Hermitian. This is in accordance with our previous statement that a symbol should satisfy Eq.~(\ref{eq:symbol-Hermitian-operator-condition}) to give rise to a Hermitian operator.
Equations~(\ref{eq:standard-quantization-symbol}) and~(\ref{eq:standard-quantization-fourier}) establish a one-to-one relation between operators and symbols~\cite{Martinez02,Zworski12}.
For a general discussion of pseudodifferential operators and their symbols, we refer to Refs.~\cite{Maslov81,Martinez02,Zworski12}. A short overview in the context of the present article can be found in Ref.~\cite{Reijnders22}.

Let us now return to the linearized Liouville-von Neumann equation~(\ref{eq:von-Neumann-linearized}). 
In order to apply the semiclassical approximation to this operator equation, one also needs to employ a semiclassical Ansatz for the induced density operator $\hat{\rho}_1$.
This Ansatz is constructed in detail in Ref.~\cite{Reijnders22}, and its symbol is expressed in terms of the amplitude $\varphi(\mathbf{x})$ and the classical action $S(\mathbf{x})$, cf. Eq.~(\ref{eq:scansatz2d}), by solving the operator equation~(\ref{eq:von-Neumann-linearized}).
The result for $\hat{\rho}_1$ is subsequently used to compute the induced electron density, defined by
\begin{align}  \label{eq:induced-density}
  n(\mathbf{x}) = g_\mathrm{s} \text{Tr}(\delta(\mathbf{x}'-\mathbf{x})\hat{\rho}_1) .
\end{align}
After some lengthy calculations, that are performed explicitly in Ref.~\cite{Reijnders22}, one finds that the induced density $n(\mathbf{x})$ can be written as
\begin{align}\label{eq:2delecdens}
  n(\mathbf{x}) = \hat{\Pi} V_\mathrm{pl} (\mathbf{x}),
\end{align}
where $V_\mathrm{pl}(\mathbf{x})$ is given by the semiclassical Ansatz~(\ref{eq:scansatz2d}).
The pseudodifferential operator $\hat{\Pi}$ is the polarization operator.
Its symbol has an expansion in powers of $\hbar$, cf. Eq.~(\ref{eq:symbol-classical-expansion}), i.e.,
\begin{align}
  \sigma(\hat{\Pi}) = \Pi(\mathbf{x},\mathbf{q},\hbar) = \Pi_0(\mathbf{x},\mathbf{q}) + \hbar \Pi_1(\mathbf{x},\mathbf{q}) + \mathcal{O}(\hbar^2) ,
\end{align}
where the principal symbol $\Pi_0(\mathbf{x},\mathbf{q})$ depends on the position $\mathbf{x}$ and the (plasmon) momentum $\mathbf{q}$. It equals
\begin{align}  \label{eq:polarization-principal}
  \Pi_0 (\mathbf{x}, \mathbf{q}) \! = \! \frac{g_\mathrm{s}}{(2 \pi \hbar)^d} \! \int \! \frac{\rho_0\left(H_0(\mathbf{x},\mathbf{p})\right)-\rho_0\left(H_0(\mathbf{x},\mathbf{p}+\mathbf{q})\right)}{H_0(\mathbf{x},\mathbf{p})-H_0(\mathbf{x},\mathbf{p}+\mathbf{q}) + E} \mathrm{d} \mathbf{p}
\end{align}
where the function $\rho_0(z)$ is the Fermi-Dirac distribution. Its argument $H_0(\mathbf{x},\mathbf{p})$ is the principal symbol of the Hamiltonian $\hat{H}_0$, which we constructed before.

Expression~(\ref{eq:polarization-principal}) shows strong resemblance to the familiar Lindhard expression for the case of a homogeneous charge density~\cite{Vonsovsky89,Giuliani05}, the main difference being that the energy eigenvalue $E_\mathbf{p}$ is replaced by the symbol $H_0(\mathbf{x},\mathbf{p})$ of the Hamiltonian. The integral can therefore be evaluated in the same way, see e.g. Ref.~\cite{Giuliani05}. Performing this evaluation, one obtains the Lindhard expression with the replacement $p_\mathrm{F} \to p_\mathrm{F}(\mathbf{x})$, see also the discussion in Ref.~\cite{Reijnders22}.
The principal symbol $\Pi_0(\mathbf{x},\mathbf{q})$ is therefore a local polarization, which is a function of the spatial coordinate $\mathbf{x}$ and the (plasmon) momentum $\mathbf{q}$. Note that it actually only depends on the length of the momentum vector $|\mathbf{q}|$, since the Hamiltonian $\hat{H}_0$ is isotropic in the momentum operator.

The subprincipal symbol $\Pi_1(\mathbf{x},\mathbf{q})$ satisfies Eq.~(\ref{eq:symbol-Hermitian-operator-condition}), indicating that the polarization operator is Hermitian. Of course, the latter statement only holds in the region where Landau damping is absent and the principal symbol $\Pi_0$ is real-valued.

Adding the standard time dependence $\exp(-i E t/\hbar)$ to $n(\mathbf{x})$, we have thus obtained an expression for the induced electron density $n(\mathbf{x},t)$. Note that the dimensionality $d$ is arbitrary in the derivation, so the result~(\ref{eq:2delecdens}) is equally valid for two and three spatial dimensions.

\subsection{Separating in-plane and out-of-plane degrees of freedom in the Poisson equation}
\label{subsec:derivation-separation}

As we previously mentioned, the electric field $\boldsymbol{\mathcal{E}}$ is a three-dimensional quantity. The three-dimensional potential $V(\mathbf{x},z,t)$, which is related to the electric field by $\boldsymbol{\mathcal{E}} = e^{-1} \nabla V$, satisfies the Poisson equation
\begin{align}  \label{eq:Poisson}
  \langle \nabla , \varepsilon(\mathbf{x},z) \nabla \rangle V(\mathbf{x},z,t) = - 4 \pi e^2 n(\mathbf{x},z,t),
\end{align}
where the inner product is taken in three-dimensional space. Note that $V$ differs from the electrostatic potential by a factor $-e$.
Since we consider electrons that are confined to the plane $z=0$, the induced density can be written as
\begin{align}  \label{eq:3d-from2d-elecdens}
  n(\mathbf{x},z,t) = n(\mathbf{x},t)\delta(z),
\end{align}
where $n(\mathbf{x},t)$ only depends on the in-plane coordinates $\mathbf{x}$, see also expression~(\ref{eq:2delecdens}). We consider a setup where the plane $z=0$ is encapsulated by two dielectric media. We allow the background dielectric constant $\varepsilon (\mathbf{x},z)$ to depend on both $\mathbf{x}$ and $z$. The latter occurs when we have different dielectrics above and below the plane, or even multiple layers of dielectrics. A dependence on $\mathbf{x}$ can for instance originate from a combination of different dielectrics below the plane~\cite{Rosner16,Jiang21}.

We would like to solve Eq.~(\ref{eq:Poisson}) within our semiclassical framework, and obtain an asymptotic solution for $V(\mathbf{x},z,t)$. In order to ensure self-consistency, we demand that this solution satisfies the condition
\begin{equation}  \label{eq:condition-self-consistency}
  \left. V(\mathbf{x},z,t) \right|_{z=0} = V_\mathrm{pl}(\mathbf{x},t) .
\end{equation}
In other words, the full three-dimensional potential should be equal to the in-plane potential~(\ref{eq:scansatz2d}), in the form of the semiclassical Ansatz, in the plane $z=0$.
Note that this condition is not necessary in this form for a three-dimensional electron density, since we automatically have $V$ on both sides of the Poisson equation in that case~\cite{Reijnders22}. Since we consider a time-independent Hamiltonian $\hat{H}_0$, the time dependence has the standard exponential form, see also expression~(\ref{eq:Vpl-time-dep}), and we omit it from here on.

At this point, one can proceed in two different ways. 
In the first approach, we note that the self-consistency condition~(\ref{eq:condition-self-consistency}) implies that $V(\mathbf{x},z)$ is proportional to $\exp(i S(\mathbf{x})/\hbar)$.
We may therefore write down an Ansatz for $V(\mathbf{x},z)$ as
\begin{equation}\label{eq:VtotalExpansion}
  V(\mathbf{x},z) = e^{i S(\mathbf{x})/\hbar} \left( V_0(\mathbf{x},z) + \hbar V_1(\mathbf{x},z) + \mathcal{O}(\hbar^2) \right),
\end{equation}
and solve the Poisson equation order by order in $\hbar$. However, this approach mixes the separation of the in-plane and out-of-plane degrees of freedom with the application of the semiclassical Ansatz.

A more elegant approach, in which these two steps are separated, may be developed by modifying the operator separation of variables technique~\cite{Berlyand87,Belov06}.
In its original formulation, this technique can be seen as a generalization of the Born-Oppenheimer approximation~\cite{Belov06}. It can be regarded as an adiabatic approximation, and can therefore be used when there are two different length scales. 
This is exactly the case in our problem, since the delta function causes a fast change in the out-of-plane coordinate $z$, while the semiclassical limit asserts slow changes in the in-plane coordinates $\mathbf{x}$.
The method then separates the motion in the fast and slow variables order by order in the small parameter $\hbar$. In what follows, we proceed with this second method. For completeness, we show the results of the first approach in appendix~\ref{ap:bruteforce}.

In the operator separation of variables method, we start from the Ansatz~\cite{Belov06}
\begin{align}  \label{eq:opsepvar}
  V(\mathbf{x},z) = (\hat{\Gamma} V_\mathrm{pl})(\mathbf{x},z) ,
\end{align}
where $\hat{\Gamma}$ is a pseudodifferential operator that transforms the in-plane potential $V_{\mathrm{pl}}(\mathbf{x})$, which is independent of $z$, into the full three-dimensional potential $V(\mathbf{x},z)$.
In the traditional Born-Oppenheimer approximation, one would consider a function $\Gamma(\mathbf{x},z)$, instead of an operator $\hat{\Gamma}$. Here we have to use a full pseudodifferential operator, since $V_\mathrm{pl}$ is a rapidly oscillating exponential, as discussed in detail in Ref.~\cite{Belov06}.
Looking at the symbol $\sigma(\hat{\Gamma})$, which can be defined through expression~(\ref{eq:standard-quantization-symbol}) and depends on $\mathbf{x}$, $\mathbf{q}$ and $z$, we may say that we have added the momentum variable to the Born-Oppenheimer Ansatz. As we will see in section~\ref{subsec:derivation-SC-Ansatz}, this momentum variable will be replaced by $\partial S/\partial \mathbf{x}$ at the end of the calculation.
The symbol $\sigma(\hat{\Gamma})$ has an asymptotic expansion in powers of $\hbar$, given by
\begin{align}  \label{eq:chi-symbol-expansion}
  \sigma \big(\hat{\Gamma}\big) = \Gamma_0 (\mathbf{x},\mathbf{q},z) + \hbar \Gamma_1 (\mathbf{x},\mathbf{q},z) +\mathcal{O}(\hbar^2) ,
\end{align}
cf. Eq.~(\ref{eq:symbol-classical-expansion}).

Inserting the Ansatz~(\ref{eq:opsepvar}) into the Poisson equation~(\ref{eq:Poisson}), and taking relations~(\ref{eq:3d-from2d-elecdens}) and~(\ref{eq:2delecdens}) into account, we obtain
\begin{align}
  \left(\hat{F} \hat{\Gamma} + 4 \pi e^2 \delta(z) \hat{\Pi} \right) V_\mathrm{pl} (\mathbf{x}) = 0  ,
\end{align}
where $\hat{F} \equiv \langle \nabla , \varepsilon(\mathbf{x},z) \nabla \rangle$. 
Since we are looking for a plasmon, we require that $V_\mathrm{pl} (\mathbf{x}) \neq 0$. We therefore obtain an operator equation, to wit,
\begin{align}  \label{eq:operator-plasmon}
  \hat{F} \hat{\Gamma} + 4 \pi e^2 \delta(z) \hat{\Pi} = 0 .
\end{align}
In what follows, we solve this equation order by order in $\hbar$.

To this end, we first compute the symbol of the operator $\hat{F}$ in Eq.~(\ref{eq:operator-plasmon}), taking only the slow variables into account~\cite{Belov06}. In other words, we compute the symbol using expression~(\ref{eq:standard-quantization-symbol}), leaving the variable $z$ out of the consideration. In this way, we obtain the operator-valued symbol 
\begin{equation}
  \sigma(\hat{F}) 
  = F_0\left(\mathbf{x},\mathbf{q},z,\frac{\partial}{\partial z}\right) 
  + \hbar F_1\left(\mathbf{x},\mathbf{q},z,\frac{\partial}{\partial z}\right)
  + \mathcal{O}(\hbar^2) ,
\end{equation}
where
\begin{align} \label{eq:leading-ordersymbolODE}
  F_0\left(\mathbf{x},\mathbf{q},z,\frac{\partial}{\partial z}\right) & = -\frac{1}{\hbar^2} \varepsilon(\mathbf{x},z) |\mathbf{q}|^2 + \frac{\partial}{\partial z} \left(\varepsilon(\mathbf{x},z)\frac{\partial}{\partial z}\right),\\
  F_1\left(\mathbf{x},\mathbf{q},z,\frac{\partial}{\partial z}\right) & = \frac{i}{\hbar^2} \left\langle\mathbf{q},\frac{\partial \varepsilon}{\partial \mathbf{x}}(\mathbf{x},z)\right\rangle .
\end{align}
Note that the derivative with respect to $z$ is included in the principal symbol. Naively, one may say this is because the derivative gives rise to a factor $1/\hbar$, as we will see shortly. More fundamentally, we previously observed that changes in the out-of-plane coordinate $z$ are fast, whereas changes in the in-plane coordinates $\mathbf{x}$ are slow. When one introduces proper dimensionless units, as we do in Sec.~\ref{subsec:derivation-applicability}, one therefore sees that the combination $\hbar/z$ is of order one, cf. the discussion in Ref.~\cite{Belov06}.

We can now compute the symbol for all terms in Eq.~(\ref{eq:operator-plasmon}). Using the expression for the symbol of an operator product~\cite{Martinez02,Zworski12}, that is,
\begin{equation}
  \sigma(\hat{a}\hat{b}) = \sigma(\hat{a}) \sigma(\hat{b}) - i\hbar \left\langle \frac{\partial \sigma(\hat{a})}{\partial \mathbf{q}} , \frac{\partial \sigma(\hat{b})}{\partial \mathbf{x}} \right\rangle + \mathcal{O}(\hbar^2) ,
\end{equation}
and gathering all terms that are of the same order in $\hbar$, we obtain two equations.
The terms of order $\hbar^0$ give
\begin{align} \label{eq:opsepvarhbar0}
  F_0\left(\mathbf{x},\mathbf{q},z,\frac{\partial}{\partial z}\right) \Gamma_0(\mathbf{x},\mathbf{q},z)  = - 4 \pi e^2 \delta(z) \Pi_0 (\mathbf{x},\mathbf{q}),
\end{align}
whilst the terms of order $\hbar^1$ give
\begin{align} 
  &F_1\bigg(\mathbf{x},\mathbf{q},z,\frac{\partial}{\partial z}\bigg) \Gamma_0(\mathbf{x},\mathbf{q},z) + F_0\bigg(\mathbf{x},\mathbf{q},z,\frac{\partial}{\partial z}\bigg) \Gamma_1(\mathbf{x},\mathbf{q},z) \nonumber \\
  &\hspace*{1.3cm} - i \left\langle\frac{\partial F_0}{\partial \mathbf{q}} \bigg(\mathbf{x},\mathbf{q},z,\frac{\partial}{\partial z}\bigg),\frac{\partial \Gamma_0}{\partial \mathbf{x}} (\mathbf{x},\mathbf{q},z)\right\rangle \nonumber \\
  &\hspace*{3.0cm} = - 4\pi e^2 \delta(z) \Pi_1 (\mathbf{x},\mathbf{q})  .
  \label{eq:opsepvarhbar1}
\end{align}
These expressions constitute two linear ordinary differential equations (ODE's) (with variable $z$) for the principal symbol $\Gamma_0$ and the subprincipal symbol $\Gamma_1$.
Loosely speaking, one can think of the symbol $\sigma(\hat{\Gamma})(\mathbf{x},\mathbf{q},z)$ as a generalization of the instantaneous eigenfunction in the adiabatic approximation. As shown by Eqs.~(\ref{eq:opsepvarhbar0}),~(\ref{eq:opsepvarhbar1}), and~(\ref{eq:opsepvar}), the principal symbol $\Gamma_0$ and subprincipal symbol $\Gamma_1$ express the $z$ dependence of the full potential for given values of $\mathbf{x}$ and $\mathbf{q}$. In the remainder of this subsection, we solve Eqs.~(\ref{eq:opsepvarhbar0}) and~(\ref{eq:opsepvarhbar1}) one by one and thereby construct the symbol $\sigma(\hat{\Gamma})$.

\subsubsection{Principal symbol $\Gamma_0$}  \label{subsubsec:derivation-separation-principal}
We first solve Eq.~(\ref{eq:opsepvarhbar0}) to obtain an expression for  $\Gamma_0$. As we said before, we assume that our two-dimensional electron gas is located in the plane $z=0$. From here on, we assert that $\varepsilon(\mathbf{x},z)$ does not depend on $z$ above ($\mathrm{A}$) and below ($\mathrm{B}$) this plane. We therefore have
\begin{equation}\label{eq:varepsilon}
  \varepsilon(\mathbf{x},z) = \varepsilon_i(\mathbf{x}) =
  \begin{cases}
    \varepsilon_\mathrm{A}(\mathbf{x})	, & z>0 \\
    \varepsilon_\mathrm{B}(\mathbf{x})	, & z<0.
  \end{cases}
\end{equation}
The electron gas is characterized by the position-dependent Fermi momentum $p_\mathrm{F}(\mathbf{x})$, which is related to the density $n^{(0)}(\mathbf{x})$ through expression~(\ref{eq:TF-approx}).

We solve the differential equation~(\ref{eq:opsepvarhbar0}) for the two regions $z<0$ and $z>0$ separately and connect the solutions of the two regions through boundary conditions at $z=0$. That is, we consider
\begin{align}\label{eq:leadingorderODE}
  \left(-\frac{1}{\hbar^2} \varepsilon_i(\mathbf{x}) |\mathbf{q}|^2 + \varepsilon_i(\mathbf{x}) \frac{\partial^2}{\partial z^2} \right)\Gamma_0(\mathbf{x},\mathbf{q},z) = 0,
\end{align}
where the subscript $i=\mathrm{A}$ $(i=\mathrm{B})$ indicates the solution above (below) the plane $z=0$. Since the dielectric constants $\varepsilon_i(\mathbf{x})$ drop out, the solution has the same functional form in both regions. The general solution is given by
\begin{align}  \label{eq:chi0-sol-general}
  \Gamma_{0,i}(\mathbf{x},\mathbf{q},z) =  c_{0,i}^{-} e^{-\frac{|z|}{\hbar}|\mathbf{q}|} + c_{0,i}^{+} e^{\frac{|z|}{\hbar}|\mathbf{q}|}.
\end{align}
To shorten the notation we omit the argument of $\Gamma_0 (\mathbf{x},\mathbf{q},z)$ from here on, and assume it is kept the same unless stated otherwise.

We now consider the boundary conditions. First, $\Gamma_0$ should vanish as $|z|$ goes to infinity, since the potential $V(x,z)$ has to vanish. We therefore have $c_{0,i}^{+} = 0$. Second, the differential equation~(\ref{eq:opsepvarhbar0}) shows us that  $\Gamma_0$ should be continuous at $z=0$, that is,
\begin{align}\label{eq:bc-continuity-zero}
  \Gamma_{0}|_{z \to 0^+} & = \Gamma_{0}|_{z \to 0^-},
\end{align}
which yields
\begin{align}  \label{eq:bc-continuity-result}
  c_{0,\mathrm{A}}^{-} =  c_{0,\mathrm{B}}^{-} \equiv c_0.
\end{align}
The boundary condition~(\ref{eq:bc-continuity-zero}) is a reflection of the electrostatic boundary condition that the potential is continuous~\cite{Jackson99}.

We obtain our last boundary condition by integrating the differential equation~(\ref{eq:opsepvarhbar0}) from $z=-\epsilon$ to $z=\epsilon$, and letting $\epsilon \to 0$. Taking the continuity of $\Gamma_0$ into account, we find the relation
\begin{equation}  \label{eq:bc-D-field-zero}
  \left.\varepsilon_\mathrm{A}(\mathbf{x})\frac{\partial \Gamma_0}{\partial z}\right|_{z \to 0^+}
  - \left. \varepsilon_\mathrm{B}(\mathbf{x})\frac{\partial \Gamma_0}{\partial z}\right|_{z \to 0^-}
  = -4 \pi e^2  \Pi_0(\mathbf{x},\mathbf{q}) .
\end{equation}
Inserting the solution for each of the two regions gives
\begin{align}  \label{eq:bc-D-field-result}
  c_0 & = \frac{1}{\varepsilon_\mathrm{A}(\mathbf{x})+\varepsilon_\mathrm{B}(\mathbf{x})}\frac{4 \pi e^2 \hbar}{|\mathbf{q}|} \Pi_0(\mathbf{x},\mathbf{q}) .
\end{align}
The boundary condition~(\ref{eq:bc-D-field-zero}) is reminiscent of the electrostatic boundary condition that relates the discontinuity of the displacement field $\boldsymbol{ \mathcal{D}}$ to a surface charge. Indeed, our two-dimensional electron gas can be considered as a surface charge, since we consider an infinitesimal layer~\cite{Jackson99}.

Substitution of~(\ref{eq:bc-continuity-result}) and~(\ref{eq:bc-D-field-result}) into~(\ref{eq:chi0-sol-general}) yields the expression for the principal symbol, namely
\begin{align} \label{eq:leadingorderchi0}
  \Gamma_{0}(\mathbf{x},\mathbf{q},z) =  \frac{2 \pi e^2 \hbar}{\varepsilon_{\mathrm{avg}} (\mathbf{x}) |\mathbf{q}|} \Pi_0(\mathbf{x},\mathbf{q})  e^{-\frac{|z|}{\hbar}|\mathbf{q}|},
\end{align}
where we set $2 \varepsilon_{\mathrm{avg}} (\mathbf{x}) \equiv \varepsilon_\mathrm{A}(\mathbf{x})+\varepsilon_\mathrm{B}(\mathbf{x})$. Notice that $\varepsilon_{\mathrm{avg}} (\mathbf{x})$, contrary to $\varepsilon (\mathbf{x},z)$, does not depend on $z$. For the sake of brevity, the arguments of the various $\varepsilon$ are omitted from here on.

As we discussed before, the principal symbol $\Gamma_0$ can be seen as a generalization of the instantaneous eigenfunction in the adiabatic approximation. It expresses the leading-order $z$ dependence of the three-dimensional potential $V(\mathbf{x},z)$ for given values of $\mathbf{x}$ and $\mathbf{q}$. We discuss this in more detail in subsection~\ref{subsec:derivation-SC-Ansatz}.

\subsubsection{Subprincipal symbol $\Gamma_1$} \label{subsubsec:derivation-separation-subprincipal}
For the subprincipal symbol $\Gamma_1$, we need to solve Eq.~(\ref{eq:opsepvarhbar1}). We follow the same steps as for the principal symbol, where we first solve the differential equation above and below the plane. After that, we connect the two solutions with the appropriate boundary conditions. Substituting the previously found principal symbol~(\ref{eq:leadingorderchi0}), we obtain
\begin{align}\label{eq:inhomogeneousdiff}
  \left(-\frac{1}{\hbar^2} \varepsilon_i |\mathbf{q}|^2 
  + \varepsilon_i \frac{\partial^2}{\partial z^2}\right)\Gamma_1 
  = - \frac{1}{\hbar^2} f_i e^{-\frac{|z|}{\hbar}|\mathbf{q}|}, 
\end{align}
where we defined
\begin{multline} \label{eq:inhomopart}
  f_i \equiv i \frac{2 \pi e^2 \hbar}{ \varepsilon_{\mathrm{avg}} |\mathbf{q}|} 
  \left(2 \varepsilon_i \left\langle\mathbf{q},\frac{\partial \Pi_0}{\partial \mathbf{x}}\right\rangle  + \Pi_0 \left\langle\mathbf{q},\frac{\partial \varepsilon_i}{\partial \mathbf{x}}\right\rangle  \right.\\
  \left. -  \Pi_0 \frac{2 \varepsilon_i }{ \varepsilon_{\mathrm{avg}}}\left\langle\mathbf{q},\frac{\partial \varepsilon_{\mathrm{avg}}}{\partial \mathbf{x}}\right\rangle  \right) .
\end{multline}
Note that $f_i$ only depends on $z$ through $\varepsilon(\mathbf{x},z)$, where we used the same convention as in Eq.~(\ref{eq:varepsilon}). Furthermore, we omitted the argument of $\Pi_0$. 

We can now solve Eq.~(\ref{eq:inhomogeneousdiff}) with standard techniques for inhomogeneous ordinary differential equations. We start with the homogeneous equation
\begin{align}
  \left(-\frac{1}{\hbar^2} \varepsilon_i |\mathbf{q}|^2 + \varepsilon_i \frac{\partial^2}{\partial z^2} \right)\Gamma_{1,\mathrm{H}} = 0,
\end{align}
which is the same equation as for the principal symbol Eq.~(\ref{eq:leadingorderODE}). We thus obtain
\begin{align}
  \Gamma_{1,i,\mathrm{H}}(\mathbf{x},\mathbf{q},z) =  c_{1,i}^{-} e^{-\frac{|z|}{\hbar}|\mathbf{q}|} + c_{1,i}^{+} e^{\frac{|z|}{\hbar}|\mathbf{q}|}.
\end{align}
We solve the inhomogeneous equation using the method of undetermined coefficients. As such, we consider the Ansatz
\begin{align}
  \Gamma_{1,i,\mathrm{P}}(\mathbf{x},\mathbf{q},z) =  \bigg(\alpha_{i} \frac{z}{\hbar} + \beta_{i} \frac{z^2}{\hbar^2} \bigg) e^{-\frac{|z|}{\hbar}|\mathbf{q}|}.
\end{align}
Substituting this Ansatz back into the differential equation~(\ref{eq:inhomogeneousdiff}), we find the constants $\alpha_i$ and $\beta_i$. Using our previous observation that $f_i$, for both regions $\mathrm{A}$ and $\mathrm{B}$, does not explicitly depend on $z$, we find that the constants $\beta_i$ equal zero. For the constants $\alpha_i$, we find
\begin{align}
  \alpha_{i}  = \frac{s_i}{2 |\mathbf{q}|}\frac{f_i}{\varepsilon_i(\mathbf{x})} ,
\end{align}
where $s_i=1$ for $i=\mathrm{A}$ and $s_i=-1$ for $i=\mathrm{B}$.
The full solution $\Gamma_{1,i}$ for each of the two regions is given by
\begin{align}
  \Gamma_{1,i}(\mathbf{x},\mathbf{q},z) =  \left(c_{1,i}^{-}+\alpha_{i} \frac{z}{\hbar} \right) e^{-\frac{|z|}{\hbar}|\mathbf{q}|}
  + c_{1,i}^{+}e^{\frac{|z|}{\hbar}|\mathbf{q}|},
\end{align}
which is the sum of the homogeneous and particular solution.

The differential equation~(\ref{eq:inhomogeneousdiff}) gives rise to the same boundary conditions as the differential equation~(\ref{eq:opsepvarhbar0}) for the principal symbol.
First of all, $\Gamma_{1}$ has to vanish as $|z|$ goes to infinity, hence  $c_{1,i}^{+} = 0$. Second, the continuity at $z=0$ implies $c^-_{1,\mathrm{A}} = c^-_{1,\mathrm{B}} \equiv c_1$. Finally, we can determine $c_1$ from the last boundary condition, which reads
\begin{equation}
  \left.\varepsilon_\mathrm{A}\frac{\partial \Gamma_{1,\mathrm{A}}}{\partial z}\right|_{z \to 0^+}
  - \left. \varepsilon_\mathrm{B}\frac{\partial \Gamma_{1,\mathrm{B}}}{\partial z}\right|_{z \to 0^-}
  = - 4 \pi e^2  \Pi_1.
\end{equation}
After substitution of $\Gamma_{1,i}$, we find
\begin{align}
  c_1 =  \frac{2 \pi e^2 \hbar}{ \varepsilon_{\mathrm{avg}} |\mathbf{q}|}  \Pi_1 + \frac{\varepsilon_\mathrm{A} \alpha_\mathrm{A} - \varepsilon_\mathrm{B} \alpha_\mathrm{B}}{2 \varepsilon_{\mathrm{avg}} |\mathbf{q}| } .
\end{align}
Using that $\varepsilon_\mathrm{A} \alpha_\mathrm{A} - \varepsilon_\mathrm{B} \alpha_\mathrm{B} = (f_\mathrm{A} + f_\mathrm{B})/(2|\mathbf{q}|)$, and our previous definition~(\ref{eq:inhomopart}), we find that
\begin{equation}  \label{eq:c1-final}
  c_1 = \frac{2 \pi e^2 \hbar}{\varepsilon_{\mathrm{avg}} |\mathbf{q}|^3} 
  \bigg( |\mathbf{q}|^2 \Pi_1 + i \bigg\langle \mathbf{q}, \frac{\partial \Pi_0}{\partial \mathbf{x}} \bigg\rangle - \frac{i \Pi_0}{2 \varepsilon_\mathrm{avg}} \bigg\langle \mathbf{q}, \frac{\partial \varepsilon_\mathrm{avg}}{\partial \mathbf{x}} \bigg\rangle \bigg).
\end{equation}
The subprincipal symbol $\Gamma_1$ is thus given by
\begin{equation}  \label{eq:chi1-sol-general}
  \Gamma_{1,i}(\mathbf{x},\mathbf{q},z) 
  = \bigg(c_1 + \frac{1}{2 |\mathbf{q}|} \frac{f_i}{\varepsilon_i(\mathbf{x})} \frac{|z|}{\hbar} \bigg) e^{-\frac{|z|}{\hbar}|\mathbf{q}|} ,
\end{equation}
where we used that $s_i z = |z|$.
Unlike the principal symbol, the subprincipal symbol has a different expression for $i=\mathrm{A}$ and $i=\mathrm{B}$. Moreover, the factor in front of the exponent depends on $z$.

We remark that $\Gamma_1$ does not have a straightforward physical interpretation. Loosely speaking, it could be seen as the first-order quantum correction to the generalized instantaneous eigenfunction. However, as we will see in Sec.~\ref{subsec:derivation-amplitude}, our main reason to construct it is purely technical, since we need it to obtain an expression for the amplitude $\varphi_0(\mathbf{x})$ in our semiclassical Ansatz~(\ref{eq:scansatz2d}).

\subsection{Semiclassical Ansatz for in-plane motion}
\label{subsec:derivation-SC-Ansatz}
In the previous subsection, we constructed the principal and subprincipal symbol of $\hat{\Gamma}$. Our next step is to compute the three-dimensional potential $V(\mathbf{x},z)$, given by Eq.~(\ref{eq:opsepvar}). We therefore consider how the operator $\hat{\Gamma}$ acts on the semiclassical Ansatz~(\ref{eq:scansatz2d}) for the induced potential $V_{\mathrm{pl}}(\mathbf{x})$.
In general, we can express the way in which a pseudodifferential operator acts on a rapidly oscillating exponential with the help of the symbol of the operator~\cite{Maslov81,Guillemin77}. We obtain
\begin{align} 
  & \hspace*{-0.25cm} (\hat{\Gamma} V_{\mathrm{pl}}) (\mathbf{x},z) = e^{ i S(\mathbf{x})/ \hbar} \Bigg(\Gamma_0  \varphi_0 + \hbar \Gamma_0  \varphi_1 -i \hbar \left\langle\frac{\partial \Gamma_0}{\partial \mathbf{q}},\frac{\partial \varphi_0}{\partial \mathbf{x}}\right\rangle    \Bigg.  \nonumber \\
  & \hspace*{0.25cm} + \Bigg. \hbar\Gamma_1  \varphi_0
  - \frac{i \hbar}{2} \sum_{j,k} \frac{\partial^2 \Gamma_0}{\partial q_j \partial q_k} \frac{\partial^2 S}{\partial x_j \partial x_k} \varphi_0 + \mathcal{O}(\hbar^2)\Bigg),
  \label{eq:totalpotential}
\end{align}
where $\Gamma_0$, $\Gamma_1$ and their derivatives should be evaluated at $\left(\mathbf{x},\partial S/\partial \mathbf{x},z\right)$. In other words, the momentum $\mathbf{q}$ becomes $\partial S/\partial \mathbf{x}$, which is typical in the application of the semiclassical approximation~\cite{Griffiths05,Heading62,Maslov81}.
Intuitively, we can understand this relation by noting that $-i \hbar \partial/\partial x_j$ is replaced by $\partial S/\partial x_j$ when a differential operator acts on the exponent $\exp( i S(\mathbf{x})/ \hbar)$.
The third term in Eq.~(\ref{eq:totalpotential}) arises when the differential operator acts on the amplitude $\varphi(\mathbf{x})$, and the fifth term when it acts on $\partial S/\partial x_j$.

At this point, we consider the self-consistency condition~(\ref{eq:condition-self-consistency}): the three-dimensional potential $V(\mathbf{x},z)$ should equal the semiclassical Ansatz $V_\mathrm{pl}(\mathbf{x})$ in the plane $z=0$.
Since both quantities are given by an asymptotic series in $\hbar$, we should satisfy this condition order by order.

We start our analysis with the terms of order $\hbar^0$, which give
\begin{align}
  \Gamma_0 \left(\mathbf{x},\frac{\partial S}{\partial \mathbf{x}},0\right) \varphi_0 e^{i S(\mathbf{x})/ \hbar} = \varphi_0 e^{ i S(\mathbf{x})/ \hbar},
\end{align}
or
\begin{align}\label{eq:sequlareq-L0}
  \left(1-\Gamma_0 \left(\mathbf{x},\frac{\partial S}{\partial \mathbf{x}},0\right)\right) \varphi_0 e^{i S(\mathbf{x})/ \hbar} = 0.
\end{align}
At first glance, one may think that we have incorrectly matched the different orders of $\hbar$ in this expression, since $\Gamma_0$ contains a factor of $\hbar$. As we explain in detail in Sec.~\ref{subsec:derivation-applicability}, this is however not the case. In short, the apparent contradiction can be resolved by noting that we should actually perform our semiclassical expansion using a dimensionless parameter, instead of $\hbar$. Introducing this dimensionless expansion parameter, one can show that $\Gamma_0$ is of order one. Alternatively, one may refer to the analogy with the homogeneous case, which we discuss shortly and which implies that the terms in Eq.~(\ref{eq:sequlareq-L0}) are of the same order.

As stated before, we require that $\varphi_0 e^{i S(\mathbf{x})/ \hbar} \neq 0$ in Eq.~(\ref{eq:sequlareq-L0}), otherwise we would not be considering plasmons. Consequently, the term in the brackets has to vanish and we obtain a Hamilton-Jacobi equation, specifically
\begin{align}\label{eq:HJequation}
  1 - \frac{2\pi e^2 \hbar}{\varepsilon_{\mathrm{avg}}(\mathbf{x}) |\partial S / \partial \mathbf{x}|} \Pi_0 \left(\mathbf{x},\frac{\partial S}{\partial \mathbf{x}}\right)  = 0.
\end{align}
Defining the effective classical Hamiltonian of the system as
\begin{align}\label{eq:effclassicalHam}
  \mathcal{H}_0(\mathbf{x},\mathbf{q}) = 1 -\frac{2\pi e^2 \hbar}{\varepsilon_{\mathrm{avg}}(\mathbf{x}) |\mathbf{q}|} \Pi_0(\mathbf{x}, \mathbf{q}), 
\end{align}
the Hamilton-Jacobi equation reads $\mathcal{H}_0\left(\mathbf{x},\frac{\partial S}{\partial \mathbf{x}}\right)=0$.
It is important to note that this classical Hamiltonian strongly resembles the Lindhard expression for the dielectric function of homogeneous two-dimensional materials~\cite{Giuliani05}, with two main differences. First of all, it includes a dependence on the coordinate $\mathbf{x}$, whereas the homogeneous dielectric function only depends on $\mathbf{q}$. The classical Hamiltonian $\mathcal{H}_0$ is therefore a function on classical phase space, rather than a function of the plasmon momentum only. As we discussed below Eq.~(\ref{eq:polarization-principal}), the position dependence of the polarization amounts to the replacement of the Fermi momentum by the local Fermi momentum $p_\mathrm{F}(\mathbf{x})$.
The second difference between the two-dimensional dielectric function and the effective Hamiltonian $\mathcal{H}_0$ is a factor of $\varepsilon_\mathrm{avg}$, since the former should evaluate to $\varepsilon_\mathrm{avg}$ in the absence of a polarization.

To avoid misunderstanding, we emphasize that $\mathcal{H}_0$ is not a classical Hamiltonian in the traditional sense, since it does not describe a classical plasma, as we just established. Instead, we use $\mathcal{H}_0$ to calculate the classical trajectories which form the basis for the construction of our asymptotic solution, as we discussed in the introduction. This process is analogous to the way in which one adds interference to the rays in geometrical optics.
Although we do not consider it explicitly in this article, we remark that we can still make the transition to a classical plasma in the traditional sense by making two changes in expression~(\ref{eq:polarization-principal}) for the polarization. First, we should replace the Fermi-Dirac distribution by the Maxwell-Boltzmann distribution, or, equivalently, consider the regime where the temperature is much larger than the chemical potential. Second, we should only keep the linear term in the expansion of the denominator, that is, $H_0(\mathbf{x},\mathbf{p}) - H_0(\mathbf{x},\mathbf{p}+\mathbf{q}) \approx -\langle \mathbf{p} , \mathbf{q} \rangle/m$. Unfortunately, this transition to a classical plasma leads to practical issues with Landau damping, cf. the discussion in Ref.~\cite{Reijnders22}.

The Hamilton-Jacobi equation~(\ref{eq:HJequation}) determines the action $S(\mathbf{x})$ in our Ansatz for the induced potential $V_\mathrm{pl}(\mathbf{x})$. 
As is well-known from classical mechanics~\cite{Arnold89,Goldstein02}, this Hamilton-Jacobi equation is equivalent to the system of Hamilton equations for the effective classical Hamiltonian $\mathcal{H}_0$, i.e.,
\begin{align}  \label{eq:Hamiltonian-system}
  \frac{\mathrm{d} \mathbf{x}}{\mathrm{d} \tau} = \frac{\partial \mathcal{H}_0}{ \partial \mathbf{q}}, \hspace{1cm}
  \frac{\mathrm{d} \mathbf{q}}{\mathrm{d} \tau} = -\frac{\partial \mathcal{H}_0}{ \partial \mathbf{x}}.
\end{align}
Of course, suitable initial conditions have to be specified and we will discuss these in more detail in Sec.~\ref{sec:scattering}. 
For now, let us assume that these conditions can be parameterized by $\alpha$ and constitute a line $\Lambda^1$. In a typical scattering setup, the parameter $\alpha$ corresponds to the coordinate along the initial wavefront.
Under these assumptions, we can denote the solutions of Hamilton's equations by $\left(\mathbf{X}(\tau, \alpha),\mathbf{Q}(\tau, \alpha)\right)$. Together, these solutions describe a two-dimensional plane $\Lambda^2$ in classical phase space, parameterized by $(\tau,\alpha)$, which is known as a Lagrangian manifold~\cite{Maslov81,Guillemin77}.
The classical action can now be written as
\begin{align} \label{eq:action-def}
  S(\mathbf{x}) = \int_{\mathbf{x}_0}^{\mathbf{x}} \left\langle\mathbf{Q} ,\mathrm{d} \mathbf{X}\right\rangle,
\end{align}
where we integrate from an initial point $\mathbf{x}_0$ to the point $\mathbf{x}$.
In passing from the Hamilton-Jacobi equation to Hamilton's equations, we have lifted the problem from configuration space, on which we deal with the coordinate $\mathbf{x}$, to phase space, which involves the coordinates $(\mathbf{x},\mathbf{q})$. The projection of the Lagrangian manifold $\Lambda^2$ onto the configuration space is (locally) invertible, as long as the Jacobian
\begin{align}\label{eq:Jacobian}
  J(\mathbf{x}) = \det \left(\frac{\partial \mathbf{X}}{\partial (\tau, \alpha)}\right)
\end{align}
is not equal to zero~\cite{Maslov81,Poston78,Arnold82}. As we will see in the next section, this Jacobian plays an important role in the construction of the amplitude.

\subsection{Amplitude of the induced potential}
\label{subsec:derivation-amplitude}
Now that we determined the action $S(\mathbf{x})$, our next step is to determine the amplitude $\varphi_0$.
To this end, we consider the terms of order $\hbar^1$ in the self-consistency condition~(\ref{eq:condition-self-consistency}).
Using Eq.~(\ref{eq:totalpotential}), we obtain
\begin{multline} \label{eq:subleadingopsep}
  e^{i S(\mathbf{x})/ \hbar} \Bigg(\Gamma_0  \varphi_1 + \Gamma_1  \varphi_0 - i  \left\langle\frac{\partial \Gamma_0}{\partial \mathbf{q}},\frac{\partial \varphi_0}{\partial \mathbf{x}}\right\rangle\Bigg.\\
  - \frac{i }{2} \sum_{j,k} \frac{\partial^2 \Gamma_0}{\partial q_j \partial q_k} \frac{\partial^2 S}{\partial x_j \partial x_k} \varphi_0 \Bigg)  = \varphi_1 e^{i S(\mathbf{x})/ \hbar} ,
\end{multline}
where $\Gamma_0$, $\Gamma_1$ and their derivatives are now evaluated at $\left(\mathbf{x},\partial S/\partial \mathbf{x},0\right)$.
We immediately see that the terms proportional to $\varphi_1$ drop out because of the Hamilton-Jacobi equation~(\ref{eq:HJequation}). Since we again require that $V_{\mathrm{pl}}\neq 0$, we cancel out the exponent. The remaining terms read
\begin{multline}  \label{eq:pre-transporteq}
  \Gamma_1  \varphi_0 
  - i  \left\langle\frac{\partial \Gamma_0}{\partial \mathbf{q}},\frac{\partial \varphi_0}{\partial \mathbf{x}}\right\rangle 
  - \frac{i }{2} \sum_{j,k} \frac{\partial^2 \Gamma_0}{\partial q_j \partial q_k} \frac{\partial^2 S}{\partial x_j \partial x_k} \varphi_0  = 0,
\end{multline}
where $\Gamma_1$ is given by Eq.~(\ref{eq:chi1-sol-general}), evaluated at $z=0$.
We now see why we had to compute $\Gamma_1$ in Sec.~\ref{subsubsec:derivation-separation-subprincipal}: it shows up in the differential equation~(\ref{eq:pre-transporteq}) that determines the amplitude $\varphi_0(\mathbf{x})$.

Let us now take a closer look at the derivatives of $\Gamma_0$ in Eq.~(\ref{eq:pre-transporteq}). Since taking the derivative of the exponent in $\Gamma_0$ brings down a factor $|z|$, this term vanishes when setting $z=0$. The only nonzero terms therefore originate from the derivatives of $c_0$, defined in Eq.~(\ref{eq:bc-D-field-result}).

Using our effective classical Hamiltonian $\mathcal{H}_0$, we can then rewrite Eq.~(\ref{eq:pre-transporteq}) as
\begin{multline}\label{eq:transporteq}
  \mathcal{H}_1\left(\mathbf{x},\frac{\partial S}{\partial \mathbf{x}}\right) \varphi_0 - i \left\langle\frac{\partial \mathcal{H}_0}{\partial \mathbf{q}}\left(\mathbf{x},\frac{\partial S}{\partial \mathbf{x}}\right), \frac{\partial \varphi_0}{\partial \mathbf{x}} \right\rangle\\
  - \frac{i}{2} \sum_{j,k} \frac{\partial^2 \mathcal{H}_0}{\partial q_j \partial q_k}\left(\mathbf{x},\frac{\partial S}{\partial \mathbf{x}}\right) \frac{\partial^2 S}{\partial x_k \partial x_j} \varphi_0   = 0,
\end{multline}
where we implicitly defined $\mathcal{H}_1 \equiv -c_1$, or, using our previous result~(\ref{eq:c1-final}),
\begin{multline}  \label{eq:defL1}
  \mathcal{H}_1(\mathbf{x},\mathbf{q}) \equiv - \frac{2 \pi e^2 \hbar}{\varepsilon_{\mathrm{avg}} |\mathbf{q}|} \Pi_1 
  + \frac{i}{2} \frac{2 \pi e^2 \hbar}{\varepsilon_{\mathrm{avg}}^2 |\mathbf{q}|^3} \Pi_0\left\langle \mathbf{q}, \frac{\partial \varepsilon_\mathrm{avg}}{\partial \mathbf{x}} \right\rangle \\
  - i \frac{2 \pi e^2 \hbar}{\varepsilon_{\mathrm{avg}} |\mathbf{q}|^3} \left\langle \mathbf{q}, \frac{\partial \Pi_0}{\partial \mathbf{x}} \right\rangle .
\end{multline}
Equation~(\ref{eq:transporteq}) has a specific form, which is known as the transport equation within the semiclassical approximation~\cite{Maslov81,Guillemin77}. As discussed in Ref.~\cite{Reijnders22}, it is quite remarkable that our formalism leads to this type of equation.

We now show how to solve the transport equation using standard semiclassical techniques, which will give us the amplitude $\varphi_0(\mathbf{x})$.
First, we look at the time evolution of the Jacobian~(\ref{eq:Jacobian}), which can be determined using the Liouville formula~\cite{Maslov81}
\begin{align}\label{eq:timeevoJ}
  \frac{\mathrm{d} }{\mathrm{d} t} \log{J} = \sum_{j,k} \frac{\partial^2 \mathcal{H}_0}{\partial q_j \partial q_k} \frac{\partial^2 S}{\partial x_j \partial x_k} + \sum_j \frac{\partial^2 \mathcal{H}_0}{\partial x_j \partial q_j}.
\end{align}
Second, we use that the time evolution of the amplitude $\varphi_0$ along the trajectories of the Hamiltonian system~(\ref{eq:Hamiltonian-system}) is given by
\begin{align}\label{eq:timeevoAmp}
  \frac{\mathrm{d} \varphi_0}{\mathrm{d} t} = \left\langle\frac{\partial \mathbf{x}}{\partial t},\frac{\partial \varphi_0}{\partial \mathbf{x}}\right\rangle = \left\langle\frac{\partial \mathcal{H}_0}{\partial \mathbf{q}},\frac{\partial \varphi_0}{\partial \mathbf{x}}\right\rangle.
\end{align}
Substitution of Eqs.~(\ref{eq:timeevoJ}) and~(\ref{eq:timeevoAmp}) in the transport equation~(\ref{eq:transporteq}) yields
\begin{equation}
  \frac{\mathrm{d} \varphi_0}{\mathrm{d} t} = - i \mathcal{H}_1 \varphi_0 -\frac{1}{2}\left( \frac{\mathrm{d} }{\mathrm{d} t} \log{J} - \sum_j \frac{\partial^2 \mathcal{H}_0}{\partial x_j \partial q_j}\right) \varphi_0 . 
\end{equation}
We now define $A_0(\mathbf{x}) \equiv \varphi_0(\mathbf{x}) \sqrt{J(\mathbf{x})}$, which transforms the above equation into
\begin{align}
  \frac{\mathrm{d} A_0}{\mathrm{d} t} + i\left(\mathcal{H}_1 + \frac{i}{2} \sum_j \frac{\partial^2 \mathcal{H}_0}{\partial x_j \partial q_j}\right)A_0 = 0.
\end{align}
Solving this equation we obtain the amplitude, specifically
\begin{align}\label{eq:amplitudesol}
  A_0 (x) = A_0^0 \exp\left(-i \int_0^t  \mathcal{H}_1 + \frac{i}{2} \sum_j \frac{\partial^2 \mathcal{H}_0}{\partial x_j \partial q_j} \mathrm{d} t'\right),
\end{align}
where the integration should be performed along the phase space trajectories of the Hamiltonian system~(\ref{eq:Hamiltonian-system}) in phase space. The constant $A_0^0$ is determined by the initial conditions.

Let us analyze the exponential factor in the amplitude~(\ref{eq:amplitudesol}) to see how it affects the induced potential.
Using Eq.~(\ref{eq:symbol-Hermitian-operator-condition}) to express $\Pi_1$ in terms of the derivatives of $\Pi_0$, and with the help of Eqs.~(\ref{eq:defL1}) and~(\ref{eq:effclassicalHam}), we obtain
\begin{multline}
  \mathcal{H}_1 + \frac{i}{2} \sum_j \frac{\partial^2 \mathcal{H}_0}{\partial x_j \partial q_j} =
  \frac{i}{2} \frac{2 \pi e^2 \hbar }{ \varepsilon_{\mathrm{avg}} |\mathbf{q}|} \left(- \frac{1}{|\mathbf{q}|^2} \left\langle \mathbf{q} ,\frac{\partial \Pi_0}{\partial \mathbf{x}} \right\rangle \right. \\
  +\left.  \frac{1}{ \varepsilon_{\mathrm{avg}}} \left\langle \frac{\partial \Pi_0}{\partial \mathbf{q}}  ,\frac{\partial \varepsilon_{\mathrm{avg}}}{\partial \mathbf{x}} \right\rangle  \right)  .
\end{multline}
In terms of the Poisson bracket
\begin{equation}  \label{eq:def-Poisson-bracket}
  \left\{f,g\right\} = \sum_{i=1}^N\left(\frac{\partial f}{\partial x_i}\frac{\partial g}{\partial q_i}-\frac{\partial f}{\partial q_i}\frac{\partial g}{\partial x_i}\right),
\end{equation}
this expression can be written as
\begin{align}
  \mathcal{H}_1 + \frac{i}{2} \sum_j \frac{\partial^2 \mathcal{H}_0}{\partial x_j \partial q_j} =\frac{i}{2} \frac{2 \pi e^2 \hbar }{ \varepsilon_{\mathrm{avg}}|\mathbf{q}|} \left\{ \ln{\left( \varepsilon_{\mathrm{avg}} |\mathbf{q}|\right)}, \Pi_0 \right\} .
\end{align}
We can write this more elegantly using the product rule for the Poisson bracket,
$
b \{a,c\}  =\{a,bc\} - \{a,b\}c.
$
When we set
\begin{align}
  a =\ln{( \varepsilon_{\mathrm{avg}} |\mathbf{q}|)},\qquad b = \frac{2 \pi e^2 \hbar }{ \varepsilon_{\mathrm{avg}} |\mathbf{q}|}, \qquad c = \Pi_0,
\end{align}
and use definition~(\ref{eq:effclassicalHam}), we obtain
\begin{align}\label{eq:poissonhaakjeopsep} 
  \mathcal{H}_1 + \frac{i}{2} \sum_j \frac{\partial^2 \mathcal{H}_0}{\partial x_j \partial q_j} =
  -\frac{i}{2}  \left\{ \ln{( \varepsilon_{\mathrm{avg}} |\mathbf{q}|)}, \mathcal{H}_0 (\mathbf{x},\mathbf{q}) \right\}.
\end{align}
Note that the Poisson bracket $\{a,b\}$ vanishes, since both parts are functions of $\varepsilon_{\mathrm{avg}}|\mathbf{q}|$.
Combining the result~(\ref{eq:poissonhaakjeopsep}) with Eq.~(\ref{eq:amplitudesol}) for the amplitude, we have
\begin{align}
  A_0 (\mathbf{x}) = A_0^0 \exp\left(-\frac{1}{2} \int_{t_0}^t   \left\{ \ln{( \varepsilon_{\mathrm{avg}} |\mathbf{q}|)} , \mathcal{H}_0 (\mathbf{x},\mathbf{q}) \right\} \mathrm{d} t'\right).
\end{align}
Since the Poisson bracket with the effective Hamiltonian is equal to the time derivative
\begin{align}
  \left\{\lambda(\mathbf{x},\mathbf{q}),\mathcal{H}_0(\mathbf{x},\mathbf{q})\right\} 
  & = \frac{\mathrm{d} \lambda}{\mathrm{d} t},
\end{align}
the time integration is straightforward, and we find
\begin{align}  \label{eq:leadingorderamplitude}
  A_0 (\mathbf{x}) = \frac{A_0^0}{\sqrt{ \varepsilon_{\mathrm{avg}} \left|\frac{\partial S}{\partial \mathbf{x}}\right|} } .
\end{align}
This expression shows that the amplitude is a local quantity. In other words, it does not depend on the trajectory of the plasmon in phase space, but only on the point $\mathbf{x}$.
From a physical point of view, it also means that a closed trajectory in phase space does not change the amplitude, in contrast to the case where a Berry phase is present.

A couple of remarks are in order at this point. First, we set $\mathbf{q}=\partial S/\partial \mathbf{x}$ in Eq.~(\ref{eq:leadingorderamplitude}), since we performed the integration along the trajectories of the Hamiltonian system, on which this relation holds. Second, we absorbed the integration constant into $A_0^0$. Finally, we note that we are free to add a function $g(\mathcal{H}_0(\mathbf{x},\mathbf{q}))$ to the first argument of the Poisson bracket for Eq.~(\ref{eq:poissonhaakjeopsep}) to still be true. However, this does not influence the final result, since we are constrained to level set $\mathcal{H}_0=0$ by the Hamilton-Jacobi equation. 

Equations~(\ref{eq:totalpotential}),~(\ref{eq:leadingorderchi0}),~(\ref{eq:HJequation}) and~(\ref{eq:leadingorderamplitude}) give us our final result for the leading-order term of the potential $V(\mathbf{x},z,t)$, namely
\begin{align}\label{eq:V0solutionamplitudeopsep}
  V(\mathbf{x},z,t) = \frac{A_0^0}{\sqrt{J}\sqrt{ \varepsilon_{\mathrm{avg}}\left|\frac{\partial S}{\partial \mathbf{x}}\right| }} e^{-\frac{|z|}{\hbar}\left|\frac{\partial S}{\partial \mathbf{x}}\right| }e^{i (S(\mathbf{x})-E t)/\hbar} ,
\end{align}
where we used the Hamilton-Jacobi equation to simplify the amplitude, and included the time dependence (\ref{eq:Vpl-time-dep}).
This expression clearly shows the wave-like character of the plasmons.
By adding the contributions of different trajectories, we obtain the quantum interference, as we will see explicitly in Sec.~\ref{sec:scattering}.

We immediately see that the asymptotic solution~(\ref{eq:V0solutionamplitudeopsep}) diverges at the points where $J(\mathbf{x})=0$. In a one-dimensional setting, these divergences are the so-called classical turning points, and Eq.~(\ref{eq:V0solutionamplitudeopsep}) cannot be used in their vicinity. This general fact is well known for the one-dimensional WKB approximation for the Schr\"odinger equation~\cite{Griffiths05}, where $J \propto \sqrt{p_x}$. In the vicinity of the turning points, one normally constructs an approximate solution in terms of Airy functions~\cite{Griffiths05, Maslov81}, see also the discussion in Ref.~\cite{Reijnders22}.  The additional factor $(\varepsilon_\mathrm{avg} |\partial S/\partial \mathbf{x}|)^{-1/2}$, on the other hand, is not commonly present in WKB-type approximations. In the next subsection, we give a physical interpretation to it with the help of the energy density of electromagnetic fields.

\subsection{Energy density}\label{subsec:EnergyDensity}
Before we consider the asymptotic solution~(\ref{eq:V0solutionamplitudeopsep}), we briefly discuss the meaning of the Jacobian $J$. When one considers the Schr\"odinger equation, the square of the absolute value of the wavefunction $\psi$ represents a probability density.
The Jacobian $J$ ensures that the expression for the probability to find a particle in a certain region $\gamma$ takes the same form in all coordinate systems, namely $\int_\gamma \mathrm{d}\mathbf{x}/|J|$. 
Technically, one observes that a coordinate change leads to two additional determinants that cancel: one coming from the change in the integration measure $\mathrm{d}\mathbf{x}$ and one coming from the change of variables in the Jacobian $J$.
In the mathematical literature, this is formalized using the concept of a half-density~\cite{Guillemin77,Zworski12}.
By changing coordinates to the parametrization of the Lagrangian manifold $(\tau,\alpha)$, one sees that the integral $\int_\gamma \mathrm{d}\mathbf{x}/|J|$ equals the area of the equivalent region on the Lagrangian manifold.

Since $V(\mathbf{x},z,t)$ is a potential that corresponds to an electric field, we cannot use the interpretation in terms of probabilities for our plasmons. If we want to consider an equivalent quantity, we should look at the energy density of the electromagnetic field, which can be derived from Poynting's theorem~\cite{Jackson99,Landau84}.
For a complex electric field $\boldsymbol{ \mathcal{E}}$ and displacement field $\boldsymbol{ \mathcal{D}}$, one has
\begin{align}\label{eq:PoyntingsTheorem}
  - \nabla \cdot \boldsymbol{ \mathcal{S}} = \frac{1}{16 \pi} \left( \boldsymbol{ \mathcal{E}}\cdot\frac{\partial  \boldsymbol{ \mathcal{D}^*}}{\partial t} + \boldsymbol{ \mathcal{E}^*}\cdot\frac{\partial  \boldsymbol{ \mathcal{D}}}{\partial t}\right) ,
\end{align}
where $\boldsymbol{ \mathcal{S}}$ is the Poynting vector. 
Note that we omitted the terms associated with the magnetic field, since they do not play a role in our problem. Moreover, we left out the terms proportional to $\boldsymbol{ \mathcal{E}} \cdot \boldsymbol{ \mathcal{D}}$ and $\boldsymbol{ \mathcal{E}^*} \cdot \boldsymbol{ \mathcal{D}}^*$, since they become zero after averaging over time. 
The energy density $\mathcal{U}$ is obtained~\cite{Jackson99,Landau84} by writing the right-hand side as a time derivative $\partial\:\!\mathcal{U}/\partial t$.

In our problem, the energy density $\mathcal{U}(\mathbf{x},z)$ depends on both the in-plane and out-of-plane coordinates.
In order to study the dependence on the in-plane coordinate $\mathbf{x}$, we integrate over $z$. We obtain the integrated energy density above the plane $\mathcal{U}_\mathrm{A}(\mathbf{x})$ by integrating over the region $z>0$, and similarly we obtain $\mathcal{U}_\mathrm{B}(\mathbf{x})$ from the region $z<0$. Finally, we consider the energy density that comes from the two-dimensional plane itself, denoted by $\mathcal{U}_\mathrm{pl}(\mathbf{x})$. Together, these contributions form the integrated energy density, $\mathcal{U}_\mathrm{I}(\mathbf{x}) = \mathcal{U}_\mathrm{A}(\mathbf{x}) + \mathcal{U}_\mathrm{B}(\mathbf{x}) + \mathcal{U}_\mathrm{pl}(\mathbf{x})$.

We obtain the leading-order term of the electric field in Eq.~(\ref{eq:PoyntingsTheorem}) using $\boldsymbol{ \mathcal{E}} = e^{-1} \nabla V$, where $V$ is given by Eq.~(\ref{eq:V0solutionamplitudeopsep}).
We have to be more careful for the displacement field, since the dielectric function changes as a function of $z$.
Above and below the plane $z=0$, we have two media without dispersion, characterized by real and positive functions $\varepsilon_\mathrm{A} (\mathbf{x})$ and $\varepsilon_\mathrm{B} (\mathbf{x})$, respectively.
Consequently, the displacement field is simply given by $\boldsymbol{ \mathcal{D}}_i (\mathbf{x},z) = \varepsilon_i (\mathbf{x}) \boldsymbol{ \mathcal{E}} (\mathbf{x},z)$, where $i = (\mathrm{A},\mathrm{B})$.
The displacement field in the plane $z=0$ is, however, much more complicated. Not only is there dispersion, but the dielectric function is also an operator. In appendix~\ref{ap:inplaneEnergyDensity}, we derive a general formula for the energy density in this case and show that $\mathcal{U}_\mathrm{pl}(\mathbf{x}) = 0$ in our example. The principal reason for this is that we are dealing with an infinitesimal layer.

Now that we have obtained both the electric and displacement fields in terms of the induced potential for the regions above and below the plane, we turn back to Eq.~(\ref{eq:PoyntingsTheorem}). Taking out the time derivative on the right-hand side, we find that the energy density is given by
\begin{align}\label{eq:energy-density-simple}
  \mathcal{U}_i(\mathbf{x},z) 
  = \frac{\varepsilon_i (\mathbf{x})}{16 \pi} \boldsymbol{ \mathcal{E}}(\mathbf{x},z) \cdot \boldsymbol{ \mathcal{E}}^{*}(\mathbf{x},z)
  = \frac{\varepsilon_i(\mathbf{x})}{16 \pi e^2} | \nabla V (\mathbf{x},z) |^2
\end{align}
where $i = (\mathrm{A},\mathrm{B})$. 
The leading-order term of the gradient of the induced potential~(\ref{eq:V0solutionamplitudeopsep}) reads
\begin{equation}
  \nabla V = \frac{A^0_0 \, e^{i ( S(\mathbf{x}) - E t)/\hbar}}{\sqrt{J} \sqrt{\varepsilon_{\mathrm{avg}}\left|\frac{\partial S}{\partial \mathbf{x}}\right|}}  e^{-\frac{|z|}{\hbar}\left|\frac{\partial S}{\partial \mathbf{x}}\right|} \left(\frac{i}{\hbar} \frac{\partial S}{\partial \mathbf{x}}, - \frac{1}{\hbar} \left| \frac{\partial S}{\partial \mathbf{x}}\right| \right),
\end{equation}
where the quantity in parentheses should be interpreted as a vector in the space $(\mathbf{x},z)$.
We remark that the term proportional to $|z|/\hbar$, coming from the derivative of the exponential factor with respect to $\mathbf{x}$ is not of leading order. As we discuss in the next subsection, the combination $z/\hbar$ is of order one when we introduce proper dimensionless units, see also Sec.~\ref{subsec:derivation-separation}. 
Thus, the energy density above and below the $z=0$ plane is given by 
\begin{align}
  \mathcal{U}_i(\mathbf{x},z) = \frac{ 1}{8 \pi e^2 \hbar^2 } \frac{\varepsilon_i }{\varepsilon_{\mathrm{avg}} } \frac{|A_0^0|^2}{|J|}  \left|\frac{\partial S}{\partial \mathbf{x}}\right| e^{-2\frac{|z|}{\hbar}\left|\frac{\partial S}{\partial \mathbf{x}}\right|} .
\end{align}
This result shows that a plasmon is more concentrated near the $z=0$ plane when the momentum $\left|\partial S/\partial \mathbf{x}\right|$ is larger, since the energy density has a higher peak at $z=0$ and drops off faster in $|z|$ due to the larger term in the exponent.

We now compute the integrated energy density above the plane by integrating $\mathcal{U}_\mathrm{A}(\mathbf{x},z)$ from $0$ to infinity. We obtain
\begin{equation}\label{eq:U-A-integrated}
  \mathcal{U}_\mathrm{A}(\mathbf{x}) = \int_{0}^{\infty} \mathcal{U}_\mathrm{A}(\mathbf{x},z) \mathrm{d} z 
  = \frac{|A_0^0|^2}{16 \pi e^2 \hbar} \frac{\varepsilon_\mathrm{A} }{\varepsilon_{\mathrm{avg}} } \frac{1}{|J|} .
\end{equation}
Here we see another reason why the term proportional to $|z|/\hbar$ in $\nabla V$ is of higher order: integration by parts leads to additional factors of $\hbar$.
In a similar way, we obtain an expression for the integrated energy density $\mathcal{U}_\mathrm{B}(\mathbf{x})$ below the plane. Adding the two contributions, we find the integrated energy density
\begin{equation}
  \mathcal{U}_\mathrm{I}(\mathbf{x}) = \frac{|A_0^0|^2}{8 \pi e^2 \hbar} \frac{1}{|J|} ,
\end{equation}
where we used that $\varepsilon_\mathrm{A}+\varepsilon_\mathrm{B} = 2 \varepsilon_{\mathrm{avg}}$. We observe that this expression does not depend on the momentum $\left|\partial S / \partial \mathbf{x}\right|$, and is also independent of the various $\varepsilon$. In fact, it is proportional to $1/|J|$, just as the probability density for the Schr\"odinger equation. We can therefore interpret the additional factor $(\varepsilon_\mathrm{avg} |\partial S/\partial \mathbf{x}|)^{-1/2}$ in $V(\mathbf{x},z,t)$ as follows: it ensures that the energy density has the correct mathematical structure discussed at the beginning of this subsection.

\subsection{Dimensionless parameters and applicability of the semiclassical approximation} \label{subsec:derivation-applicability}
So far, we have constructed our semiclassical theory for plasmons in inhomogeneous media as a power series in $\hbar$. 
However, since $\hbar$ is a constant, it is, strictly speaking, not possible to use it as a small parameter in the series expansion. The main reason for this is that it is unclear with respect to which other quantity it should be small. This led to the apparent contradiction in Sec.~\ref{subsec:derivation-SC-Ansatz}, where we combined terms with different powers of $\hbar$ in $\mathcal{H}_0$.
In order to resolve these issues, we should identify the correct dimensionless semiclassical parameter for the series expansion.
In this section, we consider the characteristic scales in the problem, and define dimensionless quantities. We also discuss the applicability regime of the semiclassical approximation.

We start by identifying the characteristic length scales involved. Since we consider an inhomogeneous medium, the first scale is the characteristic length $\ell$ over which the inhomogeneity $U(\mathbf{x})$ changes. The second length scale is the electron wavelength $\lambda_\mathrm{el}$, since plasmons are collective excitations of electrons. It is given by $\lambda_\mathrm{el} = 2 \pi \hbar/ p_0$, where $p_0$ is the typical value of the Fermi momentum, cf. Eq.~(\ref{eq:TF-approx}). Our semiclassical theory is applicable when $\lambda_\mathrm{el} \ll \ell$, which implies that the potential is locally almost homogeneous for the electrons~\cite{Griffiths05,Maslov81,Reijnders22}. We therefore introduce the dimensionless semiclassical parameter $h$, given by
\begin{equation}\label{eq:defsemiclassicalh}
  h = \frac{\hbar}{p_0 \ell} = \frac{1}{2 \pi} \frac{\lambda_\mathrm{el}}{\ell}.
\end{equation}	
The applicability criterion is then given by $h \ll 1$. Since we consider quantum plasmons, the plasmon wavelength is on the order of the electron wavelength. The condition $h \ll 1$ is therefore equivalent to our previous statement that we consider the regime where the length scale of the inhomogeneity is much larger than the plasmon wavelength.

The third length scale is the Thomas-Fermi screening length~\cite{Ando82,Giuliani05}. It characterizes the screening of the electric field of an electron by other electrons, and is given by $\lambda_\mathrm{TF} = \hbar^2 \varepsilon_{\mathrm{c}} / 2 g_\mathrm{s} m e^2$, where $\varepsilon_{\mathrm{c}}$ is a typical value of the average background dielectric constant $\varepsilon_{\mathrm{avg}} (\mathbf{x})$. Our second applicability criterion is that the characteristic length $\ell$ is much larger than this screening length~\cite{Reijnders22}. We therefore define a second dimensionless parameter $\kappa$, given by
\begin{equation}
  \kappa =  \frac{\lambda_\mathrm{TF}}{\ell}.
\end{equation}
We also require that $ \kappa \ll 1$.

Now that we have identified the characteristic length scales, we introduce the dimensionless coordinates $\tilde{\mathbf{x}}=\mathbf{x}/\ell$, and momenta $\tilde{q}=|\mathbf{q}|/p_0$. Moreover, we also define the dimensionless Fermi momentum $\tilde{p}_\mathrm{F}(\tilde{\mathbf{x}}) = p_\mathrm{F}(\mathbf{x})/p_0$, background dielectric constant $\tilde{\varepsilon}(\mathbf{\tilde{x}}) = \varepsilon_{\mathrm{avg}}(\mathbf{x}) / \varepsilon_{\mathrm{c}}$, and energy $\tilde{E} = 2 m E/p_0^2$. We substitute the dimensionless parameters in our effective Hamiltonian Eq.~(\ref{eq:effclassicalHam}), and calculate the integral of the polarization at zero temperature analogously to the homogeneous case~\cite{Giuliani05}. This yields the dimensionless effective Hamiltonian
\begin{multline}\label{eq:dimensionlessEffClassHam}
  \tilde{\mathcal{H}}_0(\tilde{\mathbf{x}},\tilde{q}) = 1 - \frac{h }{2 \tilde{\varepsilon} \kappa } \frac{\tilde{p}_\mathrm{F}}{\tilde{q}^2} \left(-\frac{\tilde{q}}{\tilde{p}_\mathrm{F}} -\mathrm{sign}\left(\mathrm{Re}(\tilde{\nu}_{-})\right)\sqrt{\tilde{\nu}_{-}^2-1}\right.\\ 
  \left.+ \, \mathrm{sign}\left(\mathrm{Re}(\tilde{\nu}_{+})\right)\sqrt{\tilde{\nu}_{+}^2-1}\right),  
\end{multline}
where $ \tilde{\nu}_{\pm} = \tilde{E} / 2 \tilde{q}\, \tilde{p}_\mathrm{F} \pm \tilde{q} / 2 \tilde{p}_\mathrm{F}$. Comparing this expression to Eq.~(\ref{eq:effclassicalHam}), we see that it contains the ratio of the two small parameters $h$ and $\kappa$. In contrast to the parameters themselves, this ratio is typically not small. This resolves the apparent contradiction, encountered in Sec.~\ref{subsec:derivation-SC-Ansatz}, regarding the different orders of $\hbar$. In Sec.~\ref{sec:numerical}, we calculate the ratio $h / \kappa$ and show that it is of order one for the systems we consider. Note that the above parameters are in Gaussian units. We can convert them to S.I. units, by making the substitution $e^2 \to e^2/4 \pi \varepsilon_0$. 

So far, we only discussed the length scales related to the in-plane coordinates $\mathbf{x}$, which we called the slow variables in Sec.~\ref{subsec:derivation-separation}. 
It does not seem very sensible to make the out-of-plane coordinate $z$, previously called the fast variable, dimensionless using the length scale $\ell$. After all, changes in the electron density $n(\mathbf{x},z)$ happen over a characteristic length $\ell_z$, which is much smaller than $\ell$, as we discussed in the introduction. In our setup, one may even say that this length scale is infinitely small, since we are considering a delta function. However, in a more realistic setup, $\ell_z$ would correspond to the thickness of the two-dimensional electron gas, cf. Refs.~\cite{Marzari12,Wehling11}.
When introducing dimensionless units, we have to look at the combination $\hbar/(p_0 z)$, cf. Eq.~(\ref{eq:leading-ordersymbolODE}) and Sec.~\ref{subsec:derivation-separation}. Rewriting this quotient as $\hbar/(p_0 \ell_z) \cdot \ell_z/z$, we see that both terms in the latter expression are of order one. In particular, the dimensionless parameter $\hbar/(p_0 \ell_z)$ is not small, because $\ell_z \ll \ell$. This is the reason why we included the derivatives with respect to $z$ in the principal symbol in Sec.~\ref{subsec:derivation-separation}, and also why we used the ratio $z/\hbar$ in our Ansatz for $\Gamma_{1,i,\mathrm{P}}$.

Using the dimensionless quantities that we defined in this section, one can show that our semiclassical approximation is an asymptotic series expansion in powers of the dimensionless semiclassical parameter $h$, cf. the discussion in Ref.~\cite{Reijnders22}.
Since we assumed that $q$ and $p_0$ are of the same order of magnitude when defining $h$ and $\kappa$, we cannot use our semiclassical scheme when $q$ is too small.

We end this subsection with a short analysis of the semiclassical parameter $h$, to get a better understanding for which electron densities the theory is applicable.
Since the criterion for $h$ reads $h = \hbar / p_0 \ell \ll 1$, we can consider for which numerical values we have $h=0.1$. As mentioned before, the parameter $p_0$ is directly related to the electron density by Eq.~(\ref{eq:TF-approx}). We readily see that for a typical metal with electron density $n = 10^{14}$~cm$^{-2}$, we have $h=0.1$ for a characteristic length of $\ell = 4$~nm. When we go to smaller densities, and for example consider a typical semiconductor with electron density around $n= 10^{11}$~cm$^{-2}$, we have $h=0.1$ for a characteristic length scale of about $120$~nm.
Whilst the above parameters give an idea of the applicability regime of the theory, we emphasize that the semiclassical approximation generally gives good results outside of this regime as well, as discussed in the introduction.

\section{Scattering}
\label{sec:scattering}
In the previous section, we developed a completely general semiclassical theory for plasmons in two-dimensional inhomogeneous media. In this section, we study a specific simple scattering experiment with this theory. Specifically, we consider a two-dimensional electron gas at $z=0$, encapsulated by two homogeneous dielectric layers. The electron density in the $z=0$ plane is constant, except near the $\mathbf{x}$-origin, where a radially symmetric inhomogeneity is placed, which locally increases or decreases the electron density. We consider the situation where a plasmon, represented as a plane wave, comes in from the left-hand side ($x \rightarrow -\infty$), and is scattered by the inhomogeneity. 
Preliminary understanding of the scattering can be gained by plotting the classical trajectories of this system, shown in Fig.~\ref{fig:clastrajec}, which can be computed using the effective classical Hamiltonian~(\ref{eq:effclassicalHam}).

In this section, we construct a semiclassical description for this experiment. Our description is based on the in-plane induced potential $V_\mathrm{pl}$, which we denote by $V$ throughout this section in order to make the notation less heavy. We first identify the phase shift of the scattered plasmons in Sec.~\ref{subsec:crossec}, which can be used to calculate the total and differential scattering cross sections. In Sec.~\ref{subsec:scatphase}, we subsequently construct a formula for this phase shift within the semiclassical approximation. Finally, we rewrite this result in a way more suitable for numerical calculations in Sec.~\ref{subsec:scattdiff}. 

We remark that an alternative method to describe scattering of plasmons was developed in Ref.~\cite{Torre17}. This method uses the analogy between the RPA equations for the electrostatic potential and the Lipmann-Schwinger formalism in conventional scattering theory~\cite{Taylor72}. Whereas the formalism in Ref.~\cite{Torre17} seems to be more general, our approach is more suitable to construct a semiclassical description.

\subsection{Scattering cross sections and phase shifts}\label{subsec:crossec}
\begin{figure}[t]
  \centering
  \includegraphics[width=0.85\linewidth]{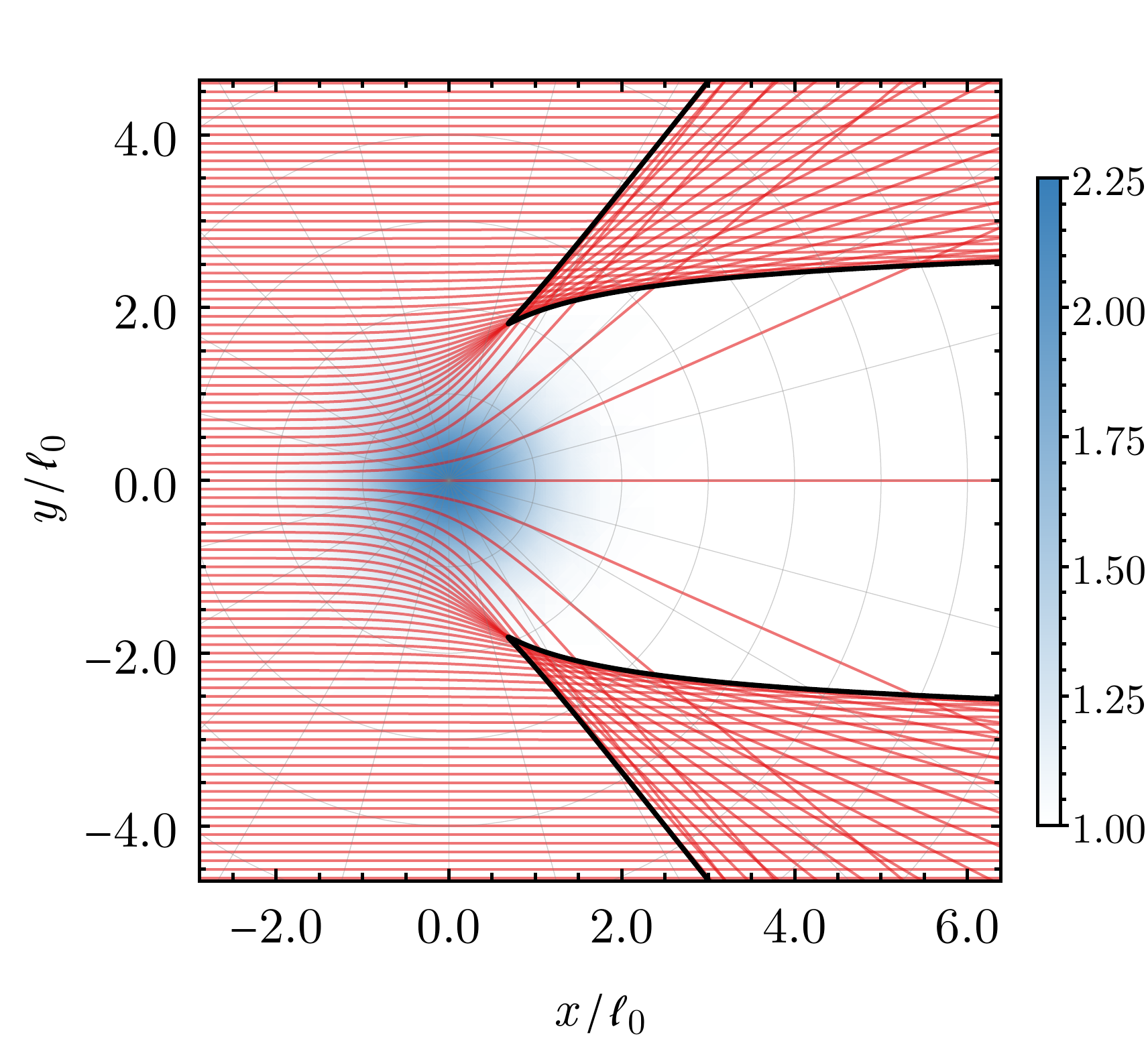}
  \caption{Classical trajectories of a plasmon, coming in from $x \rightarrow -\infty$. The maximum change in local density is $\delta n = +1.25$. The blue shading indicates the region where the electron density is increased. The solid black lines are the caustics.}  \label{fig:clastrajec}
\end{figure}

Before we define the phase shift, we first have to carefully set the scene. In order to obtain the relation between the velocity and the momentum, we consider the Hamiltonian system Eqs.~(\ref{eq:Hamiltonian-system}). A short calculation yields 

\begin{equation}  \label{eq:velocity-momentum-opposite}
  \mathcal{H}_0 \approx 1 - \frac{g_\mathrm{s} e^2 p_\mathrm{F}^2 q}{2 m \varepsilon_{\mathrm{avg}} \hbar E^2} , \quad
  \frac{\mathrm{d} \mathbf{x}}{\mathrm{d} \tau}= \frac{\partial \mathcal{H}_0}{ \partial \mathbf{q}} 
  \approx - \frac{g_\mathrm{s} e^2 p_\mathrm{F}^2}{2 m \varepsilon_{\mathrm{avg}} \hbar E^2} \frac{\mathbf{q}}{q},
\end{equation}
where we considered the limit of small $\mathbf{q}$, and $q = |\mathbf{q}|$. This counter-intuitive result shows that the (classical) velocity of the plasmon is in the opposite direction of the momentum $\mathbf{q}$.
Note that it is sufficient to calculate the velocity in the limit of small $q$, since it is not expected to change sign for larger $\mathbf{q}$.
A right-moving plasmon, propagating along the $x$ axis from negative infinity, thus corresponds to a plane wave with momentum $q_x=-|\mathbf{q}|<0$ parallel to the $x$ axis, see also Fig.~\ref{fig:clastrajec}.

In the remainder of this section, we derive the formulas that express the total and differential scattering cross section in terms of the phase shift. Readers who are familiar with this subject may directly skip to the result Eq.~(\ref{eq:diffcrosssec}).

We first take a closer look at the incoming plasmon, which is described by a plane wave, namely,
\begin{align}\label{eq:incomingwavepolarcoor}
  V (r,\theta) = A e^{-i \frac{q x}{\hbar}} = A e^{-i \frac{q r}{\hbar} \cos{\theta}} ,
\end{align}
where the amplitude $A$ is constant, we used that $q_x = -q = -|\mathbf{q}|$ and introduced polar coordinates in the last step.
Because the inhomogeneity is radially symmetric, we expand this incoming plane wave in radial waves. Using the Fourier-Bessel series~\cite{Abromowitz64}, we obtain
\begin{align}
  V (r,\theta) = A\sum_{m=-\infty}^{\infty} (-1)^m i^m J_m\left(\frac{q r}{\hbar}\right) e^{i m \theta},
\end{align}
where $J_m$ is the Bessel function of the first kind and where the factor $(-1)^m$ arises because the argument of the exponent is negative.
We subsequently rewrite each Bessel function as the sum of two Hankel functions, which represent incoming and outgoing waves.
Because we are interested in the far-field regime $r/\ell \gg 1$, far away from the inhomogeneity, we can use the asymptotic expansions of the Hankel functions~\cite{Abromowitz64} to obtain
\begin{multline} \label{eq:incoming-decomp-asymp}
  V (r,\theta) = A \sqrt{\frac{\hbar}{2 \pi q r}} \sum_{m=-\infty}^{\infty} e^{-i \frac{m \pi}{2}  -i\frac{\pi}{2} } e^{i m \theta}\\
  \times \left( e^{i\frac{q r}{\hbar} + i\frac{\pi}{4} - i \frac{m \pi }{2}} 
  - e^{-i\frac{q r}{\hbar} -i \frac{\pi}{4} + i \frac{m \pi }{2}} \right),
\end{multline}
where the first and second terms represent radially incoming and outgoing waves, respectively.

Our next step is to take a closer look at the outgoing wave.
Far away from the inhomogeneity ($r/\ell \gg 1$), the scattered plasmon is described by
\begin{align}\label{eq:scatsolSommerfeld}
  V^{\mathrm{scat}} (r,\theta) \approx A\left(e^{-i \frac{q r}{\hbar} \cos{\theta}} +  g(\theta) \frac{e^{-i\frac{q r}{\hbar} }}{\sqrt{r}}\right) .
\end{align}
The first term in this expression comes from the incoming wave Eq.~(\ref{eq:incomingwavepolarcoor}), and the second term represents the outgoing radial wave in the far field.
This asymptotic form can be understood as a extension of the Sommerfeld radiation condition for the Helmholtz equation~\cite{Vainberg89,Sommerfeld49,Schot92}.
Our goal is to find the scattering amplitude $g(\theta)$, which expresses the amplitude of the scattered wave scattered in every direction $\theta$.

It is common to express the effect of the inhomogeneity using a phase shift $\delta_m$~\cite{Griffiths05}. By definition, twice this phase shift is added to the outgoing radial components in expression~(\ref{eq:incoming-decomp-asymp}). Here we use a minus sign, mainly because it simplifies our results. On a more fundamental level, one may think that we need this minus sign because $\exp(- i q r / \hbar)$ is the outgoing wave in our problem, instead of $\exp(i q r / \hbar)$. We therefore define
\begin{multline}\label{eq:quantumscatteringsol}
  V^{\mathrm{scat}} (r,\theta) = A \sqrt{\frac{\hbar}{2 \pi q r}} \sum_{m=-\infty}^{\infty} e^{-i \frac{m \pi}{2}  -i\frac{\pi}{2} } e^{i m \theta} \\
  \times \left( e^{i\frac{q r}{\hbar} + i\frac{\pi}{4} - i \frac{m \pi }{2}} 
  - e^{-i\frac{q r}{\hbar} -i \frac{\pi}{4} +i \frac{m \pi }{2} - 2 i
    \delta_m } \right) .
\end{multline}
This definition shows that the phase shift is defined up to an integer multiple of $\pi$.
For future reference, we note that each of the terms in the series~(\ref{eq:quantumscatteringsol}) can be rewritten in terms of a sine. We have
\begin{align}\label{eq:quantumscatteringsolSine}
  V^{\mathrm{scat}}_m (r,\theta) \propto  \sin\left( \frac{q r}{\hbar} + \delta_m - \frac{m \pi}{2} + \frac{\pi}{4} \right),
\end{align}
where we left out the factors of proportionality.

We remark that in our scattering setup both the incoming wave and the inhomogeneity possess mirror symmetry around the $y=0$ axis. The scattered wave~(\ref{eq:quantumscatteringsol}) should therefore have the same symmetry. Requiring that $V(r,\theta) = V(r,-\theta)$ and reversing the summation order ($m \rightarrow -m$) on the right hand side, we find that the phase shifts with opposite $m$ have to be related by $\delta_{-m} =  \delta_m + n_m \pi$, where $n_m$ is an arbitrary integer. Since the phase shift is only defined up to an integer multiple of $\pi$, we may set $\delta_{-m} =  \delta_m$.

Comparing expressions~(\ref{eq:scatsolSommerfeld}) and~(\ref{eq:quantumscatteringsol}), we can now obtain an expression for the scattering amplitude $g(\theta)$.
Since the differential cross section is equal to the square of the absolute value of this scattering amplitude~\cite{Griffiths05}, we find
\begin{align}\label{eq:diffcrosssec}
  \frac{\mathrm{d} \sigma}{\mathrm{d} \theta} =|g(\theta)|^2 = \left|e^{-i\frac{\pi}{4}} \frac{\sqrt{2 \hbar} }{\sqrt{\pi q}} \sum_{m=-\infty}^{\infty} \sin{\delta_m} e^{ i m \theta - i \delta_m}\right|^2.
\end{align}
Integrating over all angles $\theta$ gives the total scattering cross section, specifically
\begin{align}\label{eq:totscatcross}
  \sigma = \int |g(\theta)|^2 \mathrm{d} \theta = \frac{4 \hbar}{ q} \sum_{m=-\infty}^{\infty}  \sin^2{\delta_m},
\end{align}
where we used the orthogonality of the azimuthal terms. 

At the end of this section, we briefly come back to the dimensionless parameters we introduced in the previous section. Since the scattering cross section has the dimensions of length, we can make it dimensionless by dividing by the characteristic length scale, i.e., $\tilde{\sigma} = \sigma/\ell$. This yields
\begin{align}\label{eq:totscatcrossdimen}
  \tilde{\sigma} = \frac{4 h}{\tilde{q}} \sum_{m=-\infty}^{\infty}  \sin^2{\delta_m},
\end{align}
where we used the definitions for $h$ and $\tilde{q}$ from Sec.~\ref{subsec:derivation-applicability}.

\subsection{Derivation of the semiclassical phase shift}\label{subsec:scatphase}
In the previous subsection, we showed that the phase shift is sufficient to calculate both the differential and the total scattering cross section.
In this subsection, we obtain a semiclassical expression for this phase shift. To this end, we first construct an asymptotic solution using the semiclassical Ansatz, and subsequently compare it to the solution~(\ref{eq:quantumscatteringsol}), which was derived using general quantum mechanical principles.

Our starting point is the asymptotic solution~(\ref{eq:V0solutionamplitudeopsep}), which is based on the classical trajectories. In the construction of this expression, we implicitly assumed that to each point $\mathbf{x}$ corresponds a single trajectory. However, this is generally not the case, because multiple electron trajectories may arrive at the same point, cf. Fig.~\ref{fig:clastrajec}. Physically, this is nothing but the well-known phenomenon of interference. 
We obtain the full asymptotic solution by adding the contributions of the individual trajectories~\cite{Maslov81,Guillemin77}, that is,
\begin{align} \label{eq:MaslovCanOpSolCart}
  V^{\mathrm{SC}} (\mathbf{x}) =   \sum_j \frac{A_0^0}{\sqrt{ \varepsilon_{\mathrm{avg}}\left|\frac{\partial S_j}{\partial \mathbf{x}}\right| }} \frac{1}{\sqrt{\left|J_j (\mathbf{x})\right|}}  e^{-i \frac{\pi}{2} \mu_j } e^{\frac{i}{\hbar} S_j (\mathbf{x})},
\end{align}
where $S_j$ is the action along the $j$-th trajectory. The object $\mu_j$ is the Maslov index~\cite{Maslov81,Guillemin77,Arnold89}, which expresses the complex phase of the Jacobian $J_j$ and will be discussed in more detail later on. Formally, this solution can be obtained with the so-called Maslov canonical operator~\cite{Maslov81}.

As previously mentioned, Fig.~\ref{fig:clastrajec} shows the classical trajectories that arise from the incoming plane wave~(\ref{eq:incomingwavepolarcoor}). 
In principle, we could use Eq.~(\ref{eq:MaslovCanOpSolCart}) to construct an asymptotic solution for our scattering problem based on these trajectories, see Ref.~\cite{Reijnders18} for an example.
However, as clearly explained in Ref.~\cite{Berry72}, it is not at all straightforward to relate this asymptotic solution to Eq.~(\ref{eq:quantumscatteringsol}), which takes the form of a series of radially incoming and outgoing waves labeled by the index $m$.

We therefore construct the full asymptotic solution in a different way. Keeping the asymptotic expression~(\ref{eq:quantumscatteringsol}) in mind, we start by considering radially symmetric incoming plane waves. 
In order to compute the classical trajectories, we first write down the initial conditions, which form a one-dimensional surface $\Lambda^1$ in phase space.
Since we consider radially symmetric waves, $\Lambda^1$ is a circle:
\begin{align}  \label{eq:def-Lambda1}
  \Lambda^1 = \{r = r_0, \: \theta = \alpha, \: q_r = q_{r,0}, \: q_\theta = q_{\theta,0}\} ,
\end{align}
where the parameter $\alpha \in [0,2\pi)$ corresponds to the angle of incidence. Although one should theoretically consider $r_0 \to \infty$, a numerical computation requires a finite $r_0$, and we assert that $r_0/\ell \gg 1$. Moreover, note that $q_r = q_{r,0} > 0$ gives rise to an incoming wave, cf. Eq.~(\ref{eq:velocity-momentum-opposite}).
The total momentum is given by $(q_0)^2 = (q_{r,0})^2 + \left(q_{\theta,0}/r_0\right)^2$, and is equal to $q_{r,0}$ at large distances. 

\begin{figure}[t]
  \centering
  \includegraphics[width=0.99\linewidth]{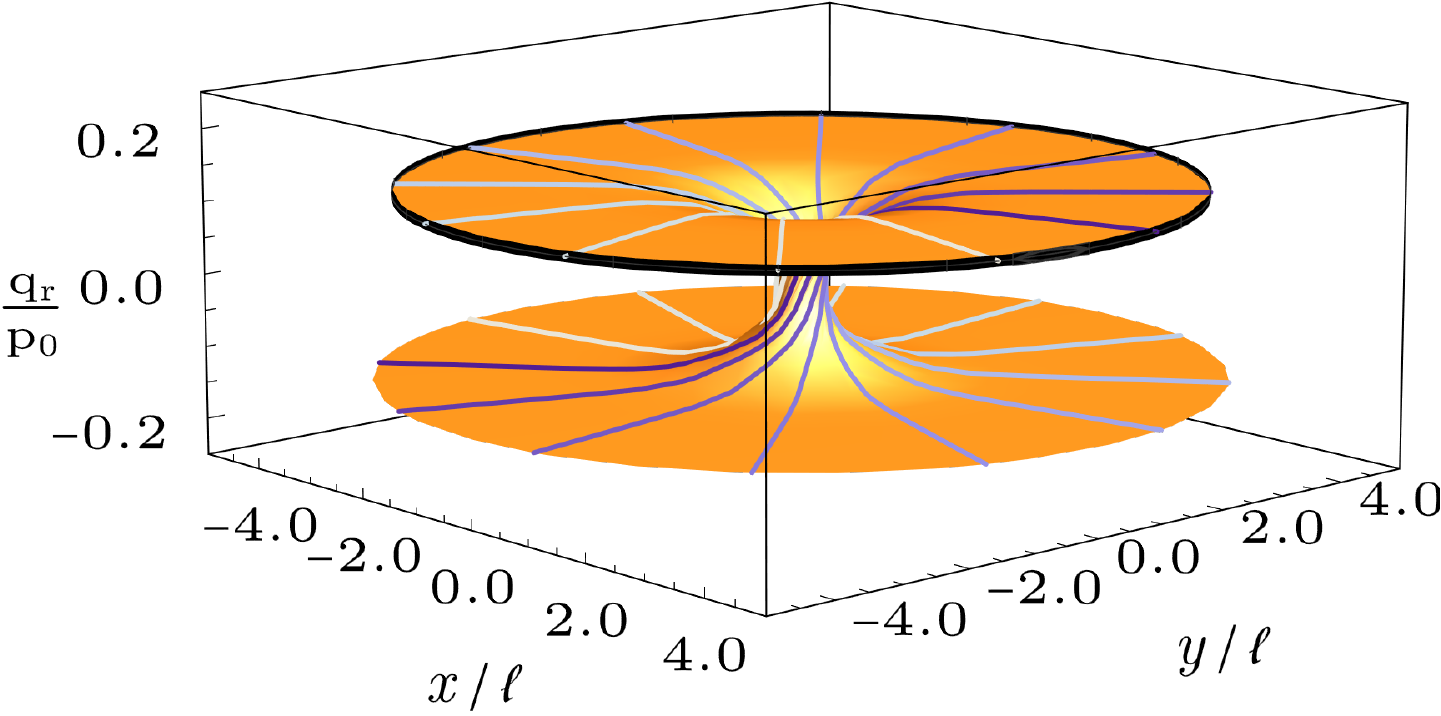}
  \caption{Schematic representation of the Lagrangian manifold $\Lambda^2$ in orange. The surface $\Lambda^1$ is depicted by the black circle. The purple, blue and grey colored lines represent plasmon trajectories, coming in from $\Lambda^1$ with momentum $q_r^0$.}
  \label{fig:Lambda2}
\end{figure}

Next, we consider the time evolution of each point on $\Lambda^1$, shown as a black circle in Fig.~\ref{fig:Lambda2}, by the Hamiltonian system $\mathrm{d}\mathbf{x}/\mathrm{d} t = \partial \mathcal{H}_0/\partial \mathbf{p}$, $\mathrm{d}\mathbf{p}/\mathrm{d} t = -\partial \mathcal{H}_0/\partial \mathbf{x}$ with the effective classical Hamiltonian $\mathcal{H}_0$, see Eq.~(\ref{eq:effclassicalHam}). This time evolution gives rise to the two-dimensional surface $\Lambda^2$, shown in orange in Fig.~\ref{fig:Lambda2}. Each point on this surface can be parameterized by the angle of incidence $\alpha$ and the time $\tau$, i.e.,
\begin{align}
  \Lambda^2 = \{&r = r(\tau, \alpha), \theta = \theta(\tau, \alpha), q_r = q_r (\tau, \alpha), \nonumber \\ 
  &  q_\theta = q_\theta(\tau, \alpha) \; | \; \alpha \in [0,2\pi), \tau \in [0, \infty) \} .
\end{align}
With a more detailed analysis, one can show that this surface is a so-called Lagrangian manifold~\cite{Maslov81,Guillemin77,Arnold82,Reijnders18,Dobrokhotov03}. An important property of such a manifold is that the action integral~(\ref{eq:action-def}) is path independent, that is, its outcome only depends on the initial and final points on $\Lambda^2$, and not on the specific path that is used to compute the integral.

Looking at Fig.~\ref{fig:Lambda2}, one sees that $\Lambda^2$ consists of two distinct leaves. In other words, when one projects $\Lambda^2$ onto the coordinate plane, each point on the coordinate plane corresponds to two points on $\Lambda^2$. The upper leaf corresponds to the incoming wave, since it consists of points with radial momentum $q_r^\mathrm{inc} > 0$. Similarly, the lower leaf corresponds to the outgoing wave, since it consists of points with $q_r^\mathrm{out} < 0$.
The two leaves join at the point $r_\mathrm{c}(q_\theta)$, where $q_r = 0$. This point is a classical turning point, as can be inferred from Eq.~(\ref{eq:velocity-momentum-opposite}).
Moreover, one can show that $q_r^2 \propto r-r_\mathrm{c}$ in the vicinity of $r_\mathrm{c}$, which implies that we are dealing with a simple turning point~\cite{Berry72,Maslov81,Poston78,Arnold82}.

The proof of this statement is simplest when there is no inhomogeneity at all, in other words, when the system is homogeneous, since in this case both $q_\theta$ and $q^2 = q_r^2(r) + q_\theta^2/r^2$ are constants of motion. Provided that $q_\theta \neq 0$, one has $r_\mathrm{c} = |q_\theta |/ q $, and a Taylor expansion of $q^2 = q_r^2(r) + q_\theta^2/r^2$ around $r_\mathrm{c}$ readily gives the result.
When there is a radially symmetric inhomogeneity present, $q_\theta$ is still a constant of motion, since the effective classical Hamiltonian $\mathcal{H}_0$ does not depend on $\theta$. When $q_\theta$ is small, one can consider the Taylor expansion~(\ref{eq:velocity-momentum-opposite}). Solving the equation $\mathcal{H}_0=0$ for the energy $E$ leads to the familiar square-root dispersion relation
\begin{equation}
  E \approx \sqrt{\frac{g_\mathrm{s} e^2 p_\mathrm{F}(r)^2}{2 m \varepsilon_\mathrm{avg} \hbar} q}
  = \sqrt{\frac{2\pi e^2 \hbar n(r)}{m \varepsilon_{\mathrm{avg}}} q} ,
\end{equation}
where $p_\mathrm{F}(r)$ and $n(r)$ depend on the radial coordinate, and we used that $n(r)=g_\mathrm{s} p_\mathrm{F}^2(r)/(4\pi\hbar^2)$ in the last equality. The turning point is subsequently defined by the relation $E^2=2\pi e^2 \hbar n(r) |q_\theta|/(m \varepsilon_{\mathrm{avg}} r_\mathrm{c})$, since it corresponds to $q_r=0$. Performing a Taylor expansion in $r$ around $r_\mathrm{c}$ then gives
\begin{equation}
  q_r^2 \approx \frac{2 q_\theta^2}{r_\mathrm{c}^2} \left(\frac{1}{r_\mathrm{c}} - \frac{n'(r_\mathrm{c})}{n(r_\mathrm{c})} \right) (r-r_\mathrm{c}) ,
\end{equation}
which indeed shows that $q_r^2 \propto r-r_\mathrm{c}$.
When $|q_\theta|$ is not small, one can use numerical methods to show that the latter proportionality continues to hold.
This result is in accordance with the general theory of caustics and singularities, which states that this proportionality is generic for effectively one-dimensional geometries~\cite{Poston78,Arnold82}.

Now that we have studied the classical trajectories and the Lagrangian manifold $\Lambda^2$, we can construct the full asymptotic solution.
Since we started with radially symmetric incoming plane waves, we parameterize the Cartesian plane with polar coordinates, i.e., $\mathbf{x} = \mathbf{x}(r,\theta)$.
As expressed by Eq.~(\ref{eq:MaslovCanOpSolCart}) and mentioned previously, we obtain the induced potential at the point $\mathbf{x}(r,\theta)$ by adding the contributions of the two points on $\Lambda^2$ that are projected onto $\mathbf{x}(r,\theta)$. Note that this construction cannot be used in the vicinity of the turning point, which is a singular point of the projection. We therefore limit ourselves to the far-field regime $r / \ell \gg 1$.

We first consider the action~(\ref{eq:action-def}), Since it is given by a line integral over a path on $\Lambda^2$, we can consider the action as a function of the coordinates $(\tau,\alpha)$ on $\Lambda^2$. The action $S_j(r,\theta)$ on a given leaf follows from this more general quantity by projection.
Switching to polar coordinates, and using the definitions of the polar momenta, i.e., $q_x = \cos\theta \, q_r  - \frac{\sin\theta}{r} q_\theta $, and $q_y = \sin\theta \, q_r + \frac{\cos\theta}{r } q_\theta$, cf. Refs.~\cite{Goldstein02,Arnold82}, we have
\begin{equation}  \label{eq:def-action-radial}
  S(\tau,\alpha) = \int_\mathcal{C} \left\langle q_\mathbf{x}, \mathrm{d}\mathbf{x} \right\rangle = \int_\mathcal{C} q_r \mathrm{d}r + q_\theta \mathrm{d}\theta .
\end{equation}
The integral is taken over the line $\mathcal{C}$ with starting point, the so-called central point, $(\tau_0,\alpha_0)$~\cite{Maslov81} and end point $(\tau,\alpha)$. For convenience, we set $\tau_0=\alpha_0=0$ from here on.

As we previously mentioned, the integral~(\ref{eq:def-action-radial}) is independent of the specific path, since $\Lambda^2$ is a Lagrangian manifold~\cite{Maslov81,Guillemin77,Arnold82,Reijnders18,Dobrokhotov03}.
Given a point $(r,\theta)$, we therefore split the integration into two parts: we first integrate along $\Lambda^1$ to the initial point of the trajectory on which the point $(r,\theta)$ lies, and then proceed the integration along the trajectory.
Figure~\ref{fig:Lambda2} shows that the two trajectories that contain the point $(r,\theta)$, corresponding to the two different leaves of $\Lambda^2$, and hence to the incoming and the outgoing waves, originate from two different angles $\theta_1=\alpha_1$ and $\theta_2=\alpha_2$.
We therefore compute the integral separately for both trajectories.
For the trajectory corresponding to the incoming wave, we have
\begin{align}  \label{eq:S1-computed}
  S_1(\tau,\alpha) &= \int_{(0,0)}^{(0,\alpha_1)} (q_r \mathrm{d}r + q_\theta \mathrm{d}\theta) + \int_{(0,\alpha_1)}^{(\tau,\alpha)} (q_r \mathrm{d}r + q_\theta \mathrm{d}\theta) , \nonumber \\
  S_1(r,\theta) &= q_\theta \theta_1 + q_\theta(\theta - \theta_1) + \int_{r_0}^r q_r^\mathrm{inc}(r') \mathrm{d} r' ,
\end{align}
where $r_0$ was defined in Eq.~(\ref{eq:def-Lambda1}). We used that $q_\theta$ is constant because $\mathcal{H}_0$ is independent of $\theta$ due to the radial symmetry, and immediately see that the terms containing $\theta_1$ drop out.

In a similar way, we can compute the action for the trajectory corresponding to the outgoing wave. Since it has passed the turning point $r_\mathrm{c}$, there is an additional radial contribution. We find
\begin{multline}
  S_2(r,\theta) = q_\theta \theta_2 + q_\theta(\theta - \theta_2) + \int_{r_0}^{r_\mathrm{c}} q_r^\mathrm{inc}(r') \mathrm{d} r' \\ + \int_{r_\mathrm{c}}^{r} q_r^\mathrm{out}(r') \mathrm{d} r' ,
\end{multline}
where the terms containing $\theta_2$ cancel as before.

At this point, we have to consider an additional constraint: the action $S(\tau,\alpha)$ should be single-valued, since $\Lambda^2$ is a Lagrangian manifold. When we consider the action along a circular path that goes around the central hole in $\Lambda^2$, it should therefore equal a multiple of $2\pi$. This is nothing but an expression of the Bohr-Sommerfeld quantization condition~\cite{Maslov81}, see also Refs.~\cite{Curtis04,Reijnders22} for similar applications. We have
\begin{align}
  \int_0^{2\pi} q_\theta \mathrm{d} \theta = 2 \pi m \hbar,
\end{align}
where $m$ is the (integer) azimuthal quantum number.
This quantization condition determines the values which $q_\theta$ can take, and shows that $q_\theta = m \hbar$. Inserting this result into the expressions for $S_1$ and $S_2$, and considering Eq.~(\ref{eq:MaslovCanOpSolCart}), we see that the angular dependence of our asymptotic solution is given by $\exp(im\theta)$, in accordance with the quantum mechanical results discussed in Sec.~\ref{subsec:crossec}.

Our next step is to determine the Jacobian in Eq.~(\ref{eq:MaslovCanOpSolCart}).
In general, it is given by
\begin{align}
  J = \det\begin{pmatrix} 
    \frac{\partial x}{\partial \tau} & \frac{\partial x}{\partial \alpha}\\
    \frac{\partial y}{\partial \tau} & \frac{\partial y}{\partial \alpha}
  \end{pmatrix}
  = \det\begin{pmatrix} 
    \frac{\partial x}{\partial r} & \frac{\partial x}{\partial \theta}\\
    \frac{\partial y}{\partial r} & \frac{\partial y}{\partial \theta}
  \end{pmatrix}
  \det\begin{pmatrix} 
    \frac{\partial r}{\partial \tau} & \frac{\partial r}{\partial \alpha}\\
    \frac{\partial \theta}{\partial \tau} & \frac{\partial \theta}{\partial \alpha}
  \end{pmatrix} ,
\end{align}
where the second equality follows from our parametrization in polar coordinates. The first determinant on the right-hand side is the usual Jacobian associated with the transformation to polar coordinates and equals $r$. In order to compute the second Jacobian, we first consider its value on $\Lambda^1$. Since $\partial r/\partial \alpha = 0$ and $\partial \theta/\partial \alpha = 1$ on $\Lambda^1$, we have
\begin{align}  \label{eq:Jacobian-radial-computed}
  J = r \frac{\partial r}{\partial \tau}.
\end{align}
Using the variational system for Hamilton's equations~\cite{Dobrokhotov03}, see also Refs.~\cite{Reijnders18,Reijnders22}, one can show that the time derivatives of $\partial r/\partial \alpha$ and $\partial \theta/\partial \alpha$ equal zero on $\Lambda^2$. The result~(\ref{eq:Jacobian-radial-computed}) is therefore valid on all of $\Lambda^2$.
Using Hamilton's equations, we see that the Jacobian can also be written as
\begin{equation}  \label{eq:Jacobian-radial-computed-plusHam}
  J = r \frac{\partial \mathcal{H}_0}{\partial q_r} = r \frac{\partial \mathcal{H}_0}{\partial q} \frac{q_r}{q} ,
\end{equation}
where $q = |\mathbf{q}|$ and we have used that the effective Hamiltonian $\mathcal{H}_0$ is a function of $|\mathbf{q}|$ only.

At a given point $(r,\theta)$, the incoming and outgoing waves have opposite radial momenta, $q_r^\mathrm{inc}(r)=-q_r^\mathrm{out}(r)$, see also Fig.~\ref{fig:Lambda2}. Equation~(\ref{eq:Jacobian-radial-computed-plusHam}) then shows that $J_1 = -J_2$. Since Eq.~(\ref{eq:MaslovCanOpSolCart}) contains the absolute value of the Jacobian, we obtain the same factor for the contribution of each leaf.
The sign of the Jacobian is, nevertheless, taken into account through the Maslov index~\cite{Maslov81,Guillemin77,Arnold67,Dobrokhotov03}. On the upper leaf, which corresponds to the incoming waves, the sign of the Jacobian is equal to its sign on $\Lambda^1$, and we set $\mu_1=0$. On the lower leaf, which corresponds to outgoing waves, the Jacobian has the opposite sign. The Maslov index now regulates the analytic continuation of the square root of this Jacobian. Computations performed explicitly in Ref.~\cite{Reijnders22}, cf. Ref.~\cite{Dobrokhotov03}, show that $\mu_2=-1$ for points on the lower leaf.

Lastly, we calculate the factor $\left|\partial S/\partial \mathbf{x}\right|$ in Eq.~(\ref{eq:MaslovCanOpSolCart}), which is part of the amplitude. In polar coordinates, we have
\begin{align}
  \left|\frac{\partial S}{\partial \mathbf{x}}\right|  = \sqrt{\left(\frac{\partial S}{\partial r}\right)^2 + \left(\frac{1}{r} \frac{\partial S}{\partial \theta}\right)^2}
  = \sqrt{q_r^2(r) + \left(\frac{q_\theta}{r}\right)^2 },
\end{align}
where the latter expression is to be understood on the Lagrangian manifold $\Lambda^2$. Since $q_r^\mathrm{inc}(r)=-q_r^\mathrm{out}(r)$, this factor is equal for both contributions to the sum~(\ref{eq:MaslovCanOpSolCart}).

We are now ready to combine all ingredients and compute the asymptotic solution~(\ref{eq:MaslovCanOpSolCart}). As a final preparatory step, we rewrite the integral from $0$ to $r$ in our expression~(\ref{eq:S1-computed}) for $S_1$ as the sum of an integral from $0$ to $r_\mathrm{c}$ and an integral from $r_\mathrm{c}$ to $r$.
Putting everything together, we obtain 
\begin{align} \label{eq:scsolution}
  V_m^{\mathrm{SC}} (r,\theta) &= \frac{A_{0,m}^0}{\varepsilon_{\mathrm{avg}}^{1/2} \left((q_r^\mathrm{inc})^2 + (\frac{q_\theta}{r})^2 \right)^{1/4}} \frac{1}{\sqrt{r |\frac{\partial \mathcal{H}_0}{\partial q} \frac{q_r^\mathrm{inc}}{q}|}} \nonumber \\
  & \;\;\; \times \left(e^{\frac{i}{\hbar} \int_{r_\mathrm{c}}^{r} q_r^{\mathrm{inc}} \mathrm{d} r' +  i \frac{\pi}{4}} - e^{-\frac{i}{\hbar} \int_{r_\mathrm{c}}^{r} q_r^{\mathrm{inc}} \mathrm{d} r' - i  \frac{\pi}{4}} \right) \nonumber \\
  & \;\;\; \times e^{i m \theta} e^{\frac{i}{\hbar} \int_{r_0}^{r_\mathrm{c}} q_r^{\mathrm{inc}} \mathrm{d} r'- i \frac{ \pi}{4}} 
  ,
\end{align} 
where we omitted the arguments of $q_r$ and used that $q_r^\mathrm{out} = -q_r^\mathrm{inc}$.
We also added an index $m$, because we constructed an asymptotic solution for a radially incoming wave with angular momentum $q_\theta = m \hbar$.

Comparing the asymptotic solution~(\ref{eq:scsolution}) with the $m$-th component of the general solution~(\ref{eq:quantumscatteringsol}), we observe that they exhibit the same asymptotic behavior. First of all, their angular dependence is the same, as they are both proportional to $\exp(i m \theta)$. Second, they both decay as $1/\sqrt{r}$ in the far field, since $q_r$ becomes constant for $r/\ell \gg 1$.

At this point, we can obtain an asymptotic solution for the original plane wave, see Fig.~\ref{fig:clastrajec}, by considering a series of incoming plane waves and matching the coefficients $A_{0,m}^0$ with the constants in front of the series in Eq.~(\ref{eq:quantumscatteringsol}). However, this is not at all necessary, since we previously established that we can express the scattering cross section in terms of the phase shift $\delta_m$. A semiclassical expression for this phase shift can be directly determined by rewriting the asymptotic solution~(\ref{eq:scsolution}) in the form of a sine, namely,
\begin{align}\label{eq:scsolutionSine}
  V_m^{\mathrm{SC}} (r,\theta) \propto  \sin\left(\frac{1}{\hbar} \int_{r_\mathrm{c}}^{r} q_r^\mathrm{inc}(r') \mathrm{d} r' + \frac{\pi}{4}\right) .
\end{align}
Comparing this result with Eq.~(\ref{eq:quantumscatteringsolSine}), we immediately obtain
\begin{align}\label{eq:scatteringphaser}
  \delta_m  = \lim_{r \to \infty} \left( \int_{r_\mathrm{c}}^{r} \frac{q_r^\mathrm{inc}(r')}{\hbar} \mathrm{d} r'  -\frac{q_\infty r}{\hbar} \right) + \frac{m \pi}{2},
\end{align}
where we used $q_\infty$ to denote the value of $|\mathbf{q}|$ at infinity where the presence of the inhomogeneity is no longer felt. By taking the limit, we can rewrite this expression as
\begin{align}\label{eq:scatteringphaseBerry}
  \delta_m  =  \int_{r_\mathrm{c}}^{\infty} \bigg( \frac{q_r^{\mathrm{inc}}(r')}{\hbar} - \frac{q_\infty}{\hbar} \bigg) \mathrm{d} r' -\frac{q_\infty r_\mathrm{c}}{\hbar} + \frac{m \pi}{2}.
\end{align}
Although this result for the semiclassical phase shift has the same form as most results in scientific literature, see e.g. Ref.~\cite{Berry72}, there is an important difference. Most semiclassical expressions for the phase shift that are commonly found in the literature are derived by first performing separation of variables in the two-dimensional differential equation and subsequently constructing an asymptotic solution for the remaining one-dimensional equation. On the other hand, we constructed an asymptotic solution for a two-dimensional pseudodifferential equation, where we carefully accounted for the contributions of the different trajectories on the Lagrangian manifold $\Lambda^2$.
An added advantage of our approach is that there is no need to perform an explicit Langer substitution Ref.~\cite{Langer37,Heading62}.
Instead, the correct expression naturally arises from the quantization of the azimuthal variable, cf. Ref.~\cite{Reijnders22}.

\subsection{Alternative expression for the phase shift} \label{subsec:scattdiff}

Unfortunately, expression~(\ref{eq:scatteringphaser}) for the scattering phase is not very convenient when one wants to compute the phase shift for a given system. If $q = |\mathbf{q}|$ is the solution of $\mathcal{H}_0(\mathbf{x},q)=0$, then $q_r^\mathrm{inc}(r) = (q^2(r)-m^2 \hbar^2/r^2)^{1/2}$ for a given value of $m$.
The integral in expression~(\ref{eq:scatteringphaser}) therefore always converges very slowly (because of the second term in $q_r^{\mathrm{inc}}$), even when $q(r)$ rapidly becomes constant. 
Hence, we need to use a large cutoff radius in a numerical implementation.
In this subsection, we use a trick from Ref.~\cite{Fluegge94} to obtain an expression for $\delta_m$ that is more suitable for practical calculations.

To this end, we consider a second system, for which $n^{(0)}$ and $\varepsilon_\mathrm{avg}$ are constant, and correspond to the values far away from the inhomogeneity in the first system.
In this second system the momentum $q$ is constant, and equal to $q_\infty>0$ which was defined in the previous subsection. The radial momentum of the incoming wave is then given by $q_r^0(r) = (q_\infty^2-m^2\hbar^2/r^2)^{1/2}$ and vanishes at the classical turning point $r_\mathrm{c}^0 = |m|\hbar/q_\infty$.
We now rewrite expression~(\ref{eq:scatteringphaser}) as
\begin{multline}  \label{eq:scatteringphase-split-limit}
  \delta_m  = \lim_{r \to \infty} \left( \int_{r_\mathrm{c}}^{r} \frac{q_r^{\mathrm{inc}}(r')}{\hbar} \mathrm{d} r'  -\int_{r^{0}_c}^{r} \frac{q^{0}_r(r')}{\hbar} \mathrm{d} r' \right) \\
  + \lim_{r \to \infty} \left(\int_{r^{0}_c}^{r} \frac{q^{0}_r(r')}{\hbar} \mathrm{d} r'   -\frac{q_\infty r}{\hbar} \right) +  \frac{m \pi}{2},
\end{multline}
where we were allowed to split the limit into two because both converge.

Using our expression for $q_r^0(r)$, the integral in the second limit can be computed explicitly~\cite{Fluegge94}. Taking the limit $r \to \infty$ in the result, we find
\begin{equation}
  \lim_{r \to \infty} \left(\int_{r^{0}_c}^{r} \frac{q^{0}_r(r')}{\hbar} \mathrm{d} r'   -\frac{q_\infty r}{\hbar} \right) = -\frac{\pi}{2} \frac{q_\infty r_\mathrm{c}^0}{\hbar}  = -\frac{|m|\pi}{2} .
\end{equation}
Inserting this result into expression~(\ref{eq:scatteringphase-split-limit}), we obtain
\begin{multline}  \label{eq:scatteringphaseFlugge-pre}
  \delta_m  = \lim_{r \to \infty} \left( \int_{r_\mathrm{c}}^{r} \frac{q_r^{\mathrm{inc}}(r')}{\hbar} \mathrm{d} r'  -\int_{r^{0}_c}^{r} \frac{q^{0}_r(r')}{\hbar} \mathrm{d} r' \right)\\
  + \frac{\pi}{2} \left(m - |m|\right).
\end{multline}
The last part of this expression equals zero for positive $m$, and $-|m|\pi$ for negative $m$. In section~\ref{subsec:crossec}, we noted that the phase shift is only defined up to an integer multiple of $\pi$. 
We can thus freely add a multiple of $\pi$ to expression~(\ref{eq:scatteringphaseFlugge-pre}) without changing the physical result. We therefore write
\begin{equation}  \label{eq:scatteringphaseFlugge}
  \delta_m  = \lim_{r \to \infty} \left( \int_{r_\mathrm{c}}^{r} \frac{q_r^{\mathrm{inc}}(r')}{\hbar} \mathrm{d} r'  -\int_{r^{0}_c}^{r} \frac{q^{0}_r(r')}{\hbar} \mathrm{d} r' \right) ,
\end{equation}
which has the property that all phase shifts vanish in the absence of an inhomogeneity, cf. Refs.~\cite{Berry72,Fluegge94}. It also satisfies the symmetry that we previously derived, $\delta_m = \delta_{-m}$.

Finally, we argue that expression~(\ref{eq:scatteringphaseFlugge}) is more suitable for practical calculations than expression~(\ref{eq:scatteringphaser}). Let us first consider an inhomogeneity with a finite range, such that $q(r) = q_\infty$ for $r>R$. We may then split the limit and obtain
\begin{multline}  \label{eq:scatteringphaseFlugge-ex}
  \delta_m  = \int_{r_\mathrm{c}}^{R} \frac{q_r^{\mathrm{inc}}(r')}{\hbar} \mathrm{d} r'  -\int_{r^{0}_c}^{R} \frac{q^{0}_r(r')}{\hbar} \mathrm{d} r' \\ + 
  \lim_{r \to \infty} \int_{R}^{r} \frac{q_r^{\mathrm{inc}}(r') - q^{0}_r(r')}{\hbar}  \mathrm{d} r' .
\end{multline}
The last part of this expression vanishes, because the behavior of the two integrands is identical for $R<r<\infty$. More colloquially, the (large $r$) tails of the integrands cancel, and as a result we only have to integrate over a finite interval. Moreover, we directly see that $\delta_m$ vanishes when both $r_\mathrm{c} > R$ and $r_\mathrm{c}^0 > R$.
Comparing this to Eq.~(\ref{eq:scatteringphaser}), we see that the integral in the latter expression converges very slowly in terms of $r$, because of the slow decay of $q_r^\mathrm{inc}(r) = (q^2(r)-m^2 \hbar^2/r^2)^{1/2}$. We therefore have to integrate over a much larger interval in order to obtain an accurate result, which is computationally more demanding.

When the inhomogeneity does not have a finite range, the difference between the two expressions is less clear cut. 
Nevertheless, one can use a similar argument to show that expression~(\ref{eq:scatteringphaseFlugge}) converges faster, in terms of $r$, than expression~(\ref{eq:scatteringphaser}) whenever $q(r)$ decays faster than $1/r$.
We therefore consider expression~(\ref{eq:scatteringphaseFlugge}) more suitable for our numerical computations in the next section.

\section{Numerical results for the scattering cross section} \label{sec:numerical}
In this section, we apply our scattering theory to a specific example. First, we introduce an explicit shape of the inhomogeneity, i.e. a change in local electron density $n^{(0)}$. Subsequently, we numerically evaluate the semiclassical phase shift~(\ref{eq:scatteringphaseFlugge}) with Wolfram Mathematica~\cite{Mathematica} and discuss the total and differential scattering cross sections, given by Eqs.~(\ref{eq:totscatcrossdimen}), and~(\ref{eq:diffcrosssec}), respectively. We show their dependence on three parameters: the change in local electron density, plasmon energy, and decay length of the inhomogeneity. In order to gain a better understanding of the system, we also discuss the classical trajectories associated with the plasmon scattering, similar to those shown in Fig.~\ref{fig:clastrajec}.

We can compute the semiclassical phase shift with different approaches. In the first approach, we numerically determine the root of the effective classical Hamiltonian $\mathcal{H}_0$ for a given coordinate $r$, to obtain the radial momentum $q_r$. Following the discussion in Sec.~\ref{subsec:scattdiff}, the phase shift can then be computed efficiently using Eq.~(\ref{eq:scatteringphaseFlugge}).
In the second approach, we numerically solve Hamilton's equations~(\ref{eq:Hamiltonian-system}), supplemented with a differential equation for the action $S$ based on Eq.~(\ref{eq:action-def}). From this we determine the phase shift directly, by noting that both terms in Eq.~(\ref{eq:scatteringphaseFlugge}) are the actions of the plasmon in the two systems discussed in Sec.~\ref{subsec:scattdiff}. Unfortunately, the latter approach is computationally expensive, because it requires small time steps in the numerical integration. We therefore use the first approach in our computation. 

In practical applications of Eq.~(\ref{eq:scatteringphaseFlugge}), we have to choose a cut-off radius $R$ for the integration. We pick $R$ in such a way that the last term in Eq.~(\ref{eq:scatteringphaseFlugge-ex}) is negligible. This choice of $R$ is, naturally, highly affected by the spatial decay of the inhomogeneity. In order to ensure rapid convergence, we therefore consider a Gaussian inhomogeneity. However, the theory is not limited to this form. Specifically, we describe the local Fermi momentum as 
\begin{align}
  p_\mathrm{F} (r) = p_0 \left(1 + \delta p_{\mathrm{F}} \, e^{-r^2/\ell^2}\right),
\end{align}
where $p_0$ is the Fermi momentum far away from the inhomogeneity. In this model for the inhomogeneity, we have two independent parameters, namely the maximum change in local Fermi momentum $\delta p_\mathrm{F}$, and the decay length $\ell$. Following Eq.~(\ref{eq:TF-approx}), the change in Fermi momentum can be directly related to the change in electron density by $\delta n = g_\mathrm{s} p_\mathrm{F}^{2} \left(2 \delta p_\mathrm{F} + \delta p_\mathrm{F}^2 \right) / 4 \pi \hbar^2 n^{(0)}$. Throughout this section, we consider changes in the local electron density instead of the Fermi momentum, since they can be more easily compared to experiments. 

\begin{figure}[t]
  \centering
  \includegraphics[width=0.9\linewidth]{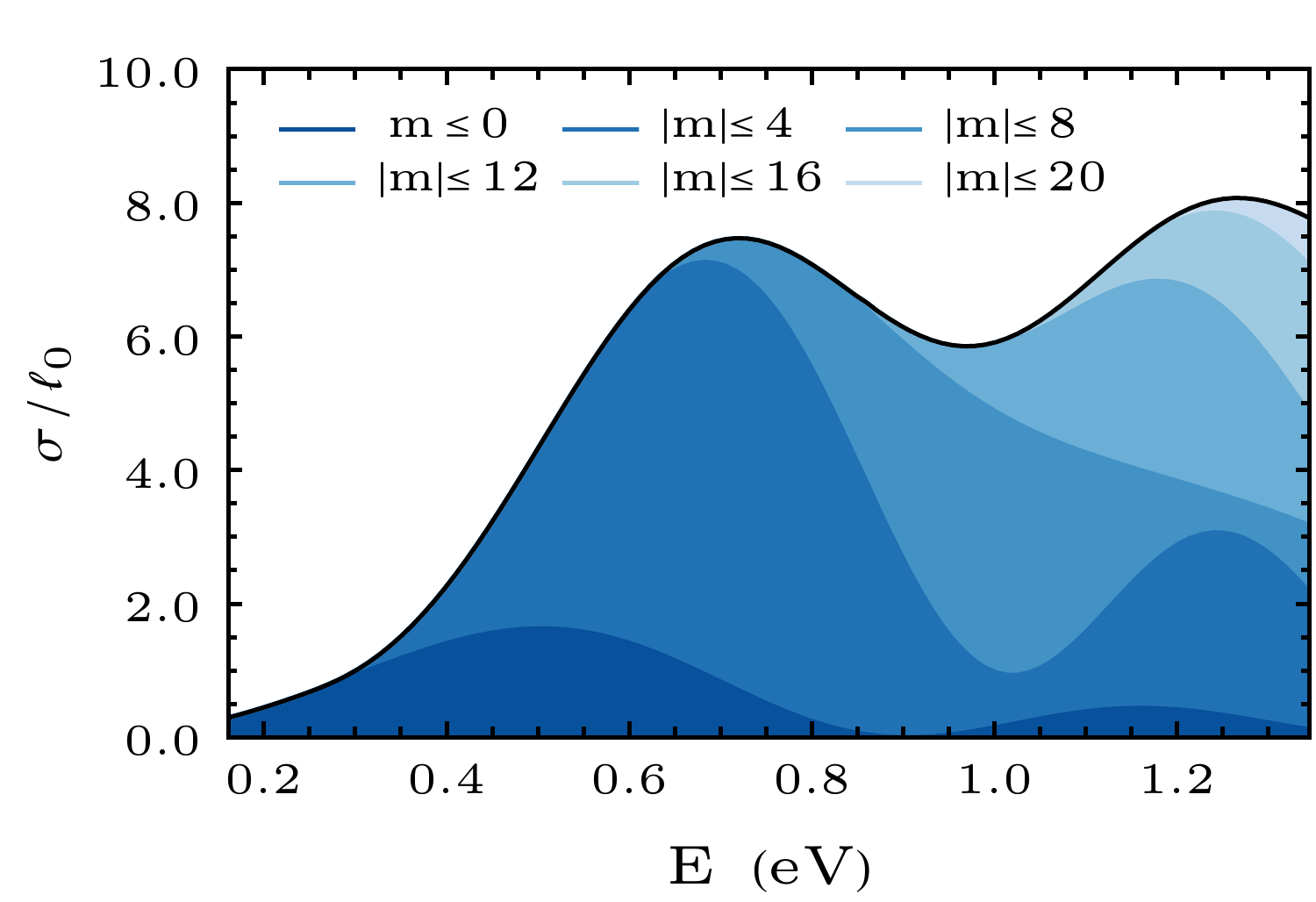}
  \caption{Numerically evaluated total scattering cross section $\sigma / \ell_0$ as function of the plasmon energy $E$. In this plot, the local density change $\delta n =  1.25$, and the decay length is $\ell_0 = 13.2$~nm. The shades of blue indicate the partial waves which have been taken into account, whilst the solid black line indicates the cross section for $|m| \leq 28$. The scattering cross section increases monotonously up to around $E = 0.7$~eV, after which it starts to oscillate.} \label{fig:sigmaenergy}
\end{figure}

In our specific example, we consider a free-electron density of $n^{(0)} = 2.25 \times 10^{15}$~cm$^{-2}$, indicative of a metallic system. The active layer is encapsulated by two dielectric materials, with $\varepsilon_\mathrm{avg} = 10\varepsilon_0$. The effective mass of the electrons is taken as $m_\mathrm{eff} = m_\mathrm{e} $. Unless stated otherwise, we consider a plasmon with an energy of $E = 0.54$~eV, a local increase in electron density $\delta n = 1.25$, and a decay length of $\ell = \ell_0 = 13.2$~nm. For these values, we have $h = 0.0064$ and $\kappa = 0.0100$, and the ratio $h / \kappa$ is of order one. We therefore satisfy the requirements given in Sec.~\ref{subsec:derivation-applicability}. For these parameters, the classical trajectories corresponding to a scattered plasmon are plotted in Fig.~\ref{fig:clastrajec}.

We first computed the total cross section as function of the plasmon energy, which is shown in Fig.~\ref{fig:sigmaenergy}. Intuitively, the total cross section increases with increasing plasmon energy or momentum. We recognize this for low energies, but for higher energies the total cross section $\sigma / \ell_0$ starts to oscillate. We attribute this oscillation to interference between different overlapping waves, as we discuss shortly.

\begin{figure}[t]
  \centering
  \includegraphics[width=0.8\linewidth]{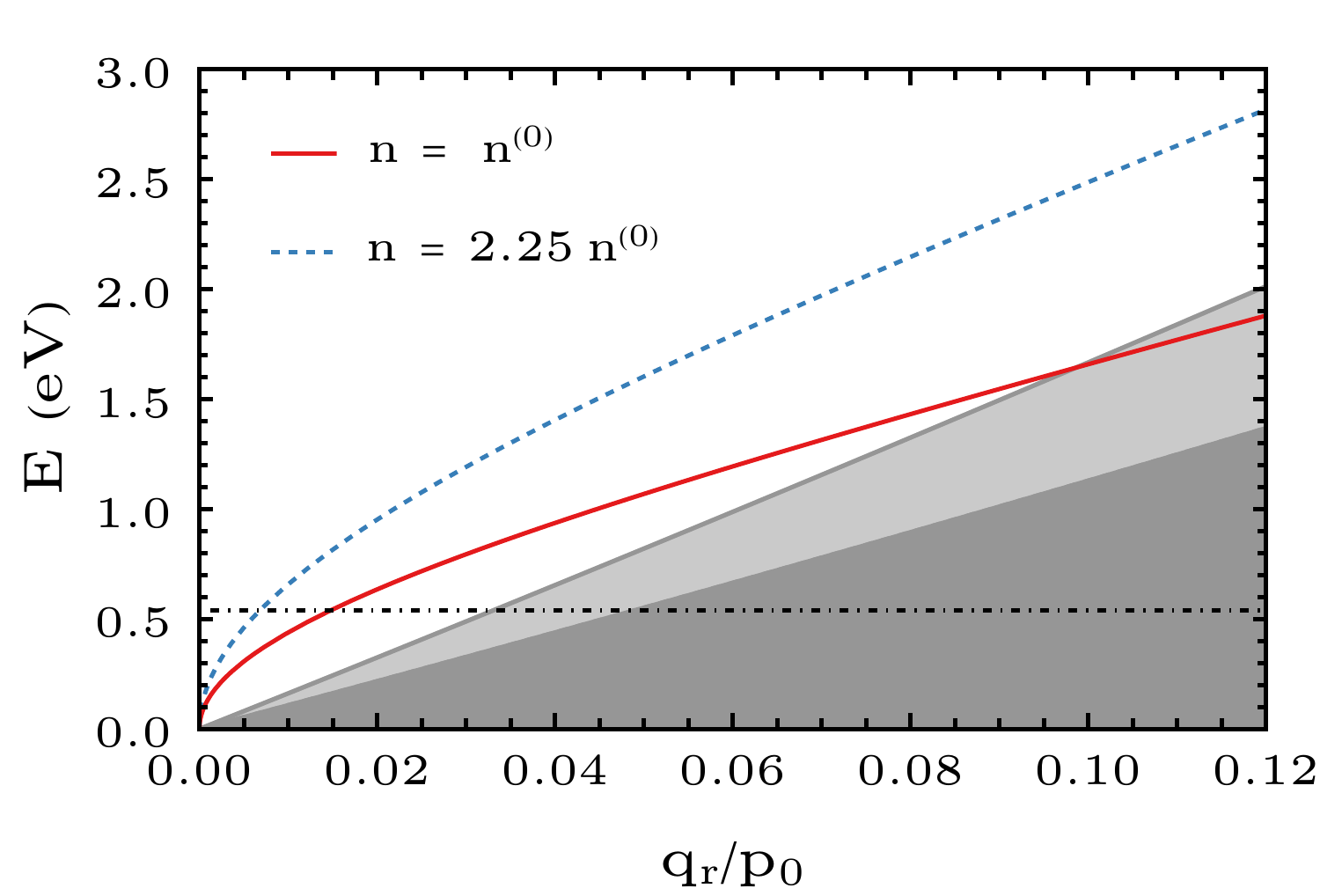}
  \caption{The dispersion relation for plasmons in a two-dimensional system for two different electron densities, namely $n^{(0)} = 2.25 \times 10^{15}$~cm$^{-2}$ (solid red line) and $n^{(0)} = 5.06 \times 10^{15}$~cm$^{-2}$ (dashed blue line). The black dash-dotted line indicates a constant energy. One can see that the plasmon momentum corresponding to a given energy decreases when the electron density increases. The light and dark gray area depict the Landau damped regions for the higher and lower electron density, respectively.} \label{fig:plasmondisp}
\end{figure}

Looking at the dispersion relation in Fig.~\ref{fig:plasmondisp}, we observe that the plasmon dispersion enters a shaded region at a certain energy. This shaded region corresponds to the region of Landau damping~\cite{Vonsovsky89,Giuliani05}, where the collective electron excitation transfers energy to incoherent electron-hole pairs. Mathematically, this corresponds to the point where $\mathcal{H}_0$, see Eq.~(\ref{eq:dimensionlessEffClassHam}), becomes complex, which leads to many complications in the application of the semiclassical approximation~\cite{Reijnders22}. When plotting the cross section in Fig.~\ref{fig:sigmaenergy}, we therefore made sure that we stayed well outside the region of Landau damping, which starts at $E = 5.7$~eV for the given parameters.

Let us now examine the interference, by taking a closer look at the classical trajectories in Fig.~\ref{fig:clastrajec}. Since the trajectories emerge from a classical picture, they do not take the wave-like character into account and therefore do not show the interference. However, we can determine the regions in which interference takes place, by looking at points that are reached by more than one trajectory. The black lines in Fig.~\ref{fig:clastrajec}, which are known as caustics~\cite{Poston78,Arnold82}, separate the regions where each point lies on a single trajectory from the regions where each point lies on multiple trajectories. It is precisely in the latter regions where interference takes place. The Jacobian~(\ref{eq:Jacobian}) vanishes on the caustics, and therefore our expressions for the induced potential and the energy density diverge. This indicates that we cannot use our expressions in the vicinity of the caustic, and implies that the energy density is larger in this region. The interference is visible in the total cross section shown in Fig.~\ref{fig:sigmaenergy}, since our semiclassical expression takes the wave-like character of the plasmons into account.

\begin{figure}[t]
  \centering
  \includegraphics[width=0.9\linewidth]{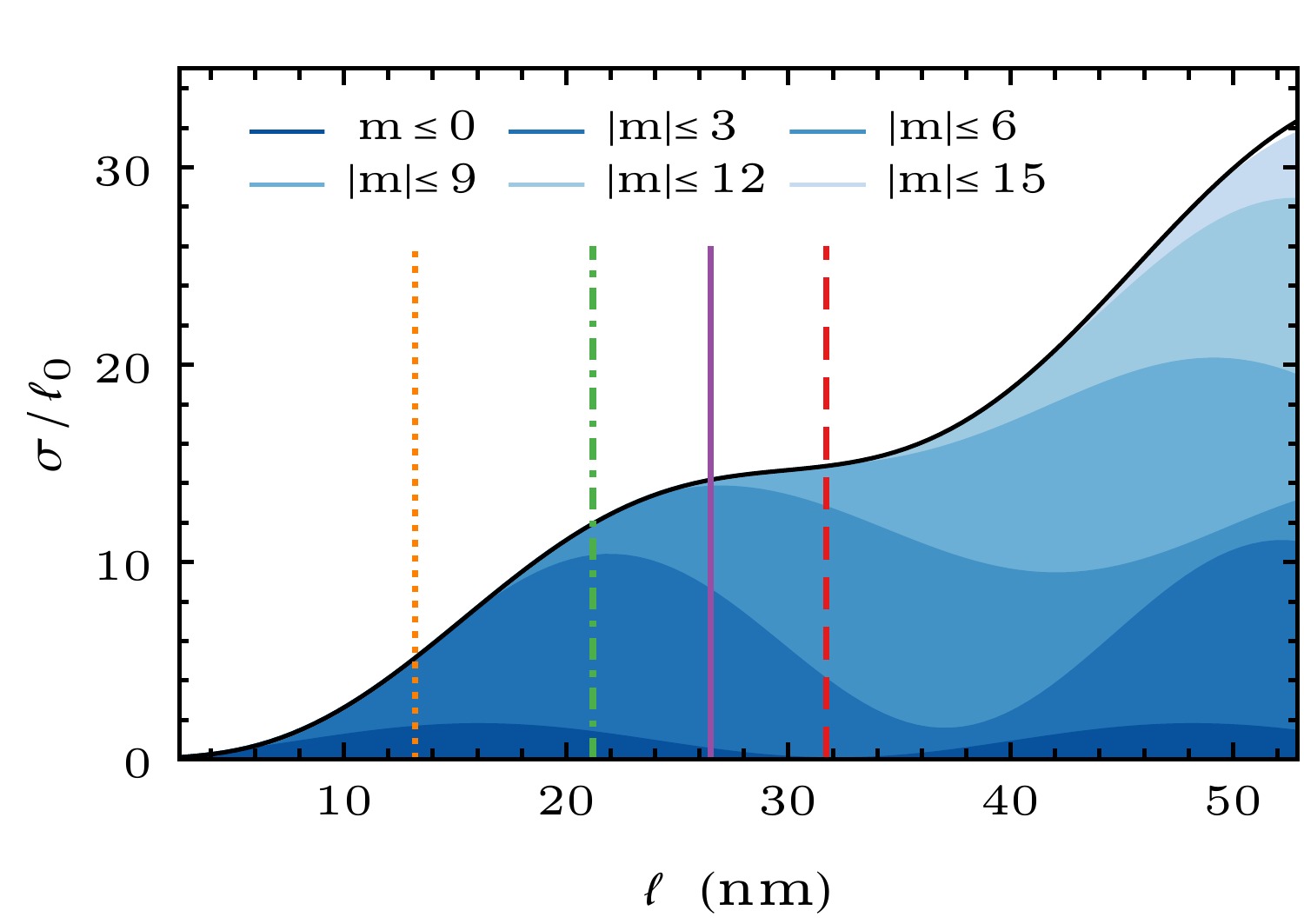}
  \caption{Numerically evaluated total scattering cross section $\sigma / \ell_0$ as function of the decay length $\ell$. In this plot, the local density change $\delta n =  1.25$, the decay length is $\ell_0 = 13.2$~nm, and the plasmon energy is $E = 0.54$~eV. The shades of blue indicate the partial waves which have been taken into account, whilst the solid black line indicates the cross section for $|m| \leq 23$. The vertical colored lines correspond to the decay lengths for which the differential cross section is plotted in Fig.~\ref{fig:dsigmadthetalength}.} \label{fig:sigmalength}
\end{figure}

\begin{figure}[t]
  \centering
  \includegraphics[width=0.7\linewidth]{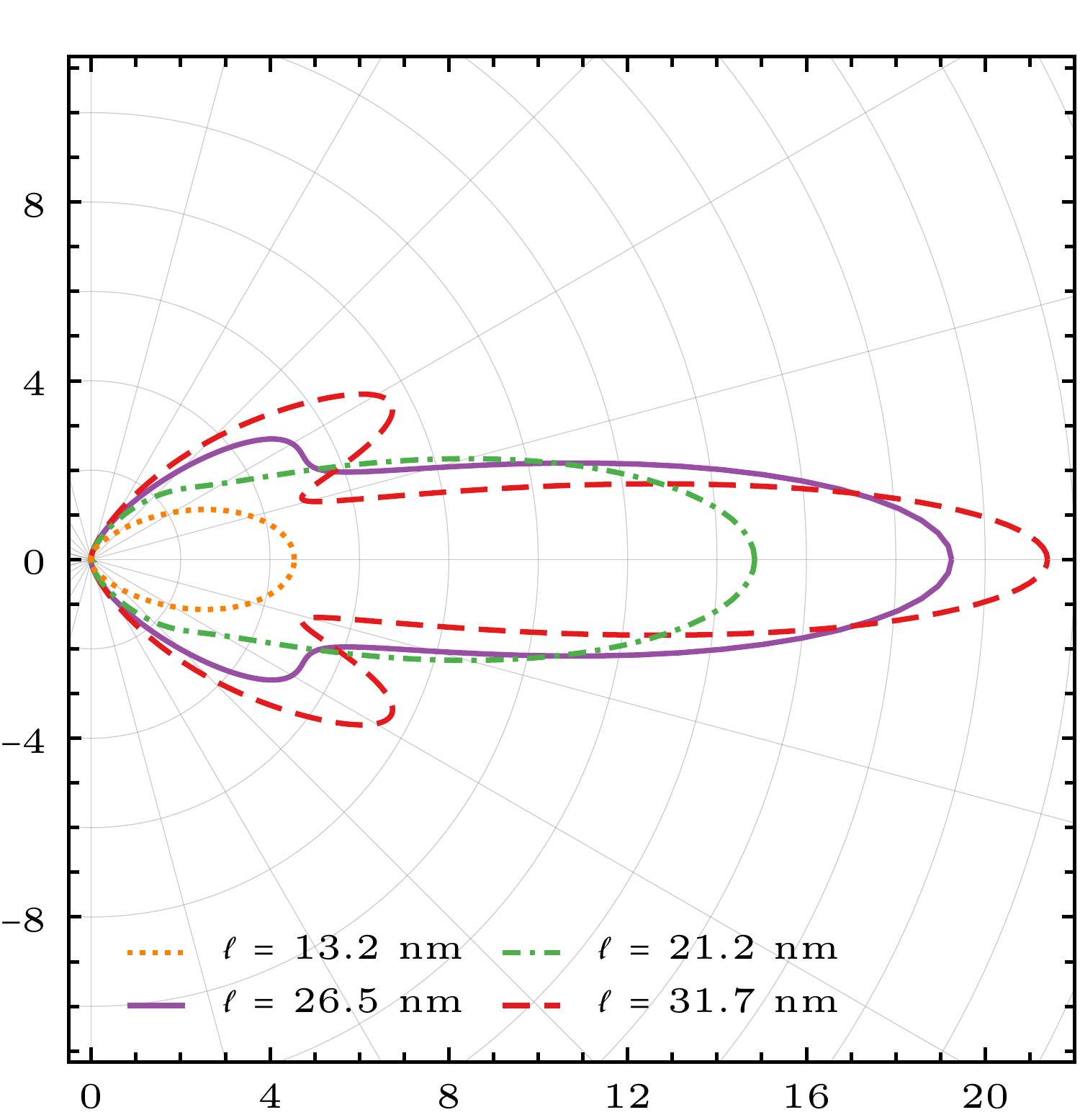}
  \caption{Numerically evaluated differential scattering cross section for different values of the decay length $\ell$. The $\ell = 13.2$~nm (dotted orange) and $\ell = 21.2$~nm (dash-dotted green) line both show mainly forward plasmon scattering. The minima and maxima in the $\ell = 26.5$~nm (solid purple) and $\ell = 31.7$~nm (dashed red) line show clear signs of  interference. The parameters used to create this plot correspond to those in Fig.~\ref{fig:sigmalength}. Partial waves up to $|m| \leq 15$ were taken into account. Note the two minima around $\pm \pi/12$ and maxima around $\pm \pi/6$ for $\ell = 26.5$~nm and $\ell = 31.7 $~nm. } \label{fig:dsigmadthetalength}
\end{figure}

A similar, but slightly different, oscillatory behavior is found in Fig.~\ref{fig:sigmalength}, where the total scattering cross section is plotted for different values of the decay length $\ell$, not to be confused with the constant $\ell_0$. The total cross section increases monotonously as a function of the decay length, however, with oscillating slope. We again attribute this behavior to interference between the overlapping plasmon trajectories. We can confirm again that this effect cannot be explained by the classical picture, by noting that the classical trajectories do not depend on the decay length $\ell$ once proper dimensionless parameters have been introduced, see Sec.~\ref{subsec:derivation-applicability}.

\begin{figure}[t]
  \centering
  \includegraphics[width=0.44\linewidth]{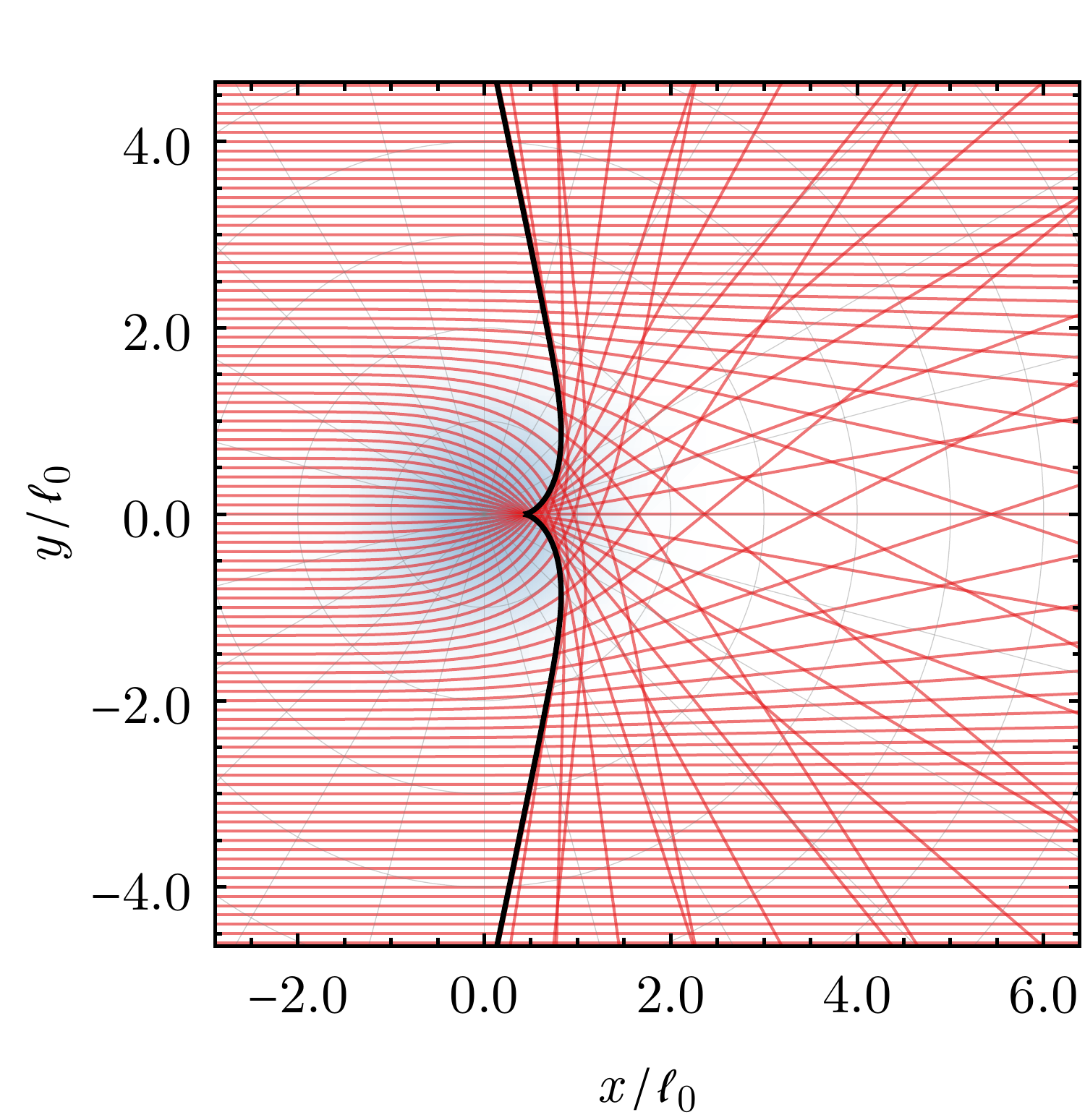}
  \includegraphics[width=0.49\linewidth]{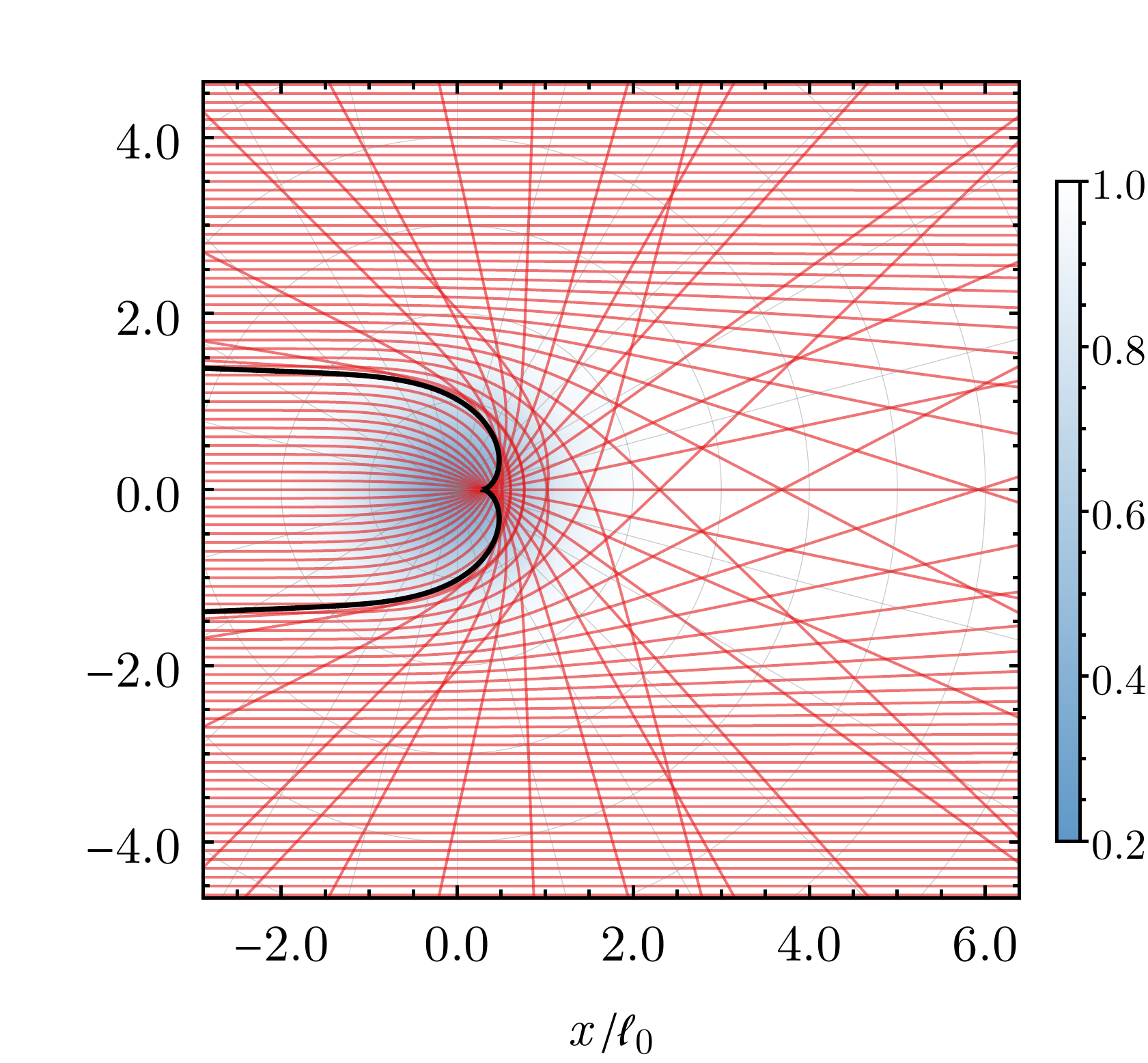}
  \caption{Classical trajectories of a plasmon, coming in from $x \rightarrow -\infty$. The maximum change in local density for the left plot is $\delta n = -0.58$, which corresponds to the first maximum in $\sigma / \ell_0$ in Fig.~\ref{fig:sigmadeltanpos}. The right plot has a maximum change in local density of $\delta n = -0.70$, which is just before the local minimum in the total scattering cross section. The solid black lines are the caustics.} \label{fig:classicalpathsnegdens}
\end{figure}

To further investigate the interference, we plot the differential cross section $d\sigma/d\theta$, given by Eq.~(\ref{eq:diffcrosssec}), which indicates the angular dependence of the plasmon scattering, for the lengths corresponding to the colored vertical lines in Fig.~\ref{fig:sigmalength}. 
We divide it by $\ell_0$ to make it a dimensionless quantity. Figure~\ref{fig:dsigmadthetalength} shows that there is strong forward scattering for all decay lengths. For larger decay lengths, in other words, for smaller values of the semiclassical parameter $h$, two additional scattering peaks appear. The new local maxima around $\pm \pi/6$ originate from constructive interference, and similarly two local minima arise around $\pm \pi/12$ due to destructive interference. Comparing Figs.~\ref{fig:dsigmadthetalength} and~\ref{fig:clastrajec}, we observe that these additional extrema are indeed located in the regions where the trajectories overlap. The interference effects thus become more clearly visible for smaller values of $h$, in other words, in the deep semiclassical limit. We see similar results when the differential cross section is plotted for increasing energies.

Lastly, we look at the effect of a change in the local electron density $\delta n$ on the plasmon scattering. In Fig.~\ref{fig:plasmondisp}, we see that, for a given energy $E$, an increase in the local electron density decreases the momentum $q$. In this sense, an increase in local electron density ($\delta n > 0$) can be considered a repelling potential, as shown in Fig.~\ref{fig:clastrajec}. In contrast, a decreasing local density ($\delta n < 0$) attracts the plasmon, as shown in Fig.~\ref{fig:classicalpathsnegdens}.

\begin{figure}[t]
  \centering
  \includegraphics[width=0.9\linewidth]{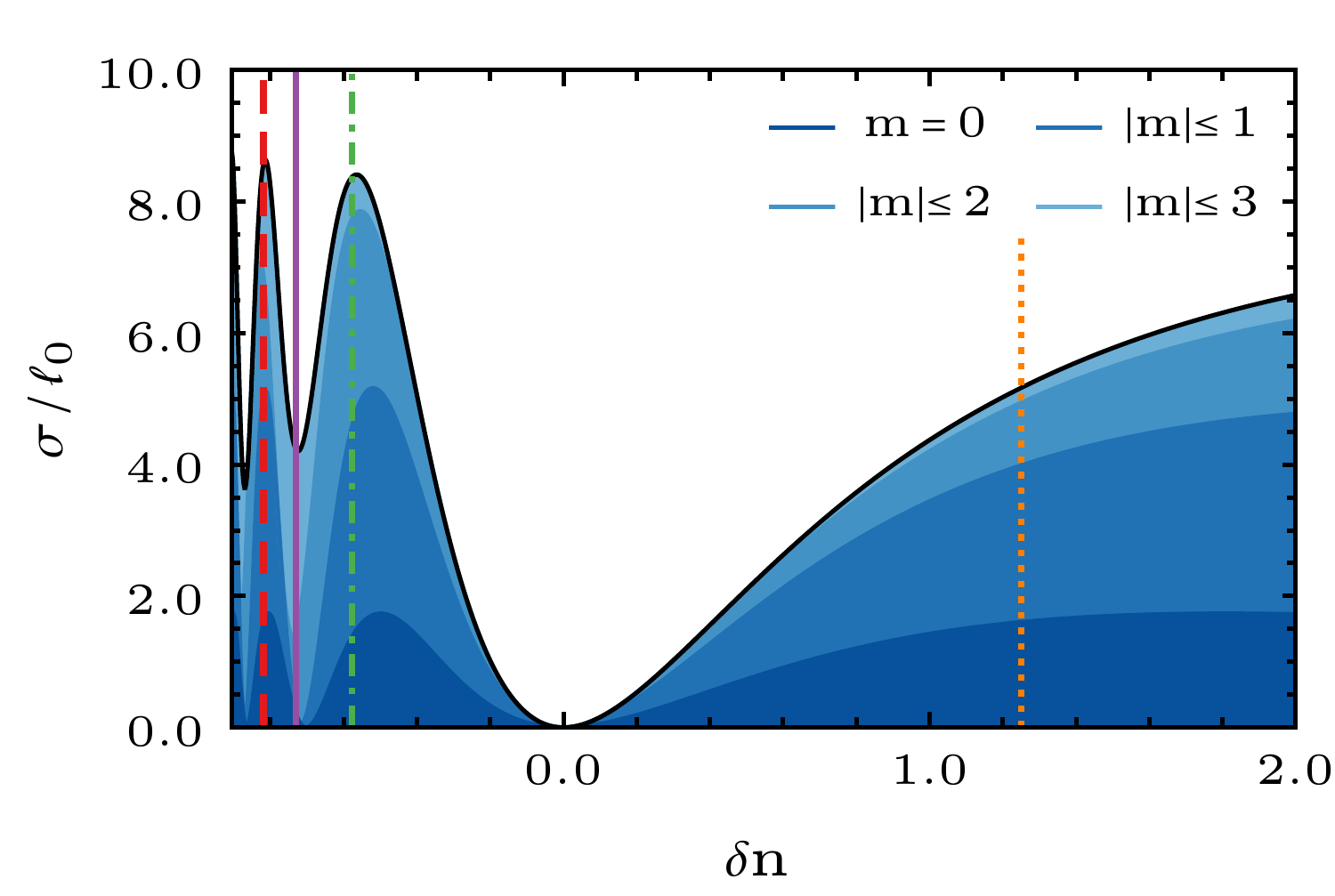}
  \caption{Numerically calculated total scattering cross section $\sigma / \ell_0$ as a function of the maximum change in local electron density. The plasmon energy is fixed at $E = 0.54$~eV, and the decay length is given by $\ell_0 = 13.2$~nm. The shades of blue indicate the partial waves which has been taken into account, whereas the solid black line indicates the cross section for $|m| \leq 6$. Note the oscillatory behavior in $\sigma / \ell_0$ for negative change in local density. The vertical colored lines correspond to the decay lengths for which the differential cross sections are plotted in Fig.~\ref{fig:dsigmadthetadeltan}.} \label{fig:sigmadeltanpos}
\end{figure}

\begin{figure}[t]
  \centering
  \includegraphics[width=0.7\linewidth]{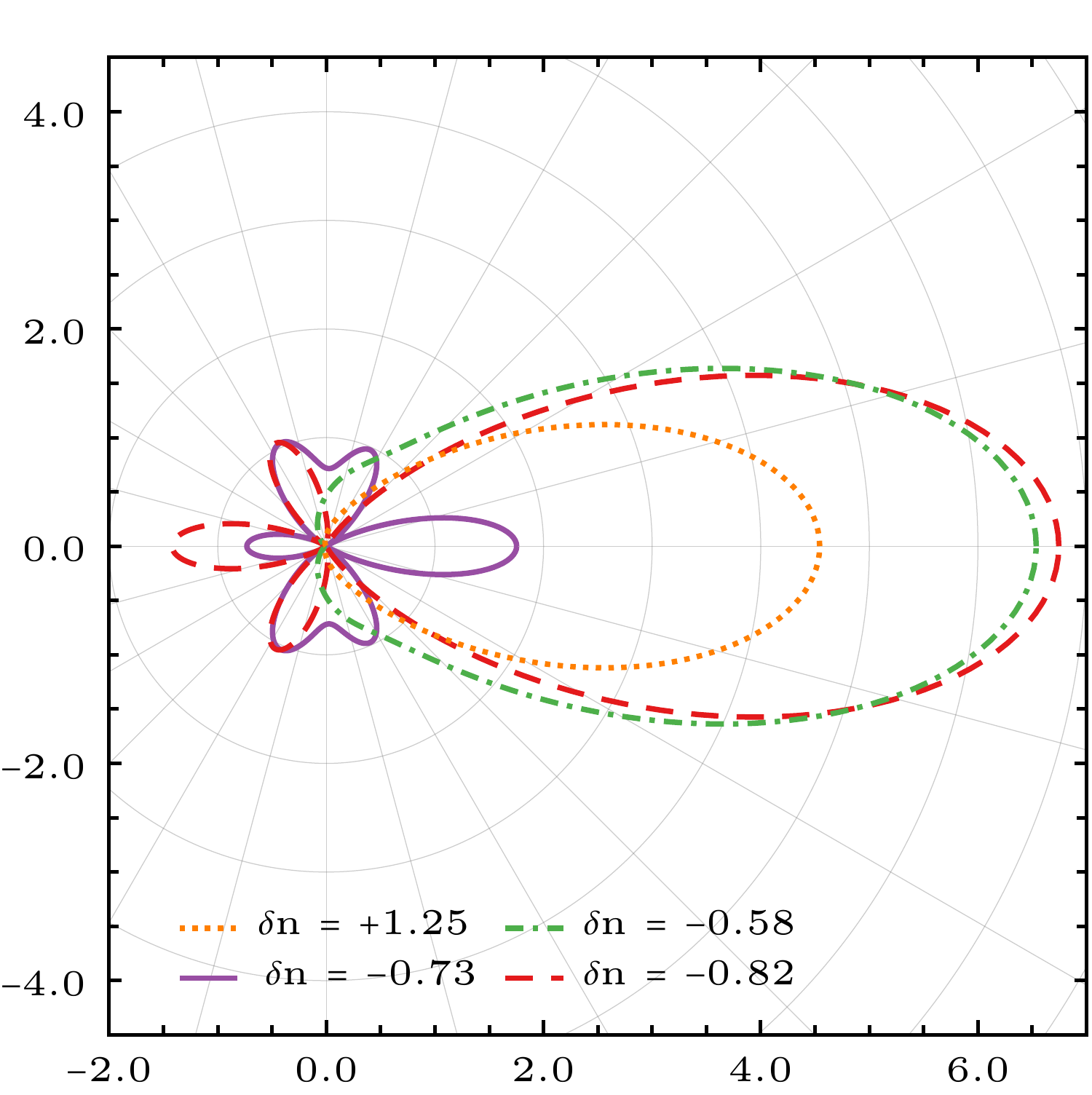}
  \caption{Numerically computed differential scattering cross section for different values of the change in local density  $\delta n$. The $\delta n = + 1.25$ (dotted orange) line is the reference used throughout the rest of the calculations, the $\delta n = -0.58$ (dash-dotted green) line corresponds to the first maximum in $\sigma / \ell_0$. The $\delta n = -0.75$ (solid purple) and $\delta n = -0.82$ (dashed red) line correspond to the next minimum and maximum in $\sigma / \ell_0$ respectively. The energy of the plasmon is $E = 0.54$~eV, and the decay length is $\ell_0 = 13.2$~nm. Partial waves up to $|m| \leq 8$ were taken into account. Note the extra maxima and minima and the backscattering peak for $\delta n = -0.75$ and $\delta n = -0.82$. } \label{fig:dsigmadthetadeltan}
\end{figure}

Note that decreasing the local electron density also lowers the energy at which the Landau damped region is reached, as can be seen in Fig.~\ref{fig:plasmondisp}. This means that for a certain energy there is a maximum to the decrease in local electron density. For $E= 0.54$~eV, the maximum becomes $\delta n_\mathrm{c} = -0.99$. For the total scattering cross section shown in Fig.~\ref{fig:sigmadeltanpos}, we stay well below this limit.

As expected, we see in Fig.~\ref{fig:sigmadeltanpos} that the total scattering cross section $\sigma / \ell_0$ increases for larger increases in the local density $|\delta n|$. The scattering cross section is asymmetric around $\delta n = 0$, which is expected since there is a maximum decrease in local electron density for the existence of plasmons, but no limit for positive values. Furthermore, the small $\mathbf{q}$ expansion given in Eq.~(\ref{eq:velocity-momentum-opposite}) for $\mathcal{H}_0 = 0$, yields the proportionality $q \propto 1/(n^{(0)} + \delta n )$, which is also not symmetric around $\delta n = 0$. 

We observe in Fig.~\ref{fig:sigmadeltanpos} that the total scattering cross section increases monotonically for $\delta n > 0$. Looking at the dotted orange line in Fig.~\ref{fig:dsigmadthetadeltan}, we see that the plasmon is mainly scattered forward for $\delta n = 1.25$, without any additional maxima and minima. Note that this differential cross section is the same as the dotted orange differential cross section in Fig.~\ref{fig:dsigmadthetalength}. On the contrary, for negative values of $\delta n$, we see that the total cross section shows maxima at approximately $\delta n = -0.58$ and $\delta n = -0.82$ and minima at $\delta n = -0.75$, which we attribute to interference. 

In Fig.~\ref{fig:dsigmadthetadeltan}, we take a closer look at these extrema by considering the differential cross section for the densities indicated by the colored vertical lines in Fig.~\ref{fig:sigmadeltanpos}. We see additional scattering directions emerging for negative $\delta n$, including a sharp backscattering peak for $\delta n = - 0.75$ and $\delta n = -0.82$. The backscattering comes from plasmons that are attracted and deflected by the lower local density, as can be seen from Fig.~\ref{fig:classicalpathsnegdens}, where the classical trajectories are plotted for $\delta n = - 0.58$ and $\delta n = - 0.70$. Lowering the local density to $\delta n = - 0.70$, which is just before the local minimum in the total scattering cross section, gradually deflects the plasmons $180^\mathrm{o}$ degrees, and ergo explains the backscattering. Lowering $\delta n$ further causes the caustics to intersect, and consequently points are reached by even more trajectories in that region. 
This in turn increases the interference effects, which is evident form the rough changes below $\delta n = -0.85$ in the total scattering cross section shown in Fig.~\ref{fig:sigmadeltanpos}. 

\section{Conclusion and outlook}
\label{sec:conclusion}
In this article, we constructed a semiclassical theory for plasmons in spatially inhomogeneous two-dimensional media. 
We extended the theory derived in Ref.~\cite{Reijnders22} to two dimensions, by carefully separating the in-plane and out-of-plane coordinates before applying the semiclassical Ansatz. More precisely, we used a modified version of the operator separation of variables technique~\cite{Belov06}.
The first result of this formalism is the Hamilton-Jacobi equation~(\ref{eq:HJequation}), 
which describes the dynamics of the quantum plasmons in classical phase space. 
Second, we derived the leading-order term (in the semiclassical parameter) of the induced potential~(\ref{eq:V0solutionamplitudeopsep}) by solving the transport equation~(\ref{eq:transporteq}).
This induced potential expresses the wave-like character of the plasmons,
and its derivative is proportional to the electric field. We used the latter to compute the energy density and showed that it has a natural interpretation in the context of the semiclassical approximation.

To illustrate our theory, we studied scattering of plasmons by a radially symmetric inhomogeneity in the background electron density. We derived a semiclassical expression for the phase shift, Eq.~(\ref{eq:scatteringphaseFlugge}), which can be used to compute the total and differential scattering cross section, given by Eqs.~(\ref{eq:totscatcross}) and~(\ref{eq:diffcrosssec}), respectively. In Sec.~\ref{sec:numerical}, we calculated these quantities for a specific model for the inhomogeneity, namely a Gaussian bump or well. We considered the dependence of the total cross section on the plasmon energy $E$, the spatial scale $\ell$ of the inhomogeneity, and the change in local density $\delta n$. We connected these results to the classical trajectories of the plasmons and discussed the effect of interference between different trajectories on both the total and differential cross section. This interference can be clearly observed in the differential cross section for different $\ell$ and $\delta n$, and is more clearly visible when the dimensionless semiclassical parameter $h$ is smaller, that is, in the deep semiclassical limit.
For a depletion in the local density, $\delta n < 0$, we observed backscattering in the differential cross section, which is in agreement with the classical trajectories.

The main assumption in our formalism is that the electron wavelength is much smaller than the length scale of the inhomogeneity, which allows for the introduction of a dimensionless semiclassical parameter $h \ll 1$. As discussed in Sec.~\ref{subsec:derivation-applicability}, this implies that the length scale of the inhomogeneity has to be on the order of a few nanometers for metallic systems, and on the order of a hundred nanometers for semiconductor systems. Since these length scales are experimentally accessible, our predictions for the total and differential cross sections can be tested in scattering experiments.
Although we performed calculations for a Gaussian inhomogeneity in Sec.~\ref{sec:numerical}, our semiclassical results are not restricted to this specific model. In fact, they can be applied to any inhomogeneity that can be described by a radially symmetric, sufficiently smooth, change in the local electron density.

In our derivations, we assumed that our two-dimensional material is infinitely thin, i.e., that the charge density is purely two-dimensional. As we discussed in the introduction, this is a good first approximation to describe van der Waals materials.
However, to construct a more accurate description of realistic systems, 
one has to introduce a finite thickness, which is on the order of a few interatomic distances, and an additional dielectric constant~\cite{Wehling11,Rosner15,Andersen15,Cho18,Jiang21}.
Both of these quantities can be extracted from first-principle calculations~\cite{Wehling11,VanSchilfgaarde11,Rosner15}. 
In particular, the finite thickness can essentially modify the electron-electron interaction within the layer, as was studied in Refs.~\cite{Asgari14,Rosner15} for the case of graphene.
An important next step would therefore be to extend our description to systems with a finite thickness.

Another possible direction would be to consider plasmonic bound states in two-dimensional systems. These arise, e.g., in plasmonic waveguides, which can be created by patterning two-dimensional materials~\cite{Nikitin11,Christensen12}.
As shown with numerical calculations in Refs.~\cite{Rosner16,Jiang21}, plasmonic waveguides can also be created in a noninvasive way, by considering a two-dimensional material with an inhomogeneous dielectric environment. 
In Ref.~\cite{Reijnders22}, the semiclassical approximation was used to study plasmonic bound states in a three-dimensional analog of these systems. It would be interesting to extend this description to two-dimensional systems.

The last research direction that we would like to mention is to extend our theory to systems with nonparabolic dispersion relations.
Indeed, many two-dimensional materials have a different dispersion relation, and plasmons in these systems have been studied extensively in recent years.
These include graphene~\cite{Fei12,Chen12,Grigorenko12}, black phosphorus~\cite{Prishchenko17,Jin15}, transition metal dichalcogenides and other van der Waals materials~\cite{Basov16,Low17}. It would therefore be very interesting, both from a theoretical and experimental perspective, to extend our description to different electronic Hamiltonians.

To conclude, we consider our semiclassical theory a valuable new tool to study plasmons in spatially inhomogeneous two-dimensional media and believe that it can be extended to describe realistic experimental setups.

\begin{acknowledgments}
  We would like to thank Malte R\"osner for helpful discussions. 
  The work was supported by the European Union's Horizon 2020 research and innovation programme under European Research Council synergy grant 854843 ``FASTCORR".
\end{acknowledgments}

\appendix

\section{Alternative derivation of $V(\mathbf{x},z)$}\label{ap:bruteforce}

In this appendix, we present an alternative derivation of Eqs.~(\ref{eq:HJequation}) and~(\ref{eq:transporteq}). 
Instead of using the operator separation technique discussed in Sec.~\ref{subsec:derivation-separation}, we use the Ansatz~(\ref{eq:VtotalExpansion}) for $V(\mathbf{x},z)$ to solve the Poisson equation~(\ref{eq:Poisson}) order by order in $\hbar$. On the one hand, we consider this method less elegant, since it mixes the separation of the in-plane and out-of-plane degrees of freedom with the application of the semiclassical Ansatz. On the other hand, it may be easier to understand for readers less familiar with operator techniques.

On the left-hand side of the Poisson equation~(\ref{eq:Poisson}), we have
\begin{align} \label{eq:appenpoissoineq}
  &\langle \nabla , \varepsilon(\mathbf{x},z) \nabla \rangle V(\mathbf{x},z) \nonumber \\
  & \hspace*{0.6cm} = \bigg( -   \frac{\varepsilon(\mathbf{x},z)}{\hbar^2} V_0(\mathbf{x},z) \left(\frac{\partial S}{\partial \mathbf{x}}\right)^2  
  +\frac{\partial}{\partial z} \varepsilon(\mathbf{x},z) \frac{\partial V_0}{\partial z} \nonumber \\
  & \hspace*{1cm} 
  - \frac{\varepsilon(\mathbf{x},z)}{\hbar} V_1(\mathbf{x},z)  \left(\frac{\partial S}{\partial \mathbf{x}}\right)^2
  + \hbar \frac{\partial}{\partial z} \varepsilon(\mathbf{x},z) \frac{\partial V_1}{\partial z} \nonumber \\
  & \hspace*{1cm}  + i\frac{2 \varepsilon(\mathbf{x},z) }{\hbar} \left\langle \frac{\partial V_0}{\partial \mathbf{x}}, \frac{\partial S}{\partial \mathbf{x}} \right\rangle 
  + i \frac{\varepsilon(\mathbf{x},z)}{\hbar} V_0(\mathbf{x},z)  \frac{\partial^2 S}{\partial \mathbf{x}^2} \nonumber \\
  & \hspace*{1cm} + \frac{i}{\hbar}  V_0(\mathbf{x},z) \left\langle \frac{\partial \varepsilon(\mathbf{x},z)}{\partial \mathbf{x}}, \frac{\partial S}{\partial \mathbf{x}} \right\rangle
  \bigg) e^{i S(\mathbf{x})/\hbar},
\end{align}
where we left out all higher-order terms. 
As discussed in Sec.~\ref{subsec:derivation-separation} and Sec.~\ref{subsec:derivation-applicability}, the second derivative of $V_0$ with respect to the out-of-plane direction $z$ belongs to the leading-order term, since the combination $\hbar/z$ is of order one when proper dimensionless parameters are introduced.

The induced electron density $n(\mathbf{x},z)$ on the right-hand side of the Poisson equation~(\ref{eq:Poisson}) is given by Eqs.~(\ref{eq:3d-from2d-elecdens}) and~(\ref{eq:2delecdens}). Using our previous Ansatz~(\ref{eq:scansatz2d}) for the induced potential $V_\mathrm{pl}(\mathbf{x})$ in the plane, as well as a relation similar to Eq.~(\ref{eq:totalpotential}), cf. Refs.~\cite{Maslov81,Guillemin77}, we obtain the first two terms in the asymptotic expansion of $n(\mathbf{x},z)$.
In the following two subsections, we analyze the terms of leading and subleading order in the Poisson equation, respectively.

\subsection{Leading-order term}
Collecting the leading-order terms in Eq.~(\ref{eq:appenpoissoineq}), we have
\begin{multline}
  \bigg(-\frac{1}{\hbar^2} \varepsilon_i(\mathbf{x}) \left(\frac{\partial S}{\partial \mathbf{x}}\right)^2  + \varepsilon_i(\mathbf{x})\frac{\partial^2 }{\partial z^2}\bigg)V_0(\mathbf{x},z)  e^{i S(\mathbf{x})/\hbar} = \\
  - 4 \pi e^2 \delta(z) \Pi_0\left(\mathbf{x},\frac{\partial S}{\partial \mathbf{x}}\right) \varphi_0(\mathbf{x}) e^{i S(\mathbf{x})/\hbar} ,
\end{multline}
where we made use of the fact that $\varepsilon_i(\mathbf{x})$ is piecewise constant, see Eq.~(\ref{eq:varepsilon}). The right-hand side contains the leading-order term of the induced electron density $n(\mathbf{x},z)$.

We proceed by noting that the action exponents on both sides cancel out, given that they do not vanish. We can then solve the remaining differential equation in the way discussed in Sec.~\ref{subsubsec:derivation-separation-principal}. We find that
\begin{equation}\label{eq:V0solutionAppendix}
  V_0(\mathbf{x},z) = c_0\left(\mathbf{x}, \frac{\partial S}{\partial \mathbf{x}}\right) e^{-\frac{|z|}{\hbar}\left|\frac{\partial S}{\partial \mathbf{x}}\right|} ,
\end{equation}
with
\begin{equation} \label{eq:def-c0-appendix}
  c_0\left(\mathbf{x}, \frac{\partial S}{\partial \mathbf{x}}\right) = \frac{2\pi e^2 \hbar}{\varepsilon_{\mathrm{avg}}(\mathbf{x}) | \partial S/ \partial \mathbf{x}|} \Pi_0\left(\mathbf{x}, \frac{\partial S}{\partial \mathbf{x}}\right) \varphi_0(\mathbf{x}) ,
\end{equation}
where we used the definition $2 \varepsilon_{\mathrm{avg}} (\mathbf{x}) \equiv \varepsilon_\mathrm{A}(\mathbf{x})+\varepsilon_\mathrm{B}(\mathbf{x})$, and $\left|\partial S/\partial \mathbf{x}\right| = \big(\sum_j \big(\partial S/\partial x_j\big)^2\big)^{1/2} $. Expression~(\ref{eq:V0solutionAppendix}) corresponds to the leading-order term of $\left(\hat{\Gamma} V_\mathrm{pl}\right)(\mathbf{x},z)$ in the main text, see Eqs.~(\ref{eq:leadingorderchi0}) and~(\ref{eq:totalpotential}).

At this point, we can directly apply the self-consistency condition~(\ref{eq:condition-self-consistency}) at $z=0$ to the leading-order terms. We have
\begin{align}
  V_0(\mathbf{x},z=0)e^{i S(\mathbf{x})/\hbar} = \varphi_0(\mathbf{x}) e^{i S(\mathbf{x})/\hbar},
\end{align}
from which we obtain the Hamilton-Jacobi equation
\begin{align}\label{eq:appenHamJac}
  \mathcal{H}_0\left(\mathbf{x},\frac{\partial S}{\partial \mathbf{x}}\right) = 1 -  \frac{2\pi e^2 \hbar}{\varepsilon_{\mathrm{avg}}(\mathbf{x}) |{\partial S/\partial \mathbf{x}}|} \Pi_0\left(\mathbf{x}, \frac{\partial S}{\partial \mathbf{x}}\right) = 0 .
\end{align}
This is the same result as Eq.~(\ref{eq:HJequation}) in the main text.
From a technical point of view, one may say that this derivation of the leading-order term is not that different from our previous derivation in Sec.~\ref{sec:derivation}. The main difference is that the derivatives with respect to $\mathbf{x}$ directly act on the semiclassical Ansatz, leading to the replacement of $\mathbf{q}$ by $\partial S/ \partial \mathbf{x}$.

\subsection{Subleading-order term}

The derivation of the subleading-order term $V_1(x,z)$ is however quite different from our derivation in the main text.
We first calculate $n_1(\mathbf{x})$, the subleading part of $(\hat{\Pi} V_\mathrm{pl})(\mathbf{x})$, using a relation similar to Eq.~(\ref{eq:totalpotential}), cf. Refs.~\cite{Maslov81,Guillemin77}.
In this way, we find
\begin{align}
  n_1(\mathbf{x}) = \,& n_{1,\mathrm{a}}(\mathbf{x}) e^{i S(\mathbf{x})/\hbar} \nonumber \\
  n_{1,\mathrm{a}}(\mathbf{x}) = \,& \Pi_0\left( \mathbf{x},\frac{\partial S}{\partial \mathbf{x}} \right) \varphi_1 (\mathbf{x}) + \Pi_1 \left( \mathbf{x},\frac{\partial S}{\partial \mathbf{x}} \right)  \varphi_0(\mathbf{x}) \nonumber \\
  &- \frac{i}{2} \sum_{jk} \frac{\partial^2 \Pi_0}{\partial q_j \partial q_k} \left( \mathbf{x},\frac{\partial S}{\partial \mathbf{x}} \right) \frac{\partial^2 S}{\partial x_j \partial x_k} \varphi_0 (\mathbf{x}) \nonumber \\
  &- i \left\langle \frac{\partial \Pi_0}{\partial \mathbf{q}} \left( \mathbf{x},\frac{\partial S}{\partial \mathbf{x}} \right) , \frac{\partial \varphi_0} {\partial \mathbf{x}}\right\rangle  ,
\end{align}
where $\Pi_1\left( \mathbf{x},\partial S/\partial \mathbf{x} \right)$ satisfies Eq.~(\ref{eq:symbol-Hermitian-operator-condition}).

Next, we consider the terms of subleading order in Eq.~(\ref{eq:appenpoissoineq}). Inserting our solution~(\ref{eq:V0solutionAppendix}) for $V_0(\mathbf{x},z)$, computing the various derivatives, and using that $\varepsilon(\mathbf{x},z)$ is piecewise constant, we find
\begin{align}\label{eq:AppenInhomogeneousdiff}
  & \bigg( -\frac{1}{\hbar^2} \varepsilon_i(\mathbf{x}) \left(\frac{\partial S}{\partial \mathbf{x}}\right)^2 + \varepsilon_i(\mathbf{x}) \frac{\partial^2}{\partial z^2} \bigg) V_1(\mathbf{x},z)\nonumber \\
  & \hspace*{1.6cm} = -\frac{i}{\hbar^2} \left( f_{1,i} (\mathbf{x}) + f_{2,i} (\mathbf{x}) \frac{|z|}{\hbar} \right) e^{-\frac{|z|}{\hbar}\left|\frac{\partial S}{\partial \mathbf{x}}\right|}\nonumber \\
  & \hspace*{2cm} - 4 \pi e^2 \delta(z) n_{1,\mathrm{a}}(\mathbf{x}) ,
\end{align}
where we canceled the exponent $\exp(i S(\mathbf{x})/\hbar)$ on both sides.
Moreover, we defined $f_{1,i}(\mathbf{x})$ and $f_{2,i}(\mathbf{x})$ by
\begin{align}
  f_{1,i} 
  &=  \left\langle \frac{\partial \varepsilon_i}{\partial \mathbf{x}} , \frac{\partial S}{\partial \mathbf{x}} \right\rangle c_0 
  + 2  \varepsilon_i \left\langle \frac{\partial c_0}{\partial \mathbf{x}} , \frac{\partial S}{\partial \mathbf{x}} \right\rangle
  + \varepsilon_i c_0 \frac{\partial^2 S}{\partial \mathbf{x}^2} , \\
  f_{2,i} 
  &= - \frac{2 \varepsilon_i c_0}{|\partial S/\partial \mathbf{x}|} \sum_{j,k} \frac{\partial S}{\partial x_j} \frac{\partial S}{\partial x_k} \frac{\partial^2 S}{\partial x_j \partial x_k} ,
\end{align}
where $c_0$ was defined in Eq.~(\ref{eq:def-c0-appendix}) and the subscript $i = A$ $(i= B)$ denotes the region above (below) the plane $z=0$, just as in Sec.~\ref{subsec:derivation-separation}.

We can solve the differential equation for $V_1$ using the methods discussed in Sec.~\ref{subsec:derivation-separation}. We first determine the solution of the homogeneous equation, given by
\begin{align}\label{eq:appenhomosol}
  V_{1,i,\mathrm{H}}(\mathbf{x},z) =  c_{1,i}^{-} e^{-\frac{|z|}{\hbar}\left|\frac{\partial S}{\partial \mathbf{x}}\right|} + c_{1,i}^{+} e^{\frac{|z|}{\hbar}\left|\frac{\partial S}{\partial \mathbf{x}}\right|}.
\end{align}
Second, we solve the inhomogeneous equation using the method of undetermined coefficients. To this end, we employ the Ansatz
\begin{align}\label{eq:appenpartsol}
  V_{1,i,\mathrm{P}}(\mathbf{x},z) =  \left(\alpha_{i} \frac{z}{\hbar} + \beta_{i} \frac{z^2}{\hbar^2} \right) e^{-\frac{|z|}{\hbar}\left|\frac{\partial S}{\partial \mathbf{x}}\right|}.
\end{align}
Inserting this Ansatz into the differential equation~(\ref{eq:AppenInhomogeneousdiff}), we obtain the constants $\alpha_i$ and $\beta_i$. After some calculus, we find that the constants $\beta_i$ are the same above and below the plane, and given by
\begin{equation}
  \beta = - \frac{i}{2} \frac{2 \pi e^2 \hbar }{\varepsilon_{\mathrm{avg}}(\mathbf{x}) } \frac{\Pi_0\left(\mathbf{x}, \frac{\partial S}{\partial \mathbf{x}}\right)}{| \partial S/ \partial \mathbf{x}|^3}  \varphi_0(\mathbf{x}) \sum_{j,k} \frac{\partial S}{\partial x_j}\frac{\partial^2 S}{\partial x_j \partial x_k} \frac{\partial S}{\partial x_k},
\end{equation}
where we dropped the subscript. Although the constant $\beta$ vanished for the operator separation, it is non-zero in this derivation. We come back to this shortly. Using our result for $\beta$, we find $\alpha_i$ as
\begin{align}
  \alpha_i 
  &= \frac{i}{2} \frac{s_i}{| \partial S/ \partial \mathbf{x}|}
  \bigg( - \frac{c_0}{ |\partial S/ \partial \mathbf{x}|^2 } \sum_{j,k} \frac{\partial S}{\partial x_j} \frac{\partial S}{\partial x_k} \frac{\partial^2 S}{\partial x_j \partial x_k} \nonumber \\
  & \hspace*{0.5cm} + \frac{c_0}{\varepsilon_i} \left\langle \frac{\partial \varepsilon_i}{\partial x} , \frac{\partial S}{\partial x} \right\rangle 
  + 2 \left\langle \frac{\partial c_0}{\partial x} , \frac{\partial S}{\partial x} \right\rangle 
  + c_0 \frac{\partial^2 S}{\partial \mathbf{x}^2} \bigg) ,  
  \label{eq:alpha-i-app}
\end{align}
where $s_i=1$ for $i=\mathrm{A}$ and $s_i=-1$ for $i=\mathrm{B}$, as in the main text.
The full solution is given by the sum of the homogeneous~(\ref{eq:appenhomosol}), and particular~(\ref{eq:appenpartsol}) solution. 

As before, the constants $c_{1,i}^{-}$, and $c_{1,i}^{+}$ are determined via the boundary conditions. First, the induced potential tends to zero as $|z| \rightarrow \infty$, which yields $c_{1,i}^{+} = 0$. Second, the potential has to be continuous at the interface $z=0$, whence $c_{1,\mathrm{A}}^{-}=c_{1,\mathrm{B}}^{-}=c_{1}$. The final boundary condition concerns the derivative of $V$ at the interface. It can either be derived from the differential equation~(\ref{eq:AppenInhomogeneousdiff}) or from the requirement that the $\boldsymbol{\mathcal{D}}$-field is discontinuous at the interface, with the discontinuity given by the surface charge determined by the induced electron density. We find that the constant $c_{1}$ is given by
\begin{equation}  \label{eq:c1-app}
  c_{1}= \frac{ 2 \pi e^2 \hbar}{\varepsilon_\mathrm{avg}(\mathbf{x})} \frac{n_1(\mathbf{x})}{\left|\partial S/\partial \mathbf{x}\right|}  + \frac{1}{ \left|\partial S/\partial \mathbf{x}\right|}\frac{\varepsilon_\mathrm{A}(\mathbf{x})\alpha_\mathrm{A} - \varepsilon_\mathrm{B}(\mathbf{x})\alpha_\mathrm{B}}{ 2\varepsilon_\mathrm{avg}(\mathbf{x})} .
\end{equation}
The total solution for $V_{1,i}$ is then given by
\begin{equation}  \label{eq:V1i-final}
  V_{1,i}(\mathbf{x},z) = \left(c_{1} +\alpha_{i} \frac{z}{\hbar} + \beta \frac{z^2}{\hbar^2} \right) e^{-\frac{|z|}{\hbar}\left|\frac{\partial S}{\partial \mathbf{x}}\right|}.
\end{equation}
We may briefly compare this result with our expression~(\ref{eq:chi1-sol-general}) for $\Gamma_{1,i}$. The most striking difference between the two is the presence of a term with $z^2$ in Eq.~(\ref{eq:V1i-final}). In the main text, this term only arises when one computes the action of the pseudodifferential operator $\hat{\Gamma}$ on the semiclassical Ansatz for $V_\mathrm{pl}$, as considered in Eq.~(\ref{eq:totalpotential}). Specifically, it arises when both derivatives in $\partial^2 \Gamma_0/\partial q_j \partial q_k$ are applied to the exponent. In both derivations, one obtains the same result for $V_{1,i}$.

\subsection{Transport equation}

Our next step is to apply the self-consistency condition~(\ref{eq:condition-self-consistency}) at $z=0$ to the subleading-order terms. In this way, we should obtain the transport equation. The condition reads
\begin{equation}
  V_{1,i}(\mathbf{x},z=0)e^{i S(\mathbf{x})/\hbar} = \varphi_1(\mathbf{x}) e^{i S(\mathbf{x})/\hbar},
\end{equation}
which directly yields $\varphi_1 - c_1 = 0$. At this point, one has to insert the expressions~(\ref{eq:alpha-i-app}) for $\alpha_i$ into Eq.~(\ref{eq:c1-app}) and compute all derivatives. One then computes the derivatives $\partial \mathcal{H}_0/\partial q_j$ and $\partial^2 \mathcal{H}_0/\partial q_j \partial q_k$, and uses them to rewrite the terms in $c_0$.
After somewhat lengthy calculations, the equation $\varphi_1 - c_1 = 0$ becomes
\begin{align}
  &\bigg(1 - \frac{ 2 \pi e^2 \hbar }{\varepsilon_\mathrm{avg} \left|\partial S/\partial \mathbf{x}\right|} \Pi_0 \bigg) \varphi_1 
  + \bigg( \! \frac{- 2 \pi i e^2 \hbar}{\varepsilon_{\mathrm{avg}} |\partial S/\partial \mathbf{x}|^3} \left\langle \frac{\partial S}{\partial \mathbf{x}}, \frac{\partial \Pi_0}{\partial \mathbf{x}} \right\rangle \nonumber \\
  &\hspace*{1.8cm} + \frac{i}{2} \frac{2 \pi e^2 \hbar}{\varepsilon_{\mathrm{avg}}^2 |\partial S/\partial \mathbf{x}|^3} \Pi_0 \left\langle \frac{\partial S}{\partial \mathbf{x}}, \frac{\partial \varepsilon_\mathrm{avg}}{\partial \mathbf{x}} \right\rangle \nonumber \\
  &\hspace*{1.8cm} - \frac{2 \pi e^2 \hbar}{\varepsilon_{\mathrm{avg}} |\partial S/\partial \mathbf{x}|} \Pi_1 \bigg) \varphi_0 
  - i \left\langle\frac{\partial \mathcal{H}_0}{\partial \mathbf{q}}, \frac{\partial \varphi_0}{\partial \mathbf{x}} \right\rangle \nonumber \\
  &\hspace*{1.8cm} - \frac{i}{2} \sum_{j,k} \frac{\partial^2 \mathcal{H}_0}{\partial q_j \partial q_k} \frac{\partial^2 S}{\partial x_k \partial x_j} \varphi_0 = 0 .
\end{align}
First, the terms in front of $\varphi_1$ cancel because of the Hamilton-Jacobi equation~(\ref{eq:appenHamJac}). Using our previous definition~(\ref{eq:defL1}) for $\mathcal{H}_1(\mathbf{x},\mathbf{q})$, we may write the remaining terms as 
\begin{multline}\label{eq:appentransporteq}
  \mathcal{H}_1\left(\mathbf{x},\frac{\partial S}{\partial \mathbf{x}}\right) \varphi_0 - i \left\langle\frac{\partial \mathcal{H}_0}{\partial \mathbf{q}}\left(\mathbf{x},\frac{\partial S}{\partial \mathbf{x}}\right), \frac{\partial \varphi_0}{\partial \mathbf{x}} \right\rangle\\
  - \frac{i}{2} \sum_{j,k} \frac{\partial^2 \mathcal{H}_0}{\partial q_j \partial q_k}\left(\mathbf{x},\frac{\partial S}{\partial \mathbf{x}}\right) \frac{\partial^2 S}{\partial x_k \partial x_j} \varphi_0   = 0,
\end{multline}
which exactly coincides with our previous transport equation~(\ref{eq:transporteq}).

We have thus shown that the Ansatz~(\ref{eq:VtotalExpansion}) for $V(\mathbf{x},z)$ leads to the same results as the operator separation discussed in Sec.~\ref{subsec:derivation-separation}, namely the Hamilton-Jacobi equation~(\ref{eq:HJequation}) and the transport equation~(\ref{eq:transporteq}). However, the calculations are more tedious since this method mixes the separation of the in-plane and out-of-plane degrees of freedom with the application of the semiclassical Ansatz.

\section{Energy density in the plane $z=0$}\label{ap:inplaneEnergyDensity}

In Sec.~\ref{subsec:EnergyDensity}, we computed the integrated energy density $\mathcal{U}_\mathrm{I}(\mathbf{x})$ for our two-dimensional problem. In the derivation, we set the contribution $\mathcal{U}_\mathrm{pl}(\mathbf{x})$ of the plane $z=0$ to zero. In this appendix, we derive a general formula for the leading-order term of the semiclassical energy density in a medium with plasmons. Moreover, we show that $\mathcal{U}_\mathrm{pl}(\mathbf{x}) = 0$ in our example, that is, when the layer is infinitely thin.

As in the main text, the relation between the electric field and the induced potential is given by $\boldsymbol{  \mathcal{E}} = e^{-1} \nabla V$. However, the relation between the displacement field and the electric field is more complicated than in the main text.
We start from the general relation between the electric field and the displacement field, namely
\begin{align}
  \boldsymbol{ \mathcal{D}} ( \mathbf{x} , t) = \int \mathrm{d} t'\int \mathrm{d} \mathbf{x}' \varepsilon (\mathbf{x},\mathbf{x}',t-t') \boldsymbol{ \mathcal{E}} (\mathbf{x}',t').
\end{align}
We perform a Fourier transform with respect to time, i.e.,
\begin{align}  \label{eq:D-E-FT}
  \boldsymbol{ \mathcal{D}} ( \mathbf{x} , \omega) = \int \mathrm{d} \mathbf{x}' \varepsilon (\mathbf{x},\mathbf{x}',\omega) \boldsymbol{ \mathcal{E}} (\mathbf{x}',\omega).
\end{align}
Throughout this appendix, we use $\omega$ instead of $E$, not only to make a more explicit connection with the conventions in the literature~\cite{Jackson99,Landau84}, but also to create a clear distinction in notation with the electric field.
We can rewrite Eq.~(\ref{eq:D-E-FT}) as
\begin{equation}  \label{eq:D-E-operator}
  \boldsymbol{ \mathcal{D}} ( \mathbf{x} , \omega) = (\hat{\varepsilon} \boldsymbol{ \mathcal{E}}) (\mathbf{x},\omega) ,
\end{equation}
where $\hat{\varepsilon}$ is an operator that corresponds to the dielectric function. We now need to extract the proper definition of the operator $\hat{\varepsilon}$ from our semiclassical analysis.

Since plasmons are self-sustained oscillations, they are defined by the vanishing of the displacement field. For homogeneous systems~\cite{Vonsovsky89,Giuliani05}, this requirement translates to the secular equation $\varepsilon(\mathbf{q},\omega) V(\mathbf{q}) = 0$, generally speaking. When we consider inhomogeneous systems and apply the semiclassical approximation~\cite{Reijnders22}, the secular equation becomes the operator equation $\hat{\mathcal{H}} V = 0$, where $\hat{\mathcal{H}}$ is a pseudodifferential operator. Indeed, we can write the result of our derivation in Sec.~\ref{sec:derivation} as $\hat{\mathcal{H}} V_\mathrm{pl} = 0$, where $V_\mathrm{pl}(\mathbf{x},\omega) = \varphi_0 (\mathbf{x}) \exp(i S(\mathbf{x}) / \hbar )$ and the principal symbol $\mathcal{H}_0(\mathbf{x},\mathbf{q},\omega)$ of $\hat{\mathcal{H}}$ is the effective classical Hamiltonian given by Eq.~(\ref{eq:effclassicalHam}). Comparing this result to Eq.~(\ref{eq:D-E-operator}), we may say that $\hat{\varepsilon}$ is a pseudodifferential operator. It is, however, not equal to $\hat{\mathcal{H}}$, since its principal symbol should equal $\varepsilon_\mathrm{avg}$ in the absence of a polarization. We therefore argue that the principal symbol $\varepsilon_0 (\mathbf{x},\mathbf{q},\omega)$ of $\hat{\varepsilon}$ equals
\begin{align}  \label{eq:epsilon-principal}
  \varepsilon_0 (\mathbf{x},\mathbf{q},\omega) = \varepsilon_\mathrm{avg} (\mathbf{x}) - \frac{2 \pi e^2 \hbar}{|\mathbf{q}|}   \Pi_0 (\mathbf{x},\mathbf{q},\omega) ,
\end{align}
which gives the correct result in the homogeneous case.

Equations~(\ref{eq:D-E-operator}) and~(\ref{eq:epsilon-principal}) allow us to derive the leading-order term of the displacement field $\boldsymbol{ \mathcal{D}}(\mathbf{x},\omega)$. Since $V(\mathbf{x},\omega)$ has the form of a semiclassical Ansatz, so does $\boldsymbol{ \mathcal{E}}(\mathbf{x},\omega)$. We can therefore use the general formula for the commutation of a pseudo-differential operator with a rapidly oscillating exponent~\cite{Maslov81,Guillemin77}, and write
\begin{multline}\label{eq:appendixDisplacementField}
  \boldsymbol{\mathcal{D}}(\mathbf{x},\omega) 
  = (\hat{\varepsilon}\boldsymbol{ \mathcal{E}})(\mathbf{x},\omega) \\
  = \varepsilon_0\left(\mathbf{x},\frac{\partial S}{\partial \mathbf{x}},\omega\right) \boldsymbol{ \mathcal{E}}(\mathbf{x},\omega) (1 + \mathcal{O}(\hbar)) ,
\end{multline}
cf. Eq.~(\ref{eq:totalpotential}). From here on, we only consider the leading-order term, and therefore omit the $\mathcal{O}(\hbar)$.

We can now obtain an expression for the energy density by repeating the derivation in Ref.~\cite{Landau84}, almost verbatim. 
Since $\varepsilon_0\left(\mathbf{x},\partial S / \partial \mathbf{x} ,\omega\right)$ depends on $\omega$, the medium is dispersive and one cannot consider a purely monochromatic field.
Instead, we let $\boldsymbol{ \mathcal{E}}(\mathbf{x},t) = \boldsymbol{ \mathcal{E}}_c (\mathbf{x},t) \exp(-i \omega_c t)$, where $\boldsymbol{ \mathcal{E}}_c(\mathbf{x},t)$ varies only slowly with time. 
Hence, when we write down the Fourier expansion of $\boldsymbol{ \mathcal{E}} (\mathbf{x},t)$, namely,
\begin{align}
  \boldsymbol{ \mathcal{E}}(\mathbf{x},t) = \int \frac{\mathrm{d} \omega}{2 \pi} \boldsymbol{ \mathcal{E}}_{c} (\mathbf{x}, \omega) e^{-i(\omega_c + \omega)t},
\end{align}
only the components $\boldsymbol{ \mathcal{E}}_{c} (\mathbf{x}, \omega)$ with $\omega \ll \omega_c$ are significant.
Expanding the displacement field $\boldsymbol{\mathcal{D}}(\mathbf{x},t)$ in Fourier components up to leading order, and using Eq.~(\ref{eq:appendixDisplacementField}), we find that
\begin{align}
  \frac{\partial \boldsymbol{ \mathcal{D}}}{\partial t} = \int \frac{\mathrm{d} \omega}{2 \pi} f(\mathbf{x}, \omega_c +\omega) \boldsymbol{ \mathcal{E}}_{c} (\mathbf{x}, \omega) e^{-i(\omega_c + \omega)t},
\end{align}
where $f(\mathbf{x}, \omega) = - i \omega \varepsilon_0 \left(\mathbf{x}, \partial S / \partial \mathbf{x}, \omega \right)$, cf. Ref.~\cite{Landau84}. Since the Fourier components $\boldsymbol{ \mathcal{E}}_{c} (\mathbf{x}, \omega)$ are very small for $\omega\gg\omega_c$, we can expand $f(\mathbf{x}, \omega)$ to first order in $\omega$ around $\omega_c$, that is,
\begin{multline}
  \frac{\partial \boldsymbol{ \mathcal{D}}}{\partial t} = \int \frac{\mathrm{d} \omega}{2 \pi} \left(f(\mathbf{x}, \omega_c) + \omega \frac{\partial f}{\partial \omega} (\mathbf{x},\omega_c)\right)\\
  \times \boldsymbol{ \mathcal{E}}_{c} (\mathbf{x}, \omega) e^{-i(\omega_c + \omega)t}.
\end{multline}
Using the definition of the Fourier transform, the right-hand side can be rewritten as
\begin{multline}  \label{eq:D-deriv-expanded}
  \frac{\partial \boldsymbol{ \mathcal{D}}}{\partial t} =   - i \omega \varepsilon_0 \left(\mathbf{x},\frac{\partial S}{\partial \mathbf{x}}, \omega\right)  \boldsymbol{ \mathcal{E}}_c (\mathbf{x},t) e^{-i \omega t}  \\
  + \frac{\partial}{\partial \omega} \left( \omega \varepsilon_0 \left(\mathbf{x}, \frac{\partial S}{\partial \mathbf{x}}, \omega\right) \right) \frac{\partial \boldsymbol{ \mathcal{E}}_{c}}{\partial t} (\mathbf{x}, t) e^{-i\omega t},
\end{multline}
where we now omitted the subscript $c$ on $\omega$.

We now substitute the result~(\ref{eq:D-deriv-expanded}) in the Poynting theorem~(\ref{eq:PoyntingsTheorem}). Assuming that $\varepsilon_0$ is a real function, we find the leading-order term of the semiclassical energy density by taking out the time derivative. We obtain
\begin{align}
  \mathcal{U}^{\mathrm{SC}}(\mathbf{x}) = \frac{1}{16 \pi}  \frac{\partial}{\partial \omega} \left( \omega \varepsilon_0 \left(\mathbf{x},\frac{\partial S}{\partial \mathbf{x}}, \omega\right) \right)  \boldsymbol{ \mathcal{E}} ( \mathbf{x} , t)  \cdot  \boldsymbol{ \mathcal{E}}^* ( \mathbf{x} , t).
\end{align}
Note that this expression reduces to the much simpler result~(\ref{eq:energy-density-simple}) when we consider a function $\varepsilon(\mathbf{x})$ that does not depend on $\omega$. 
When we consider plasmons, we can simplify this expression by noting that $\varepsilon_0 \left(\mathbf{x},\partial S/\partial \mathbf{x}, \omega\right)$ vanishes by virtue of the Hamilton-Jacobi equation. We obtain
\begin{align}  \label{eq:U-SC-density-plasmons}
  \mathcal{U}^{\mathrm{SC}}(\mathbf{x}) = \frac{\omega}{16 \pi e^2}  \frac{\partial \varepsilon_0}{\partial \omega}  \left(\mathbf{x},\frac{\partial S}{\partial \mathbf{x}}, \omega\right) |\nabla V( \mathbf{x} , t)|^2 . 
\end{align}
where we also used the relation between the electric field and the induced potential.

In deriving the result~(\ref{eq:U-SC-density-plasmons}), we did not specify the number of dimensions, which means that it is equally valid in three and two dimensions. When we consider our two-dimensional problem, both $\boldsymbol{\mathcal{D}}$ and $\boldsymbol{\mathcal{E}}$ gain an additional coordinate $z$, and $\mathbf{x}$ becomes two-dimensional. Throughout this article, we considered an infinitely thin charge layer at $z=0$. 
However, one may intuitively argue that we can use expression~(\ref{eq:U-SC-density-plasmons}) between $z=-\epsilon$ and $z=\epsilon$, where $\epsilon$ is a small number that is determined by the requirement that the induced potential $V(\mathbf{x},z,t)$ has not yet decayed significantly. Looking at Eq.~(\ref{eq:V0solutionamplitudeopsep}), we observe that this is equivalent to $\epsilon \left|\partial S / \partial \mathbf{x}\right| / \hbar \ll 1 $, meaning that $\epsilon$ should be much smaller than the plasmon wavelength, in accordance with the derivation in Sec.~\ref{subsec:derivation-separation}.

We obtain an expression for $\mathcal{U}_\mathrm{pl}(\mathbf{x})$, the energy density that comes from the two-dimensional plane at $z=0$, by integrating Eq.~(\ref{eq:U-SC-density-plasmons}) from $z=-\epsilon$ to $z=\epsilon$. Using Eq.~(\ref{eq:V0solutionamplitudeopsep}) for $V(\mathbf{x},z,t)$, we have
\begin{multline}
  \mathcal{U}_\mathrm{pl}(\mathbf{x}) =\frac{1}{8 \pi e^2 \hbar } \frac{|A_0^0|^2}{|J|}   \frac{ \omega}{\varepsilon_{\mathrm{avg}}}\frac{\partial \varepsilon_0}{\partial \omega} \left(\mathbf{x},\frac{\partial S }{\partial \mathbf{x}} , \omega\right) \\
  \times \left(1 -e^{-2\frac{\epsilon}{\hbar}\left|\frac{\partial S}{\partial \mathbf{x}}\right|} \right).
\end{multline}
Since we previously required that $\epsilon \left|\partial S / \partial \mathbf{x}\right| / \hbar \ll 1 $, we can Taylor expand the exponent up to first order, which yields
\begin{align}
  \mathcal{U}_\mathrm{pl}(\mathbf{x}) =\frac{1}{8 \pi e^2 \hbar } \frac{|A_0^0|^2}{|J|}   \frac{ \omega}{\varepsilon_{\mathrm{avg}}}\frac{\partial \varepsilon_0}{\partial \omega} \left(\mathbf{x},\frac{\partial S }{\partial \mathbf{x}} , \omega\right) \times 2\frac{\epsilon}{\hbar}\left|\frac{\partial S}{\partial \mathbf{x}}\right|.
\end{align}
Comparing this to the result~(\ref{eq:U-A-integrated}) for $\mathcal{U}_\mathrm{A}(\mathbf{x})$, we observe that it is of higher order. Because we set out to derive the leading-order term of the integrated energy density $\mathcal{U}_\mathrm{I}(\mathbf{x})$, we conclude that the contribution of the two-dimensional plane is effectively zero, i.e. $\mathcal{U}_\mathrm{pl}(\mathbf{x})=0$.
The derivation shows that the same result likely holds for layers with a finite thickness, cf. the discussion in the introduction and the conclusion, provided that $\epsilon \left|\partial S / \partial \mathbf{x}\right| / \hbar \ll 1 $.


\begin{thebibliography}{66}%
  \makeatletter
  \providecommand \@ifxundefined [1]{%
    \@ifx{#1\undefined}
  }%
  \providecommand \@ifnum [1]{%
    \ifnum #1\expandafter \@firstoftwo
    \else \expandafter \@secondoftwo
    \fi
  }%
  \providecommand \@ifx [1]{%
    \ifx #1\expandafter \@firstoftwo
    \else \expandafter \@secondoftwo
    \fi
  }%
  \providecommand \natexlab [1]{#1}%
  \providecommand \enquote  [1]{``#1''}%
  \providecommand \bibnamefont  [1]{#1}%
  \providecommand \bibfnamefont [1]{#1}%
  \providecommand \citenamefont [1]{#1}%
  \providecommand \href@noop [0]{\@secondoftwo}%
  \providecommand \href [0]{\begingroup \@sanitize@url \@href}%
  \providecommand \@href[1]{\@@startlink{#1}\@@href}%
  \providecommand \@@href[1]{\endgroup#1\@@endlink}%
  \providecommand \@sanitize@url [0]{\catcode `\\12\catcode `\$12\catcode
    `\&12\catcode `\#12\catcode `\^12\catcode `\_12\catcode `\%12\relax}%
  \providecommand \@@startlink[1]{}%
  \providecommand \@@endlink[0]{}%
  \providecommand \url  [0]{\begingroup\@sanitize@url \@url }%
  \providecommand \@url [1]{\endgroup\@href {#1}{\urlprefix }}%
  \providecommand \urlprefix  [0]{URL }%
  \providecommand \Eprint [0]{\href }%
  \providecommand \doibase [0]{https://doi.org/}%
  \providecommand \selectlanguage [0]{\@gobble}%
  \providecommand \bibinfo  [0]{\@secondoftwo}%
  \providecommand \bibfield  [0]{\@secondoftwo}%
  \providecommand \translation [1]{[#1]}%
  \providecommand \BibitemOpen [0]{}%
  \providecommand \bibitemStop [0]{}%
  \providecommand \bibitemNoStop [0]{.\EOS\space}%
  \providecommand \EOS [0]{\spacefactor3000\relax}%
  \providecommand \BibitemShut  [1]{\csname bibitem#1\endcsname}%
  \let\auto@bib@innerbib\@empty
  \bibitem [{\citenamefont {Vonsovsky}\ and\ \citenamefont
    {Katsnelson}(1989)}]{Vonsovsky89}%
  \BibitemOpen
  \bibfield  {author} {\bibinfo {author} {\bibfnamefont {S.~V.}\ \bibnamefont
      {Vonsovsky}}\ and\ \bibinfo {author} {\bibfnamefont {M.~I.}\ \bibnamefont
      {Katsnelson}},\ }\href@noop {} {\emph {\bibinfo {title} {Quantum solid-state
        physics}}}\ (\bibinfo  {publisher} {Springer-Verlag, Berlin Heidelberg},\
  \bibinfo {year} {1989})\BibitemShut {NoStop}%
  \bibitem [{\citenamefont {Giuliani}\ and\ \citenamefont
    {Vignale}(2005)}]{Giuliani05}%
  \BibitemOpen
  \bibfield  {author} {\bibinfo {author} {\bibfnamefont {G.~F.}\ \bibnamefont
      {Giuliani}}\ and\ \bibinfo {author} {\bibfnamefont {G.}~\bibnamefont
      {Vignale}},\ }\href@noop {} {\emph {\bibinfo {title} {Quantum theory of the
        electron liquid}}}\ (\bibinfo  {publisher} {Cambridge University Press,
    Cambridge},\ \bibinfo {year} {2005})\BibitemShut {NoStop}%
  \bibitem [{\citenamefont {Platzman}\ and\ \citenamefont
    {Wolff}(1973)}]{Platzman73}%
  \BibitemOpen
  \bibfield  {author} {\bibinfo {author} {\bibfnamefont {P.~M.}\ \bibnamefont
      {Platzman}}\ and\ \bibinfo {author} {\bibfnamefont {P.~A.}\ \bibnamefont
      {Wolff}},\ }\href@noop {} {\emph {\bibinfo {title} {Waves and Interactions in
        Solid State Plasmas}}}\ (\bibinfo  {publisher} {Academic Press, New York},\
  \bibinfo {year} {1973})\BibitemShut {NoStop}%
  \bibitem [{\citenamefont {Nozi\`eres}\ and\ \citenamefont
    {Pines}(1999)}]{Nozieres99}%
  \BibitemOpen
  \bibfield  {author} {\bibinfo {author} {\bibfnamefont {P.}~\bibnamefont
      {Nozi\`eres}}\ and\ \bibinfo {author} {\bibfnamefont {D.}~\bibnamefont
      {Pines}},\ }\href@noop {} {\emph {\bibinfo {title} {Theory of Quantum
        Liquids}}}\ (\bibinfo  {publisher} {Hachette, United Kingdom},\ \bibinfo
  {year} {1999})\BibitemShut {NoStop}%
  \bibitem [{\citenamefont {Pines}\ and\ \citenamefont {Bohm}(1952)}]{Pines52}%
  \BibitemOpen
  \bibfield  {author} {\bibinfo {author} {\bibfnamefont {D.}~\bibnamefont
      {Pines}}\ and\ \bibinfo {author} {\bibfnamefont {D.}~\bibnamefont {Bohm}},\
  }\href@noop {} {\bibfield  {journal} {\bibinfo  {journal} {Phys. Rev.}\
    }\textbf {\bibinfo {volume} {85}},\ \bibinfo {pages} {338} (\bibinfo {year}
    {1952})}\BibitemShut {NoStop}%
  \bibitem [{\citenamefont {Barnes}\ \emph {et~al.}(2003)\citenamefont {Barnes},
    \citenamefont {Dereux},\ and\ \citenamefont {Ebbesen}}]{Barnes03}%
  \BibitemOpen
  \bibfield  {author} {\bibinfo {author} {\bibfnamefont {W.~L.}\ \bibnamefont
      {Barnes}}, \bibinfo {author} {\bibfnamefont {A.}~\bibnamefont {Dereux}},\
    and\ \bibinfo {author} {\bibfnamefont {T.~W.}\ \bibnamefont {Ebbesen}},\
  }\href@noop {} {\bibfield  {journal} {\bibinfo  {journal} {Nature}\ }\textbf
    {\bibinfo {volume} {424}},\ \bibinfo {pages} {824} (\bibinfo {year}
    {2003})}\BibitemShut {NoStop}%
  \bibitem [{\citenamefont {Tame}\ \emph {et~al.}(2013)\citenamefont {Tame},
    \citenamefont {McEnery}, \citenamefont {\"Ozdemir}, \citenamefont {Lee},
    \citenamefont {Maier},\ and\ \citenamefont {Kim}}]{Tame13}%
  \BibitemOpen
  \bibfield  {author} {\bibinfo {author} {\bibfnamefont {M.~S.}\ \bibnamefont
      {Tame}}, \bibinfo {author} {\bibfnamefont {K.~R.}\ \bibnamefont {McEnery}},
    \bibinfo {author} {\bibfnamefont {{\relax \c{S}}.~K.}\ \bibnamefont
      {\"Ozdemir}}, \bibinfo {author} {\bibfnamefont {J.}~\bibnamefont {Lee}},
    \bibinfo {author} {\bibfnamefont {S.~A.}\ \bibnamefont {Maier}},\ and\
    \bibinfo {author} {\bibfnamefont {M.~S.}\ \bibnamefont {Kim}},\ }\href@noop
  {} {\bibfield  {journal} {\bibinfo  {journal} {Nat. Phys.}\ }\textbf
    {\bibinfo {volume} {9}},\ \bibinfo {pages} {329} (\bibinfo {year}
    {2013})}\BibitemShut {NoStop}%
  \bibitem [{\citenamefont {Fitzgerald}\ \emph {et~al.}(2016)\citenamefont
    {Fitzgerald}, \citenamefont {Narang}, \citenamefont {Craster}, \citenamefont
    {Maier},\ and\ \citenamefont {Giannini}}]{Fitzgerald16}%
  \BibitemOpen
  \bibfield  {author} {\bibinfo {author} {\bibfnamefont {J.~M.}\ \bibnamefont
      {Fitzgerald}}, \bibinfo {author} {\bibfnamefont {P.}~\bibnamefont {Narang}},
    \bibinfo {author} {\bibfnamefont {R.~V.}\ \bibnamefont {Craster}}, \bibinfo
    {author} {\bibfnamefont {S.~A.}\ \bibnamefont {Maier}},\ and\ \bibinfo
    {author} {\bibfnamefont {V.}~\bibnamefont {Giannini}},\ }\href@noop {}
  {\bibfield  {journal} {\bibinfo  {journal} {Proc. IEEE}\ }\textbf {\bibinfo
      {volume} {104}},\ \bibinfo {pages} {2307} (\bibinfo {year}
    {2016})}\BibitemShut {NoStop}%
  \bibitem [{\citenamefont {Scholl}\ \emph {et~al.}(2012)\citenamefont {Scholl},
    \citenamefont {Koh},\ and\ \citenamefont {Dionne}}]{Scholl12}%
  \BibitemOpen
  \bibfield  {author} {\bibinfo {author} {\bibfnamefont {J.~A.}\ \bibnamefont
      {Scholl}}, \bibinfo {author} {\bibfnamefont {A.~L.}\ \bibnamefont {Koh}},\
    and\ \bibinfo {author} {\bibfnamefont {J.~A.}\ \bibnamefont {Dionne}},\
  }\href@noop {} {\bibfield  {journal} {\bibinfo  {journal} {Nature}\ }\textbf
    {\bibinfo {volume} {483}},\ \bibinfo {pages} {421} (\bibinfo {year}
    {2012})}\BibitemShut {NoStop}%
  \bibitem [{\citenamefont {Grigorenko}\ \emph {et~al.}(2012)\citenamefont
    {Grigorenko}, \citenamefont {Polini},\ and\ \citenamefont
    {Novoselov}}]{Grigorenko12}%
  \BibitemOpen
  \bibfield  {author} {\bibinfo {author} {\bibfnamefont {A.~N.}\ \bibnamefont
      {Grigorenko}}, \bibinfo {author} {\bibfnamefont {M.}~\bibnamefont {Polini}},\
    and\ \bibinfo {author} {\bibfnamefont {K.~S.}\ \bibnamefont {Novoselov}},\
  }\href@noop {} {\bibfield  {journal} {\bibinfo  {journal} {Nat. Photonics}\
    }\textbf {\bibinfo {volume} {6}},\ \bibinfo {pages} {749} (\bibinfo {year}
    {2012})}\BibitemShut {NoStop}%
  \bibitem [{\citenamefont {Basov}\ \emph {et~al.}(2016)\citenamefont {Basov},
    \citenamefont {Fogler},\ and\ \citenamefont {Garc{\'\i}a~de
      Abajo}}]{Basov16}%
  \BibitemOpen
  \bibfield  {author} {\bibinfo {author} {\bibfnamefont {D.~N.}\ \bibnamefont
      {Basov}}, \bibinfo {author} {\bibfnamefont {M.~M.}\ \bibnamefont {Fogler}},\
    and\ \bibinfo {author} {\bibfnamefont {F.~J.}\ \bibnamefont {Garc{\'\i}a~de
        Abajo}},\ }\href@noop {} {\bibfield  {journal} {\bibinfo  {journal}
      {Science}\ }\textbf {\bibinfo {volume} {354}},\ \bibinfo {pages} {aag1992}
    (\bibinfo {year} {2016})}\BibitemShut {NoStop}%
  \bibitem [{\citenamefont {Low}\ \emph {et~al.}(2017)\citenamefont {Low},
    \citenamefont {Chaves}, \citenamefont {Caldwell}, \citenamefont {Kumar},
    \citenamefont {Fang}, \citenamefont {Avouris}, \citenamefont {Heinz},
    \citenamefont {Guinea}, \citenamefont {Martin-Moreno},\ and\ \citenamefont
    {Koppens}}]{Low17}%
  \BibitemOpen
  \bibfield  {author} {\bibinfo {author} {\bibfnamefont {T.}~\bibnamefont
      {Low}}, \bibinfo {author} {\bibfnamefont {A.}~\bibnamefont {Chaves}},
    \bibinfo {author} {\bibfnamefont {J.~D.}\ \bibnamefont {Caldwell}}, \bibinfo
    {author} {\bibfnamefont {A.}~\bibnamefont {Kumar}}, \bibinfo {author}
    {\bibfnamefont {N.~X.}\ \bibnamefont {Fang}}, \bibinfo {author}
    {\bibfnamefont {P.}~\bibnamefont {Avouris}}, \bibinfo {author} {\bibfnamefont
      {T.~F.}\ \bibnamefont {Heinz}}, \bibinfo {author} {\bibfnamefont
      {F.}~\bibnamefont {Guinea}}, \bibinfo {author} {\bibfnamefont
      {L.}~\bibnamefont {Martin-Moreno}},\ and\ \bibinfo {author} {\bibfnamefont
      {F.}~\bibnamefont {Koppens}},\ }\href@noop {} {\bibfield  {journal} {\bibinfo
      {journal} {Nat. Mater.}\ }\textbf {\bibinfo {volume} {16}},\ \bibinfo
    {pages} {182} (\bibinfo {year} {2017})}\BibitemShut {NoStop}%
  \bibitem [{\citenamefont {Geim}\ and\ \citenamefont
    {Grigorieva}(2013)}]{Geim13}%
  \BibitemOpen
  \bibfield  {author} {\bibinfo {author} {\bibfnamefont {A.~K.}\ \bibnamefont
      {Geim}}\ and\ \bibinfo {author} {\bibfnamefont {I.~V.}\ \bibnamefont
      {Grigorieva}},\ }\href@noop {} {\bibfield  {journal} {\bibinfo  {journal}
      {Nature}\ }\textbf {\bibinfo {volume} {499}},\ \bibinfo {pages} {419}
    (\bibinfo {year} {2013})}\BibitemShut {NoStop}%
  \bibitem [{\citenamefont {Fei}\ \emph {et~al.}(2012)\citenamefont {Fei},
    \citenamefont {Rodin}, \citenamefont {Andreev}, \citenamefont {Bao},
    \citenamefont {McLeod}, \citenamefont {Wagner}, \citenamefont {Zhang},
    \citenamefont {Zhao}, \citenamefont {Thiemens}, \citenamefont {Dominguez},
    \citenamefont {Fogler}, \citenamefont {Castro~Neto}, \citenamefont {Lau},
    \citenamefont {Keilmann},\ and\ \citenamefont {Basov}}]{Fei12}%
  \BibitemOpen
  \bibfield  {author} {\bibinfo {author} {\bibfnamefont {Z.}~\bibnamefont
      {Fei}}, \bibinfo {author} {\bibfnamefont {A.~S.}\ \bibnamefont {Rodin}},
    \bibinfo {author} {\bibfnamefont {G.~O.}\ \bibnamefont {Andreev}}, \bibinfo
    {author} {\bibfnamefont {W.}~\bibnamefont {Bao}}, \bibinfo {author}
    {\bibfnamefont {A.~S.}\ \bibnamefont {McLeod}}, \bibinfo {author}
    {\bibfnamefont {M.}~\bibnamefont {Wagner}}, \bibinfo {author} {\bibfnamefont
      {L.~M.}\ \bibnamefont {Zhang}}, \bibinfo {author} {\bibfnamefont
      {Z.}~\bibnamefont {Zhao}}, \bibinfo {author} {\bibfnamefont {M.}~\bibnamefont
      {Thiemens}}, \bibinfo {author} {\bibfnamefont {G.}~\bibnamefont {Dominguez}},
    \bibinfo {author} {\bibfnamefont {M.~M.}\ \bibnamefont {Fogler}}, \bibinfo
    {author} {\bibfnamefont {A.~H.}\ \bibnamefont {Castro~Neto}}, \bibinfo
    {author} {\bibfnamefont {C.~N.}\ \bibnamefont {Lau}}, \bibinfo {author}
    {\bibfnamefont {F.}~\bibnamefont {Keilmann}},\ and\ \bibinfo {author}
    {\bibfnamefont {D.~N.}\ \bibnamefont {Basov}},\ }\href@noop {} {\bibfield
    {journal} {\bibinfo  {journal} {Nature}\ }\textbf {\bibinfo {volume} {487}},\
    \bibinfo {pages} {82} (\bibinfo {year} {2012})}\BibitemShut {NoStop}%
  \bibitem [{\citenamefont {Chen}\ \emph {et~al.}(2012)\citenamefont {Chen},
    \citenamefont {Badioli}, \citenamefont {Alonso-Gonz{\'a}lez}, \citenamefont
    {Thongrattanasiri}, \citenamefont {Huth}, \citenamefont {Osmond},
    \citenamefont {Spasenovi{\'c}}, \citenamefont {Centeno}, \citenamefont
    {Pesquera}, \citenamefont {Godignon}, \citenamefont {Zurutuza~Elorza},
    \citenamefont {Camara}, \citenamefont {de~Abajo}, \citenamefont
    {Hillenbrand},\ and\ \citenamefont {Koppens}}]{Chen12}%
  \BibitemOpen
  \bibfield  {author} {\bibinfo {author} {\bibfnamefont {J.}~\bibnamefont
      {Chen}}, \bibinfo {author} {\bibfnamefont {M.}~\bibnamefont {Badioli}},
    \bibinfo {author} {\bibfnamefont {P.}~\bibnamefont {Alonso-Gonz{\'a}lez}},
    \bibinfo {author} {\bibfnamefont {S.}~\bibnamefont {Thongrattanasiri}},
    \bibinfo {author} {\bibfnamefont {F.}~\bibnamefont {Huth}}, \bibinfo {author}
    {\bibfnamefont {J.}~\bibnamefont {Osmond}}, \bibinfo {author} {\bibfnamefont
      {M.}~\bibnamefont {Spasenovi{\'c}}}, \bibinfo {author} {\bibfnamefont
      {A.}~\bibnamefont {Centeno}}, \bibinfo {author} {\bibfnamefont
      {A.}~\bibnamefont {Pesquera}}, \bibinfo {author} {\bibfnamefont
      {P.}~\bibnamefont {Godignon}}, \bibinfo {author} {\bibfnamefont
      {A.}~\bibnamefont {Zurutuza~Elorza}}, \bibinfo {author} {\bibfnamefont
      {N.}~\bibnamefont {Camara}}, \bibinfo {author} {\bibfnamefont {F.~J.~G.}\
      \bibnamefont {de~Abajo}}, \bibinfo {author} {\bibfnamefont {R.}~\bibnamefont
      {Hillenbrand}},\ and\ \bibinfo {author} {\bibfnamefont {F.~H.~L.}\
      \bibnamefont {Koppens}},\ }\href@noop {} {\bibfield  {journal} {\bibinfo
      {journal} {Nature}\ }\textbf {\bibinfo {volume} {487}},\ \bibinfo {pages}
    {77} (\bibinfo {year} {2012})}\BibitemShut {NoStop}%
  \bibitem [{\citenamefont {R{\"o}sner}\ \emph {et~al.}(2016)\citenamefont
    {R{\"o}sner}, \citenamefont {Steinke}, \citenamefont {Lorke}, \citenamefont
    {Gies}, \citenamefont {Jahnke},\ and\ \citenamefont {Wehling}}]{Rosner16}%
  \BibitemOpen
  \bibfield  {author} {\bibinfo {author} {\bibfnamefont {M.}~\bibnamefont
      {R{\"o}sner}}, \bibinfo {author} {\bibfnamefont {C.}~\bibnamefont {Steinke}},
    \bibinfo {author} {\bibfnamefont {M.}~\bibnamefont {Lorke}}, \bibinfo
    {author} {\bibfnamefont {C.}~\bibnamefont {Gies}}, \bibinfo {author}
    {\bibfnamefont {F.}~\bibnamefont {Jahnke}},\ and\ \bibinfo {author}
    {\bibfnamefont {T.~O.}\ \bibnamefont {Wehling}},\ }\href@noop {} {\bibfield
    {journal} {\bibinfo  {journal} {Nano Lett.}\ }\textbf {\bibinfo {volume}
      {16}},\ \bibinfo {pages} {2322} (\bibinfo {year} {2016})}\BibitemShut
  {NoStop}%
  \bibitem [{\citenamefont {Jiang}\ \emph {et~al.}(2021)\citenamefont {Jiang},
    \citenamefont {Haas},\ and\ \citenamefont {R\"osner}}]{Jiang21}%
  \BibitemOpen
  \bibfield  {author} {\bibinfo {author} {\bibfnamefont {Z.}~\bibnamefont
      {Jiang}}, \bibinfo {author} {\bibfnamefont {S.}~\bibnamefont {Haas}},\ and\
    \bibinfo {author} {\bibfnamefont {M.}~\bibnamefont {R\"osner}},\ }\href@noop
  {} {\bibfield  {journal} {\bibinfo  {journal} {2D Mater.}\ }\textbf {\bibinfo
      {volume} {8}},\ \bibinfo {pages} {035037} (\bibinfo {year}
    {2021})}\BibitemShut {NoStop}%
  \bibitem [{\citenamefont {Westerhout}\ \emph {et~al.}(2018)\citenamefont
    {Westerhout}, \citenamefont {van Veen}, \citenamefont {Katsnelson},\ and\
    \citenamefont {Yuan}}]{Westerhout18}%
  \BibitemOpen
  \bibfield  {author} {\bibinfo {author} {\bibfnamefont {T.}~\bibnamefont
      {Westerhout}}, \bibinfo {author} {\bibfnamefont {E.}~\bibnamefont {van
        Veen}}, \bibinfo {author} {\bibfnamefont {M.~I.}\ \bibnamefont
      {Katsnelson}},\ and\ \bibinfo {author} {\bibfnamefont {S.}~\bibnamefont
      {Yuan}},\ }\href@noop {} {\bibfield  {journal} {\bibinfo  {journal} {Phys.
        Rev. B}\ }\textbf {\bibinfo {volume} {97}},\ \bibinfo {pages} {205434}
    (\bibinfo {year} {2018})}\BibitemShut {NoStop}%
  \bibitem [{\citenamefont {Westerhout}\ \emph {et~al.}(2021)\citenamefont
    {Westerhout}, \citenamefont {Katsnelson},\ and\ \citenamefont
    {R\"osner}}]{Westerhout21}%
  \BibitemOpen
  \bibfield  {author} {\bibinfo {author} {\bibfnamefont {T.}~\bibnamefont
      {Westerhout}}, \bibinfo {author} {\bibfnamefont {M.~I.}\ \bibnamefont
      {Katsnelson}},\ and\ \bibinfo {author} {\bibfnamefont {M.}~\bibnamefont
      {R\"osner}},\ }\href@noop {} {\bibfield  {journal} {\bibinfo  {journal} {2D
        Mater.}\ }\textbf {\bibinfo {volume} {9}},\ \bibinfo {pages} {014004}
    (\bibinfo {year} {2021})}\BibitemShut {NoStop}%
  \bibitem [{\citenamefont {Reijnders}\ \emph {et~al.}(2022)\citenamefont
    {Reijnders}, \citenamefont {Tudorovskiy},\ and\ \citenamefont
    {Katsnelson}}]{Reijnders22}%
  \BibitemOpen
  \bibfield  {author} {\bibinfo {author} {\bibfnamefont {K.~J.~A.}\
      \bibnamefont {Reijnders}}, \bibinfo {author} {\bibfnamefont {T.}~\bibnamefont
      {Tudorovskiy}},\ and\ \bibinfo {author} {\bibfnamefont {M.~I.}\ \bibnamefont
      {Katsnelson}},\ }\href@noop {} {\bibfield  {journal} {\bibinfo  {journal}
      {Ann. Phys.}\ }\textbf {\bibinfo {volume} {446}},\ \bibinfo {pages} {169116}
    (\bibinfo {year} {2022})}\BibitemShut {NoStop}%
  \bibitem [{\citenamefont {Ishmukhametov}(1971)}]{Ishmukhametov71}%
  \BibitemOpen
  \bibfield  {author} {\bibinfo {author} {\bibfnamefont {B.~{\relax Kh}.}\
      \bibnamefont {Ishmukhametov}},\ }\href@noop {} {\bibfield  {journal}
    {\bibinfo  {journal} {Phys. Status Solidi (B)}\ }\textbf {\bibinfo {volume}
      {45}},\ \bibinfo {pages} {669} (\bibinfo {year} {1971})}\BibitemShut
  {NoStop}%
  \bibitem [{\citenamefont {Ishmukhametov}\ and\ \citenamefont
    {Katsnelson}(1975)}]{Ishmukhametov75}%
  \BibitemOpen
  \bibfield  {author} {\bibinfo {author} {\bibfnamefont {B.~{\relax Kh}.}\
      \bibnamefont {Ishmukhametov}}\ and\ \bibinfo {author} {\bibfnamefont {M.~I.}\
      \bibnamefont {Katsnelson}},\ }\href@noop {} {\bibfield  {journal} {\bibinfo
      {journal} {Fiz. Met. Metalloved.}\ }\textbf {\bibinfo {volume} {40}},\
    \bibinfo {pages} {736} (\bibinfo {year} {1975})}.\ \bibinfo {note} {(This
    reference is not available digitally and very hard to come by. The most
    essential parts are reproduced in Appendix D of
    Ref.~\cite{Reijnders22}.)}\BibitemShut {Stop}%
  \bibitem [{\citenamefont {Maslov}\ and\ \citenamefont
    {Fedoryuk}(1981)}]{Maslov81}%
  \BibitemOpen
  \bibfield  {author} {\bibinfo {author} {\bibfnamefont {V.~P.}\ \bibnamefont
      {Maslov}}\ and\ \bibinfo {author} {\bibfnamefont {M.~V.}\ \bibnamefont
      {Fedoryuk}},\ }\href@noop {} {\emph {\bibinfo {title} {Semi-classical
        approximation in quantum mechanics}}}\ (\bibinfo  {publisher} {Reidel,
    Dordrecht},\ \bibinfo {year} {1981})\BibitemShut {NoStop}%
  \bibitem [{\citenamefont {Guillemin}\ and\ \citenamefont
    {Sternberg}(1977)}]{Guillemin77}%
  \BibitemOpen
  \bibfield  {author} {\bibinfo {author} {\bibfnamefont {V.}~\bibnamefont
      {Guillemin}}\ and\ \bibinfo {author} {\bibfnamefont {S.}~\bibnamefont
      {Sternberg}},\ }\href@noop {} {\emph {\bibinfo {title} {Geometric
        asymptotics}}}\ (\bibinfo  {publisher} {American Mathematical Society,
    Providence, Rhode Island},\ \bibinfo {year} {1977})\BibitemShut {NoStop}%
  \bibitem [{\citenamefont {Martinez}(2002)}]{Martinez02}%
  \BibitemOpen
  \bibfield  {author} {\bibinfo {author} {\bibfnamefont {A.}~\bibnamefont
      {Martinez}},\ }\href@noop {} {\emph {\bibinfo {title} {An introduction to
        semiclassical and microlocal analysis}}}\ (\bibinfo  {publisher}
  {Springer-Verlag, New York},\ \bibinfo {year} {2002})\BibitemShut {NoStop}%
  \bibitem [{\citenamefont {Zworski}(2012)}]{Zworski12}%
  \BibitemOpen
  \bibfield  {author} {\bibinfo {author} {\bibfnamefont {M.}~\bibnamefont
      {Zworski}},\ }\href@noop {} {\emph {\bibinfo {title} {Semiclassical
        analysis}}}\ (\bibinfo  {publisher} {American Mathematical Society,
    Providence, Rhode Island},\ \bibinfo {year} {2012})\BibitemShut {NoStop}%
  \bibitem [{\citenamefont {Griffiths}(2005)}]{Griffiths05}%
  \BibitemOpen
  \bibfield  {author} {\bibinfo {author} {\bibfnamefont {D.~J.}\ \bibnamefont
      {Griffiths}},\ }\href@noop {} {\emph {\bibinfo {title} {Introduction to
        quantum mechanics}}},\ \bibinfo {edition} {2nd}\ ed.\ (\bibinfo  {publisher}
  {Pearson Prentice Hall, Upper Saddle River},\ \bibinfo {year}
  {2005})\BibitemShut {NoStop}%
  \bibitem [{\citenamefont {Heading}(1962)}]{Heading62}%
  \BibitemOpen
  \bibfield  {author} {\bibinfo {author} {\bibfnamefont {J.}~\bibnamefont
      {Heading}},\ }\href@noop {} {\emph {\bibinfo {title} {An Introduction to
        Phase-Integral Methods}}}\ (\bibinfo  {publisher} {Methuen, London},\
  \bibinfo {year} {1962})\BibitemShut {NoStop}%
  \bibitem [{\citenamefont {Benedikter}(2022)}]{Benedikter22}%
  \BibitemOpen
  \bibfield  {author} {\bibinfo {author} {\bibfnamefont {N.}~\bibnamefont
      {Benedikter}},\ }\href@noop {} {\bibfield  {journal} {\bibinfo  {journal}
      {Journal of Mathematical Physics}\ }\textbf {\bibinfo {volume} {63}},\
    \bibinfo {pages} {081101} (\bibinfo {year} {2022})}\BibitemShut {NoStop}%
  \bibitem [{\citenamefont {Marzari}\ \emph {et~al.}(2012)\citenamefont
    {Marzari}, \citenamefont {Mostofi}, \citenamefont {Yates}, \citenamefont
    {Souza},\ and\ \citenamefont {Vanderbilt}}]{Marzari12}%
  \BibitemOpen
  \bibfield  {author} {\bibinfo {author} {\bibfnamefont {N.}~\bibnamefont
      {Marzari}}, \bibinfo {author} {\bibfnamefont {A.~A.}\ \bibnamefont
      {Mostofi}}, \bibinfo {author} {\bibfnamefont {J.~R.}\ \bibnamefont {Yates}},
    \bibinfo {author} {\bibfnamefont {I.}~\bibnamefont {Souza}},\ and\ \bibinfo
    {author} {\bibfnamefont {D.}~\bibnamefont {Vanderbilt}},\ }\href@noop {}
  {\bibfield  {journal} {\bibinfo  {journal} {Rev. Mod. Phys.}\ }\textbf
    {\bibinfo {volume} {84}},\ \bibinfo {pages} {1419} (\bibinfo {year}
    {2012})}\BibitemShut {NoStop}%
  \bibitem [{\citenamefont {Wehling}\ \emph {et~al.}(2011)\citenamefont
    {Wehling}, \citenamefont {{\c{S}}a{\c{s}}{\i}o{\u{g}}lu}, \citenamefont
    {Friedrich}, \citenamefont {Lichtenstein}, \citenamefont {Katsnelson},\ and\
    \citenamefont {Bl{\"u}gel}}]{Wehling11}%
  \BibitemOpen
  \bibfield  {author} {\bibinfo {author} {\bibfnamefont {T.~O.}\ \bibnamefont
      {Wehling}}, \bibinfo {author} {\bibfnamefont {E.}~\bibnamefont
      {{\c{S}}a{\c{s}}{\i}o{\u{g}}lu}}, \bibinfo {author} {\bibfnamefont
      {C.}~\bibnamefont {Friedrich}}, \bibinfo {author} {\bibfnamefont {A.~I.}\
      \bibnamefont {Lichtenstein}}, \bibinfo {author} {\bibfnamefont {M.~I.}\
      \bibnamefont {Katsnelson}},\ and\ \bibinfo {author} {\bibfnamefont
      {S.}~\bibnamefont {Bl{\"u}gel}},\ }\href@noop {} {\bibfield  {journal}
    {\bibinfo  {journal} {Phys. Rev. Lett.}\ }\textbf {\bibinfo {volume} {106}},\
    \bibinfo {pages} {236805} (\bibinfo {year} {2011})}\BibitemShut {NoStop}%
  \bibitem [{\citenamefont {Fei}\ \emph {et~al.}(2011)\citenamefont {Fei},
    \citenamefont {Andreev}, \citenamefont {Bao}, \citenamefont {Zhang},
    \citenamefont {McLeod}, \citenamefont {Wang}, \citenamefont {Stewart},
    \citenamefont {Zhao}, \citenamefont {Dominguez}, \citenamefont {Thiemens},
    \citenamefont {Fogler}, \citenamefont {Tauber}, \citenamefont {Castro-Neto},
    \citenamefont {Lau}, \citenamefont {Keilmann},\ and\ \citenamefont
    {Basov}}]{Fei11}%
  \BibitemOpen
  \bibfield  {author} {\bibinfo {author} {\bibfnamefont {Z.}~\bibnamefont
      {Fei}}, \bibinfo {author} {\bibfnamefont {G.~O.}\ \bibnamefont {Andreev}},
    \bibinfo {author} {\bibfnamefont {W.}~\bibnamefont {Bao}}, \bibinfo {author}
    {\bibfnamefont {L.~M.}\ \bibnamefont {Zhang}}, \bibinfo {author}
    {\bibfnamefont {A.~S.}\ \bibnamefont {McLeod}}, \bibinfo {author}
    {\bibfnamefont {C.}~\bibnamefont {Wang}}, \bibinfo {author} {\bibfnamefont
      {M.~K.}\ \bibnamefont {Stewart}}, \bibinfo {author} {\bibfnamefont
      {Z.}~\bibnamefont {Zhao}}, \bibinfo {author} {\bibfnamefont {G.}~\bibnamefont
      {Dominguez}}, \bibinfo {author} {\bibfnamefont {M.}~\bibnamefont {Thiemens}},
    \bibinfo {author} {\bibfnamefont {M.~M.}\ \bibnamefont {Fogler}}, \bibinfo
    {author} {\bibfnamefont {M.~J.}\ \bibnamefont {Tauber}}, \bibinfo {author}
    {\bibfnamefont {A.~H.}\ \bibnamefont {Castro-Neto}}, \bibinfo {author}
    {\bibfnamefont {C.~N.}\ \bibnamefont {Lau}}, \bibinfo {author} {\bibfnamefont
      {F.}~\bibnamefont {Keilmann}},\ and\ \bibinfo {author} {\bibfnamefont
      {D.~N.}\ \bibnamefont {Basov}},\ }\href@noop {} {\bibfield  {journal}
    {\bibinfo  {journal} {Nano Letters}\ }\textbf {\bibinfo {volume} {11}},\
    \bibinfo {pages} {4701} (\bibinfo {year} {2011})}\BibitemShut {NoStop}%
  \bibitem [{\citenamefont {Liu}\ \emph {et~al.}(2008)\citenamefont {Liu},
    \citenamefont {Willis}, \citenamefont {Emtsev},\ and\ \citenamefont
    {Seyller}}]{Liu08}%
  \BibitemOpen
  \bibfield  {author} {\bibinfo {author} {\bibfnamefont {Y.}~\bibnamefont
      {Liu}}, \bibinfo {author} {\bibfnamefont {R.~F.}\ \bibnamefont {Willis}},
    \bibinfo {author} {\bibfnamefont {K.~V.}\ \bibnamefont {Emtsev}},\ and\
    \bibinfo {author} {\bibfnamefont {{\relax Th}.}~\bibnamefont {Seyller}},\
  }\href@noop {} {\bibfield  {journal} {\bibinfo  {journal} {Phys. Rev. B}\
    }\textbf {\bibinfo {volume} {78}},\ \bibinfo {pages} {201403(R)} (\bibinfo
    {year} {2008})}\BibitemShut {NoStop}%
  \bibitem [{\citenamefont {Berlyand}\ and\ \citenamefont
    {Dobrokhotov}(1987)}]{Berlyand87}%
  \BibitemOpen
  \bibfield  {author} {\bibinfo {author} {\bibfnamefont {L.~V.}\ \bibnamefont
      {Berlyand}}\ and\ \bibinfo {author} {\bibfnamefont {S.~{\relax Yu}.}\
      \bibnamefont {Dobrokhotov}},\ }\href@noop {} {\bibfield  {journal} {\bibinfo
      {journal} {Dokl. Akad. Nauk SSSR}\ }\textbf {\bibinfo {volume} {296}},\
    \bibinfo {pages} {80} (\bibinfo {year} {1987})} \bibinfo {note} {[Sov. Phys. Dokl. \textbf{32}, 714 (1987)]}\BibitemShut {NoStop}%
  \bibitem [{\citenamefont {Belov}\ \emph {et~al.}(2006)\citenamefont {Belov},
    \citenamefont {Dobrokhotov},\ and\ \citenamefont {Tudorovskiy}}]{Belov06}%
  \BibitemOpen
  \bibfield  {author} {\bibinfo {author} {\bibfnamefont {V.~V.}\ \bibnamefont
      {Belov}}, \bibinfo {author} {\bibfnamefont {S.~{\relax Yu}.}\ \bibnamefont
      {Dobrokhotov}},\ and\ \bibinfo {author} {\bibfnamefont {T.~{\relax Ya}.}\
      \bibnamefont {Tudorovskiy}},\ }\href@noop {} {\bibfield  {journal} {\bibinfo
      {journal} {J. Eng. Math.}\ }\textbf {\bibinfo {volume} {55}},\ \bibinfo
    {pages} {183} (\bibinfo {year} {2006})}\BibitemShut {NoStop}%
  \bibitem [{\citenamefont {Arnold}\ \emph {et~al.}(1982)\citenamefont {Arnold},
    \citenamefont {Gusein-Zade},\ and\ \citenamefont {Varchenko}}]{Arnold82}%
  \BibitemOpen
  \bibfield  {author} {\bibinfo {author} {\bibfnamefont {V.~I.}\ \bibnamefont
      {Arnold}}, \bibinfo {author} {\bibfnamefont {S.~M.}\ \bibnamefont
      {Gusein-Zade}},\ and\ \bibinfo {author} {\bibfnamefont {A.~N.}\ \bibnamefont
      {Varchenko}},\ }\href@noop {} {\emph {\bibinfo {title} {Singularities of 
      Differentiable Maps, Volume 1, The Classification of Critical Points,
      Caustics and Wave Fronts}}}\ (\bibinfo  {publisher}
  {Birkh\"auser, Basel},\ \bibinfo {year} {1985})\BibitemShut {NoStop}%
  \bibitem [{\citenamefont {Reijnders}\ \emph {et~al.}(2018)\citenamefont
    {Reijnders}, \citenamefont {Minenkov}, \citenamefont {Katsnelson},\ and\
    \citenamefont {Dobrokhotov}}]{Reijnders18}%
  \BibitemOpen
  \bibfield  {author} {\bibinfo {author} {\bibfnamefont {K.~J.~A.}\
      \bibnamefont {Reijnders}}, \bibinfo {author} {\bibfnamefont {D.~S.}\
      \bibnamefont {Minenkov}}, \bibinfo {author} {\bibfnamefont {M.~I.}\
      \bibnamefont {Katsnelson}},\ and\ \bibinfo {author} {\bibfnamefont
      {S.~{\relax {Yu}}.}\ \bibnamefont {Dobrokhotov}},\ }\href@noop {} {\bibfield
    {journal} {\bibinfo  {journal} {Ann. Phys.}\ }\textbf {\bibinfo {volume}
      {397}},\ \bibinfo {pages} {65} (\bibinfo {year} {2018})}\BibitemShut
  {NoStop}%
  \bibitem [{\citenamefont {Dobrokhotov}\ and\ \citenamefont
    {Zhevandrov}(2003)}]{Dobrokhotov03}%
  \BibitemOpen
  \bibfield  {author} {\bibinfo {author} {\bibfnamefont {S.~{\relax Yu}.}\
      \bibnamefont {Dobrokhotov}}\ and\ \bibinfo {author} {\bibfnamefont {P.~N.}\
      \bibnamefont {Zhevandrov}},\ }\href@noop {} {\bibfield  {journal} {\bibinfo
      {journal} {Russ. J. Math. Phys.}\ }\textbf {\bibinfo {volume} {10}},\
    \bibinfo {pages} {1} (\bibinfo {year} {2003})}\BibitemShut {NoStop}%
  \bibitem [{\citenamefont {Lieb}(1981)}]{Lieb81}%
  \BibitemOpen
  \bibfield  {author} {\bibinfo {author} {\bibfnamefont {E.~H.}\ \bibnamefont
      {Lieb}},\ }\href@noop {} {\bibfield  {journal} {\bibinfo  {journal} {Rev.
        Mod. Phys.}\ }\textbf {\bibinfo {volume} {53}},\ \bibinfo {pages} {603}
    (\bibinfo {year} {1981})}\BibitemShut {NoStop}%
  \bibitem [{\citenamefont {Jackson}(1999)}]{Jackson99}%
  \BibitemOpen
  \bibfield  {author} {\bibinfo {author} {\bibfnamefont {J.~D.}\ \bibnamefont
      {Jackson}},\ }\href@noop {} {\emph {\bibinfo {title} {Classical
        electrodynamics}}}\ (\bibinfo  {publisher} {Wiley, New York},\ \bibinfo
  {year} {1999})\BibitemShut {NoStop}%
  \bibitem [{\citenamefont {Arnold}(1989)}]{Arnold89}%
  \BibitemOpen
  \bibfield  {author} {\bibinfo {author} {\bibfnamefont {V.~I.}\ \bibnamefont
      {Arnold}},\ }\href@noop {} {\emph {\bibinfo {title} {Mathematical methods of
        classical mechanics}}},\ \bibinfo {edition} {2nd}\ ed.\ (\bibinfo
  {publisher} {Springer, New York},\ \bibinfo {year} {1989})\BibitemShut
  {NoStop}%
  \bibitem [{\citenamefont {Goldstein}\ \emph {et~al.}(2002)\citenamefont
    {Goldstein}, \citenamefont {Poole},\ and\ \citenamefont
    {Safko}}]{Goldstein02}%
  \BibitemOpen
  \bibfield  {author} {\bibinfo {author} {\bibfnamefont {H.}~\bibnamefont
      {Goldstein}}, \bibinfo {author} {\bibfnamefont {C.~P.}\ \bibnamefont
      {Poole}},\ and\ \bibinfo {author} {\bibfnamefont {J.~L.}\ \bibnamefont
      {Safko}},\ }\href@noop {} {\emph {\bibinfo {title} {Classical Mechanics}}},\
  \bibinfo {edition} {3rd}\ ed.\ (\bibinfo  {publisher} {Addison Wesley, San
    Fransisco},\ \bibinfo {year} {2002})\BibitemShut {NoStop}%
  \bibitem [{\citenamefont {Poston}\ and\ \citenamefont
    {Stewart}(1978)}]{Poston78}%
  \BibitemOpen
  \bibfield  {author} {\bibinfo {author} {\bibfnamefont {T.}~\bibnamefont
      {Poston}}\ and\ \bibinfo {author} {\bibfnamefont {I.~N.}\ \bibnamefont
      {Stewart}},\ }\href@noop {} {\emph {\bibinfo {title} {Catastrophe theory and
        its applications}}}\ (\bibinfo  {publisher} {Pitman, Boston},\ \bibinfo
  {year} {1978})\BibitemShut {NoStop}%
  \bibitem [{\citenamefont {Landau}\ \emph {et~al.}(1984)\citenamefont {Landau},
    \citenamefont {Lifshitz},\ and\ \citenamefont {Pitaevskii}}]{Landau84}%
  \BibitemOpen
  \bibfield  {author} {\bibinfo {author} {\bibfnamefont {L.~D.}\ \bibnamefont
      {Landau}}, \bibinfo {author} {\bibfnamefont {E.~M.}\ \bibnamefont
      {Lifshitz}},\ and\ \bibinfo {author} {\bibfnamefont {L.~P.}\ \bibnamefont
      {Pitaevskii}},\ }\href@noop {} {\emph {\bibinfo {title} {Electrodynamics of
        continuous media}}}\ (\bibinfo  {publisher} {Pergamon Press, Oxford},\
  \bibinfo {year} {1984})\BibitemShut {NoStop}%
  \bibitem [{\citenamefont {Ando}\ \emph {et~al.}(1982)\citenamefont {Ando},
    \citenamefont {Fowler},\ and\ \citenamefont {Stern}}]{Ando82}%
  \BibitemOpen
  \bibfield  {author} {\bibinfo {author} {\bibfnamefont {T.}~\bibnamefont
      {Ando}}, \bibinfo {author} {\bibfnamefont {A.~B.}\ \bibnamefont {Fowler}},\
    and\ \bibinfo {author} {\bibfnamefont {F.}~\bibnamefont {Stern}},\
  }\href@noop {} {\bibfield  {journal} {\bibinfo  {journal} {Rev. Mod. Phys.}\
    }\textbf {\bibinfo {volume} {54}},\ \bibinfo {pages} {437} (\bibinfo {year}
    {1982})}\BibitemShut {NoStop}%
  \bibitem [{\citenamefont {Torre}\ \emph {et~al.}(2017)\citenamefont {Torre},
    \citenamefont {Katsnelson}, \citenamefont {Diaspro}, \citenamefont
    {Pellegrini},\ and\ \citenamefont {Polini}}]{Torre17}%
  \BibitemOpen
  \bibfield  {author} {\bibinfo {author} {\bibfnamefont {I.}~\bibnamefont
      {Torre}}, \bibinfo {author} {\bibfnamefont {M.~I.}\ \bibnamefont
      {Katsnelson}}, \bibinfo {author} {\bibfnamefont {A.}~\bibnamefont {Diaspro}},
    \bibinfo {author} {\bibfnamefont {V.}~\bibnamefont {Pellegrini}},\ and\
    \bibinfo {author} {\bibfnamefont {M.}~\bibnamefont {Polini}},\ }\href@noop {}
  {\bibfield  {journal} {\bibinfo  {journal} {Phys. Rev. B}\ }\textbf {\bibinfo
      {volume} {96}},\ \bibinfo {pages} {035433} (\bibinfo {year}
    {2017})}\BibitemShut {NoStop}%
  \bibitem [{\citenamefont {Taylor}(1972)}]{Taylor72}%
  \BibitemOpen
  \bibfield  {author} {\bibinfo {author} {\bibfnamefont {J.~R.}\ \bibnamefont
      {Taylor}},\ }\href@noop {} {\emph {\bibinfo {title} {Scattering theory: the
        quantum theory of nonrelativistic collisions}}}\ (\bibinfo  {publisher}
  {Dover publications, New York},\ \bibinfo {year} {1972})\BibitemShut
  {NoStop}%
  \bibitem [{\citenamefont {Abromowitz}\ and\ \citenamefont
    {Stegun}(1964)}]{Abromowitz64}%
  \BibitemOpen
  \bibfield  {author} {\bibinfo {author} {\bibfnamefont {M.}~\bibnamefont
      {Abromowitz}}\ and\ \bibinfo {author} {\bibfnamefont {I.}~\bibnamefont
      {Stegun}},\ }\href@noop {} {\emph {\bibinfo {title} {Handbook of Mathematical
        Functions}}}\ (\bibinfo  {publisher} {National Bureau of Standards, New
    York},\ \bibinfo {year} {1964})\ pp.\ \bibinfo {pages} {358--364}\BibitemShut
  {NoStop}%
  \bibitem [{\citenamefont {Vainberg}(1989)}]{Vainberg89}%
  \BibitemOpen
  \bibfield  {author} {\bibinfo {author} {\bibfnamefont {B.}~\bibnamefont
      {Vainberg}},\ }\href@noop {} {\emph {\bibinfo {title} {Asymptotic methods in
        equations of mathematical physics}}}\ (\bibinfo  {publisher} {Gordon and
    Breach Science Publishers, New York},\ \bibinfo {year} {1989})\BibitemShut
  {NoStop}%
  \bibitem [{\citenamefont {Sommerfeld}(1949)}]{Sommerfeld49}%
  \BibitemOpen
  \bibfield  {author} {\bibinfo {author} {\bibfnamefont {A.}~\bibnamefont
      {Sommerfeld}},\ }\href@noop {} {\emph {\bibinfo {title} {Partial differential
        equations in physics}}}\ (\bibinfo  {publisher} {Academic press, New York},\
  \bibinfo {year} {1949})\BibitemShut {NoStop}%
  \bibitem [{\citenamefont {Schot}(1992)}]{Schot92}%
  \BibitemOpen
  \bibfield  {author} {\bibinfo {author} {\bibfnamefont {S.~H.}\ \bibnamefont
      {Schot}},\ }\href@noop {} {\bibfield  {journal} {\bibinfo  {journal} {Hist.
        Math.}\ }\textbf {\bibinfo {volume} {19}},\ \bibinfo {pages} {385} (\bibinfo
    {year} {1992})}\BibitemShut {NoStop}%
  \bibitem [{\citenamefont {Berry}\ and\ \citenamefont {Mount}(1972)}]{Berry72}%
  \BibitemOpen
  \bibfield  {author} {\bibinfo {author} {\bibfnamefont {M.~V.}\ \bibnamefont
      {Berry}}\ and\ \bibinfo {author} {\bibfnamefont {K.~E.}\ \bibnamefont
      {Mount}},\ }\href@noop {} {\bibfield  {journal} {\bibinfo  {journal} {Rep.
        Progr. Phys.}\ }\textbf {\bibinfo {volume} {35}},\ \bibinfo {pages} {315}
    (\bibinfo {year} {1972})}\BibitemShut {NoStop}%
  \bibitem [{\citenamefont {Curtis}\ and\ \citenamefont
    {Ellis}(2004)}]{Curtis04}%
  \BibitemOpen
  \bibfield  {author} {\bibinfo {author} {\bibfnamefont {L.~J.}\ \bibnamefont
      {Curtis}}\ and\ \bibinfo {author} {\bibfnamefont {D.~G.}\ \bibnamefont
      {Ellis}},\ }\href@noop {} {\bibfield  {journal} {\bibinfo  {journal} {Am. J.
        Phys.}\ }\textbf {\bibinfo {volume} {72}},\ \bibinfo {pages} {1521} (\bibinfo
    {year} {2004})}\BibitemShut {NoStop}%
  \bibitem [{\citenamefont {Arnold}(1967)}]{Arnold67}%
  \BibitemOpen
  \bibfield  {author} {\bibinfo {author} {\bibfnamefont {V.~I.}\ \bibnamefont
      {Arnold}},\ }\href@noop {} {\bibfield  {journal} {\bibinfo  {journal} {Funct.
        Anal. Appl.}\ }\textbf {\bibinfo {volume} {1}},\ \bibinfo {pages} {1}
    (\bibinfo {year} {1967})}\BibitemShut {NoStop}%
  \bibitem [{\citenamefont {Langer}(1937)}]{Langer37}%
  \BibitemOpen
  \bibfield  {author} {\bibinfo {author} {\bibfnamefont {R.~E.}\ \bibnamefont
      {Langer}},\ }\href@noop {} {\bibfield  {journal} {\bibinfo  {journal} {Phys.
        Rev.}\ }\textbf {\bibinfo {volume} {51}},\ \bibinfo {pages} {669} (\bibinfo
    {year} {1937})}\BibitemShut {NoStop}%
  \bibitem [{\citenamefont {Fl\"ugge}(1994)}]{Fluegge94}%
  \BibitemOpen
  \bibfield  {author} {\bibinfo {author} {\bibfnamefont {S.}~\bibnamefont
      {Fl\"ugge}},\ }\href@noop {} {\emph {\bibinfo {title} {Practical Quantum
        Mechanics}}}\ (\bibinfo  {publisher} {Springer-Verlag, Berlin Heidelberg},\
  \bibinfo {year} {1994})\BibitemShut {NoStop}%
  \bibitem [{Mat(2023)}]{Mathematica}%
  \BibitemOpen
  \href@noop {} {\bibinfo {title} {Wolfram {M}athematica}} (\bibinfo {year}
  {1988--2023}),\ \bibinfo {note} {\url{www.wolfram.com/mathematica}. The
    computations in this article were performed with version
    13.0.0.0}\BibitemShut {NoStop}%
  \bibitem [{\citenamefont {R{\"o}sner}\ \emph {et~al.}(2015)\citenamefont
    {R{\"o}sner}, \citenamefont {{\c{S}}a{\c{s}}{\i}o{\u{g}}lu}, \citenamefont
    {Friedrich}, \citenamefont {Bl{\"u}gel},\ and\ \citenamefont
    {Wehling}}]{Rosner15}%
  \BibitemOpen
  \bibfield  {author} {\bibinfo {author} {\bibfnamefont {M.}~\bibnamefont
      {R{\"o}sner}}, \bibinfo {author} {\bibfnamefont {E.}~\bibnamefont
      {{\c{S}}a{\c{s}}{\i}o{\u{g}}lu}}, \bibinfo {author} {\bibfnamefont
      {C.}~\bibnamefont {Friedrich}}, \bibinfo {author} {\bibfnamefont
      {S.}~\bibnamefont {Bl{\"u}gel}},\ and\ \bibinfo {author} {\bibfnamefont
      {T.~O.}\ \bibnamefont {Wehling}},\ }\href@noop {} {\bibfield  {journal}
    {\bibinfo  {journal} {Phys. Rev. B}\ }\textbf {\bibinfo {volume} {92}},\
    \bibinfo {pages} {085102} (\bibinfo {year} {2015})}\BibitemShut {NoStop}%
  \bibitem [{\citenamefont {Andersen}\ \emph {et~al.}(2015)\citenamefont
    {Andersen}, \citenamefont {Latini},\ and\ \citenamefont
    {Thygesen}}]{Andersen15}%
  \BibitemOpen
  \bibfield  {author} {\bibinfo {author} {\bibfnamefont {K.}~\bibnamefont
      {Andersen}}, \bibinfo {author} {\bibfnamefont {S.}~\bibnamefont {Latini}},\
    and\ \bibinfo {author} {\bibfnamefont {K.~S.}\ \bibnamefont {Thygesen}},\
  }\href@noop {} {\bibfield  {journal} {\bibinfo  {journal} {Nano Lett.}\
    }\textbf {\bibinfo {volume} {15}},\ \bibinfo {pages} {4616} (\bibinfo {year}
    {2015})}\BibitemShut {NoStop}%
  \bibitem [{\citenamefont {Cho}\ and\ \citenamefont {Berkelbach}(2018)}]{Cho18}%
  \BibitemOpen
  \bibfield  {author} {\bibinfo {author} {\bibfnamefont {Y.}~\bibnamefont
      {Cho}}\ and\ \bibinfo {author} {\bibfnamefont {T.~C.}\ \bibnamefont
      {Berkelbach}},\ }\href@noop {} {\bibfield  {journal} {\bibinfo  {journal}
      {Phys. Rev. B}\ }\textbf {\bibinfo {volume} {97}},\ \bibinfo {pages}
    {041409(R)} (\bibinfo {year} {2018})}\BibitemShut {NoStop}%
  \bibitem [{\citenamefont {van Schilfgaarde}\ and\ \citenamefont
    {Katsnelson}(2011)}]{VanSchilfgaarde11}%
  \BibitemOpen
  \bibfield  {author} {\bibinfo {author} {\bibfnamefont {M.}~\bibnamefont {van
        Schilfgaarde}}\ and\ \bibinfo {author} {\bibfnamefont {M.~I.}\ \bibnamefont
      {Katsnelson}},\ }\href@noop {} {\bibfield  {journal} {\bibinfo  {journal}
      {Phys. Rev. B}\ }\textbf {\bibinfo {volume} {83}},\ \bibinfo {pages}
    {081409(R)} (\bibinfo {year} {2011})}\BibitemShut {NoStop}%
  \bibitem [{\citenamefont {Asgari}\ \emph {et~al.}(2014)\citenamefont {Asgari},
    \citenamefont {Katsnelson},\ and\ \citenamefont {Polini}}]{Asgari14}%
  \BibitemOpen
  \bibfield  {author} {\bibinfo {author} {\bibfnamefont {R.}~\bibnamefont
      {Asgari}}, \bibinfo {author} {\bibfnamefont {M.~I.}\ \bibnamefont
      {Katsnelson}},\ and\ \bibinfo {author} {\bibfnamefont {M.}~\bibnamefont
      {Polini}},\ }\href@noop {} {\bibfield  {journal} {\bibinfo  {journal} {Ann.
        Phys. (Berl.)}\ }\textbf {\bibinfo {volume} {526}},\ \bibinfo {pages} {359}
    (\bibinfo {year} {2014})}\BibitemShut {NoStop}%
  \bibitem [{\citenamefont {Nikitin}\ \emph {et~al.}(2011)\citenamefont
    {Nikitin}, \citenamefont {Guinea}, \citenamefont {Garc\'{\i}a-Vidal},\ and\
    \citenamefont {Mart\'{\i}n-Moreno}}]{Nikitin11}%
  \BibitemOpen
  \bibfield  {author} {\bibinfo {author} {\bibfnamefont {A.~Y.}\ \bibnamefont
      {Nikitin}}, \bibinfo {author} {\bibfnamefont {F.}~\bibnamefont {Guinea}},
    \bibinfo {author} {\bibfnamefont {F.~J.}\ \bibnamefont {Garc\'{\i}a-Vidal}},\
    and\ \bibinfo {author} {\bibfnamefont {L.}~\bibnamefont
      {Mart\'{\i}n-Moreno}},\ }\href@noop {} {\bibfield  {journal} {\bibinfo
      {journal} {Phys. Rev. B}\ }\textbf {\bibinfo {volume} {84}},\ \bibinfo
    {pages} {161407(R)} (\bibinfo {year} {2011})}\BibitemShut {NoStop}%
  \bibitem [{\citenamefont {Christensen}\ \emph {et~al.}(2012)\citenamefont
    {Christensen}, \citenamefont {Manjavacas}, \citenamefont {Thongrattanasiri},
    \citenamefont {Koppens},\ and\ \citenamefont {García~de
      Abajo}}]{Christensen12}%
  \BibitemOpen
  \bibfield  {author} {\bibinfo {author} {\bibfnamefont {J.}~\bibnamefont
      {Christensen}}, \bibinfo {author} {\bibfnamefont {A.}~\bibnamefont
      {Manjavacas}}, \bibinfo {author} {\bibfnamefont {S.}~\bibnamefont
      {Thongrattanasiri}}, \bibinfo {author} {\bibfnamefont {F.~H.~L.}\
      \bibnamefont {Koppens}},\ and\ \bibinfo {author} {\bibfnamefont {F.~J.}\
      \bibnamefont {García~de Abajo}},\ }\href@noop {} {\bibfield  {journal}
    {\bibinfo  {journal} {ACS Nano}\ }\textbf {\bibinfo {volume} {6}},\ \bibinfo
    {pages} {431} (\bibinfo {year} {2012})}\BibitemShut {NoStop}%
  \bibitem [{\citenamefont {Prishchenko}\ \emph {et~al.}(2017)\citenamefont
    {Prishchenko}, \citenamefont {Mazurenko}, \citenamefont {Katsnelson},\ and\
    \citenamefont {Rudenko}}]{Prishchenko17}%
  \BibitemOpen
  \bibfield  {author} {\bibinfo {author} {\bibfnamefont {D.~A.}\ \bibnamefont
      {Prishchenko}}, \bibinfo {author} {\bibfnamefont {V.~G.}\ \bibnamefont
      {Mazurenko}}, \bibinfo {author} {\bibfnamefont {M.~I.}\ \bibnamefont
      {Katsnelson}},\ and\ \bibinfo {author} {\bibfnamefont {A.~N.}\ \bibnamefont
      {Rudenko}},\ }\href@noop {} {\bibfield  {journal} {\bibinfo  {journal} {2D
        Mater.}\ }\textbf {\bibinfo {volume} {4}},\ \bibinfo {pages} {025064}
    (\bibinfo {year} {2017})}\BibitemShut {NoStop}%
  \bibitem [{\citenamefont {Jin}\ \emph {et~al.}(2015)\citenamefont {Jin},
    \citenamefont {Rold\'an}, \citenamefont {Katsnelson},\ and\ \citenamefont
    {Yuan}}]{Jin15}%
  \BibitemOpen
  \bibfield  {author} {\bibinfo {author} {\bibfnamefont {F.}~\bibnamefont
      {Jin}}, \bibinfo {author} {\bibfnamefont {R.}~\bibnamefont {Rold\'an}},
    \bibinfo {author} {\bibfnamefont {M.~I.}\ \bibnamefont {Katsnelson}},\ and\
    \bibinfo {author} {\bibfnamefont {S.}~\bibnamefont {Yuan}},\ }\href@noop {}
  {\bibfield  {journal} {\bibinfo  {journal} {Phys. Rev. B}\ }\textbf {\bibinfo
      {volume} {92}},\ \bibinfo {pages} {115440} (\bibinfo {year}
    {2015})}\BibitemShut {NoStop}%
\end{thebibliography}
\end{document}